\documentclass{article}

\usepackage{amsfonts,amssymb}
\usepackage{amsmath}
\usepackage{mathtools}
\usepackage{epsfig}
\usepackage{float}
\usepackage{indentfirst}
\usepackage{color}
\usepackage{caption}
\usepackage{graphicx, subfigure}
\usepackage{multirow}
\usepackage[table]{xcolor}
\usepackage{rotating}
\usepackage{ulem}
\usepackage{xcolor,cancel}
\usepackage{tikz}
\usepackage{appendix}
\usepackage{comment}
\usepackage{yfonts}

\makeatletter
  \@addtoreset{section}{part}
  \@addtoreset{@ppsaveapp}{part}
\makeatother

\numberwithin{equation}{section}

\begin{document}

\begin{center}
{\bf ACTIVE CLOAKING OF RESONANT COATED INCLUSIONS FOR WAVES IN MEMBRANES AND KIRCHHOFF PLATES}
\end{center}

\begin{center}
J. O'NEILL$^1$, \"O. SELSIL$^1$, R.C. MCPHEDRAN$^{1,2}$, A.B. MOVCHAN$^1$,  N.V. MOVCHAN$^1$
\end{center}

\vspace{-0.8cm}

\begin{center}
and 
\end{center}

\vspace{-0.8cm}

\begin{center}
C. HENDERSON MOGGACH$^1$
\end{center}

\begin{center} 
$^1$ Department of Mathematical Sciences, University of Liverpool, Liverpool L69 7ZL, UK
\end{center}

\vspace{-0.8cm}

\begin{center}
$^2$ CUDOS ARC Centre of Excellence, School of Physics, University of Sydney, Sydney, \\New South Wales 2006, Australia
\end{center}

\begin{abstract}

The dynamic response of a coated inclusion is considered in the context of active cloaking. The active cloak is achieved for a  coated inclusion in the presence of membrane and  flexural waves.
In this paper, we 
investigate the design of an active cloak for  a coated inclusion in three frequency regimes: the very low frequency (monopole dominated) range, the intermediate range, and the higher frequency range in which scattering resonances occur.  
In the first of these ranges, we validate previous work, which resulted in a simple mass-compensation design for the monopole scatterer, while in the second and third ranges, a combination of the use of
an appropriate coating and the appropriate choice of the amplitudes of the active cloaking sources is necessary. We show that such cloaking can indeed be effective in the region of strong scattering resonances.  
We give closed form analytic expressions for the required amplitudes of the active cloaking sources in the three frequency regions and provide asymptotic estimates and numerical illustrations. 

\end{abstract}

\section{Introduction}
\label{Intro}

 Cloaking of objects from detection by waves governed by the Helmholtz equation is nowadays regarded as an interesting and important problem (see, for example, \cite{UL}, \cite{JBP_DS_DRS}). The cloaking methods, well-developed for the Helmholtz operator, have also been extended to advanced structured media in electromagnetism,  elasticity, as well as waves in fluids (see 
 \cite{ANN_FAA_WJP1}--
\cite{DJC_ISJ_NVM_ABM_MB_RCM}).
 The cloaking system can be understood as a by-pass structure, which routes a probe wave around an object, reducing the scattering by the object to a minimum.

 We make a distinction between passive and active cloaking.  Passive cloaks do not change their properties and remain unchanged for all frequencies and all types of incident waves. The drawback of a passive cloak is that its practical implementation can only be effective over a limited range of frequencies (see, for example, \cite{Alu_DoCloaks}, \cite{P-YC_CA_AA}). Active cloaks on the other hand can easily overcome this limitation in their effective frequency range, and can also adapt to various types of incident waves.  Active cloaking  relies on being able to  radiate compensating fields
 which cancel out perturbations of the incident or probe field by the object to be cloaked \cite{MS_GVE}. 
  
 In a set of papers \cite{GVF_GWM_DO_2009a}--\cite{GVF_GWM_DO_2011b},  
 a continuous model of an active cloak for the Laplace and Helmholtz equations was developed. The model was set as a variational problem and was analysed mathematically in depth, and the effectiveness of the active cloaking was demonstrated. Finite cloaking devices in the continuous framework were introduced to shield any objects within the given ``shielded'' region. 

In a recent paper \cite{JO_OS_RCM_ABM_NVM}, an analytical model of an approximate cloak for flexural waves was developed by introducing a finite number of active sources, whose complex amplitudes were chosen to suppress the scattered field resulting from an interaction of an incident wave with a finite obstacle. The multipole representation was used for the solution, and the required number of multipoles in the scattered field were constrained to be zero.
Although the truncation of the multipole representation introduces an approximation error, this approximation was shown to be entirely adequate for a sufficiently wide range of frequencies. 
 
In this paper, we investigate passive scattering reduction and active cloaking in the context of both the Helmhotz equation (applying to waves in elastic membranes) and the fourth-order Kirchhoff plate theory (applying to flexural waves in thin elastic plates).  
We study these phenomena in three frequency ranges: in the first the frequency is sufficiently low, so that wave scattering is dominated by the monopole term, for both membranes and Kirchhoff plates. We show that the following mass-compensation formula 
\begin{equation}
\rho_c = \frac{\rho_e - \rho_i(a_i/a_c)^2}{1 - (a_i/a_c)^2}, \label{mass_balance}
\end{equation}
is applicable to reduction of scattering by an appropriately chosen coating around the inclusion. Here  $\rho_i, \rho_c, \rho_e$ are mass densities of the inclusion,  coating, and ambient matrix, respectively, and $a_i, a_c$ are the interior and the exterior radii of the inclusion coating.
This classical formula \cite{W_Voigt} has been recently used \cite{MF_PYC_HB_SE_SG_AA} in the context of scattering reduction for coated inclusions at low frequencies for which the dominance of the monopole term guarantees the applicability of the criterion (\ref{mass_balance}) for both the membrane and Kirchhoff plate waves. We show here that, while this formula is effective for membrane waves in reducing the scattering from a second-order to a fourth-order effect, the same is not in general true for Kirchhoff waves, where for only very specific choices of material parameters, does the formula (\ref{mass_balance}) deliver the reduction of scattering to fourth-order. The failure of (\ref{mass_balance}) to deliver the reduction to fourth order scattering occurs because in general, an appropriate balance between the Helmholtz and modified Helmholtz parts of the Kirchhoff wave expansion requires suitable choices for flexural rigidities, not just densities. We also show that, because scattering in the low frequency range is very weak, active cloaking is not necessary. 

As the frequency increases dipole and higher order terms become significant. We show that the choice of coating parameters is not sufficient to generate effective scattering reduction in this intermediate frequency range. A small number of control sources then delivers effective cloaking.

As the frequency increases further, we enter the resonant scattering region. These resonances result in strongly enhanced scattering amplitudes, which change rapidly with frequency. This causes difficulties to arise for the application of active cloaking techniques, as suggested by the following argument. 

One strategy by which active cloaking might be defeated would be for the probe beam to be swept in frequency, so that the compensating fields would also have to be frequency-swept in corresponding fashion. A circumstance under which this would be difficult occurs when the object to be cloaked has resonant scattering properties. In the presence of resonances, it is difficult to avoid time lags between the change of the probe frequency and the corresponding changes required in the compensating fields around the cloaked object. These time lags inevitably increase with the sharpness (or Q factor) of the resonance. Thus, for active cloaking to work well near resonant frequencies, one needs to be able to modify the configuration  of the cloaked object, to avoid rapid changes in scattering properties with frequency. The configurational change we investigate here is the use of the coating around the inclusion
to deliver  a flat reflectance as a function of frequency, rather than a very low reflectance. (The latter is inevitably narrow-band at any but the lowest frequency range.) This we show enables the active cloaking of the coated inclusion to operate effectively over the frequency region of flat reflectance. Up to twelve active sources may be required, but good quality cloaking is always possible by this method.

 The paper is structured as follows. Section \ref{prob_form} includes the governing equations and the multipole representations of flexural waves near a coated inclusion. An illustration of resonance regimes and active cloaking by point sources is given in Section \ref{membrane} for the case of membrane waves, as a  simple precursor to the more involved discussion necessary for flexural waves. In Section \ref{flex_res}, it is shown  that active cloaking is still effective in resonance regimes at higher frequencies, as a result of the strategy described above of using an appropriate coating to move the inclusion resonances into the desired frequency range. A closed-form expression is given controlling the choice of inclusion and coating parameters, which delivers a zero monopole scattering coefficient. This generalises the classical result of Konenkov \cite{YuKK} to coated inclusions and active cloaking.  
 The Supplementary Material 
 gives a full analytic description of our cloaking method for both membranes and Kirchhoff plates.

\section{Problem formulation: scattering of flexural waves by a coated inclusion in a Kirchhoff plate}
\label{prob_form}

The out-of-plane elastic displacement $W^{(k)}({\bf x};t)$ in a Kirchhoff plate satisfies the equation of motion
\begin{equation}
\Delta^2 W^{(k)} ({\bf x}; t) + \frac{\rho_k h}{D_k} \ddot{W}^{(k)}({\bf x}; t) + {\cal F}({\bf x}; t)= 0, \quad {\bf x}=(x_1,x_2) \in {\Omega}_k, \quad k=i, c, e\label{gov_eqn},
\end{equation}
where ${\cal F}({\bf x}; t)$ represents active point sources placed in the exterior of the coated inclusion; the explicit representation for ${\cal F}$ and positioning of the point sources are discussed in Section \ref{active_section}.
We note that the super/sub-script $k$ is to be replaced by $i, c$ or $e$, depending on whether we are considering the inclusion, coating or exterior, respectively. In equation (\ref{gov_eqn}), $\Delta^2$ is the biharmonic operator, a dot on the variable denotes the derivative with respect to time $t$, $\rho$ is the mass density, $h$ is the thickness of the plate,
${\Omega}_k$ is the corresponding region, $D_k=E_kh^3/[12(1-\nu_k^2)]$ is the flexural rigidity, with $E_k$ the Young's modulus and $\nu_k$ the Poisson's ratio of the corresponding elastic material.

Assuming time-harmonic vibrations, that is $W^{(k)}({\bf x}; t) = w^{(k)}({\bf x}) \,{\rm exp (i\omega t)}$, outside the support of the function ${\cal F}({\bf x}; t)$ the governing equation (\ref{gov_eqn}) can be reduced to
\begin{equation}
\left(\Delta^2  - \frac{\rho_k h \,\omega^2}{D_k}\right) w^{(k)}({\bf x}) = \left(\Delta + \beta_k^2\right)\left(\Delta - \beta_k^2 \right) w^{(k)} ({\bf x})= 0, \quad {\bf x} \in {\Omega}_k, \quad k=i, c, e. \label{gov_eqn_simp}
\end{equation}
Here $\beta_k^2 = \omega\sqrt{\rho_k h / D_k}$ is the spectral parameter. Consequently, $w^{(k)} ({\bf x})$ can be written as the sum of solutions of the Helmholtz and modified Helmholtz equations
\begin{equation}
w^{(k)} ({\bf x})=  w^{(k)}_H ({\bf x}) + w^{(k)}_M ({\bf x}),
\label{w_eq_wHh_wMHh}
\end{equation}
where 
\begin{equation}
\Delta w^{(k)}_H ({\bf x}) + \beta_k^2 w^{(k)}_H ({\bf x})= 0, \quad \Delta w^{(k)}_M ({\bf x}) - \beta_k^2 w^{(k)}_M ({\bf x})= 0.
\label{Hh_MHh}
\end{equation}

\begin{figure}[H]
\begin{center}
\includegraphics[scale=1.3]{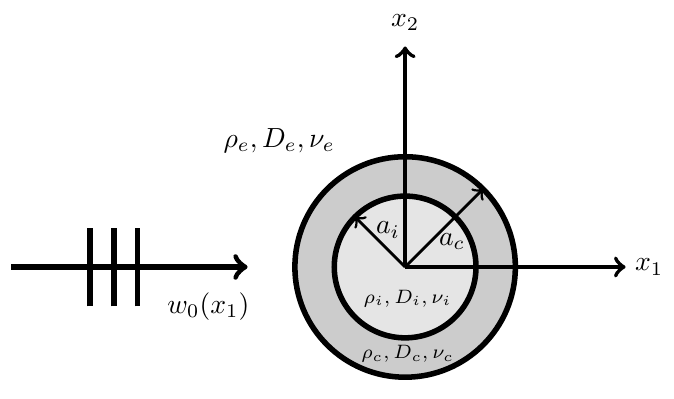}
\caption{A plane wave $w_0(x_1)$ is incident on the coated inclusion along the $x_1$-axis. The inclusion, coating and exterior are associated with the subscripts $i$, $c$ and $e$, respectively. 
}
\label{setting}
\end{center}
\end{figure}

In the case of a circular geometry, the inclusion, the coating and the exterior are defined as follows:
\begin{equation}
\Omega_i=\{ {\bf x}: x_1^2 + x_2^2 \leq a_i^2 \}, \quad \Omega_c=\{ {\bf x}: a_i^2 \leq x_1^2 + x_2^2 \leq a_c^2 \}, \quad \Omega_e = \mathbb{R}^2 \setminus \overline{\Omega_i \cup \Omega_c}, \label{domains}
\end{equation}
as shown in Fig. \ref{setting}.

On the interface boundaries $r=a_i$ where the inclusion and coating meet, and $r=a_c$ where the coating and exterior meet, the conditions for perfect bonding, that is the continuity of the displacement, its normal derivative, the moment and the transverse force,  lead to
\begin{equation}
w^{(c)}({\bf x}) = w^{(l)}({\bf x}), \quad \frac{\partial w^{(c)}({\bf x})}{\partial r}  = \frac{\partial w^{(l)}({\bf x})}{\partial r}, \label{disp_and_normal_der_disp_cont}
\end{equation}
and, with spatial dependence suppressed,
\begin{align}
D_c \left[ \frac{\partial^2 w^{(c)}}{\partial r^2} + \frac{\nu_c}{r} \left( \frac{\partial w^{(c)}}{\partial r} + \frac{1}{r} \frac{\partial^2 w^{(c)}}{\partial \theta^2}\right) \right] &= D_l \left[ \frac{\partial^2 w^{(l)}}{\partial r^2} + \frac{\nu_l}{r} \left( \frac{\partial w^{(l)}}{\partial r} + \frac{1}{r} \frac{\partial^2 w^{(l)}}{\partial \theta^2}\right) \right] , \label{moment_cont}\\ \nonumber \\
D_c \left[\frac{\partial}{\partial r} \Delta_{r\theta} w^{(c)} +  \frac{1-\nu_c}{r^2}  \frac{\partial^2}{\partial \theta^2} \left(\frac{\partial w^{(c)}}{\partial r} - \frac{w^{(c)}}{r}  \right) \right] &= 
D_l \left[\frac{\partial}{\partial r} \Delta_{r\theta} w^{(l)} +  \frac{1-\nu_l}{r^2}  \frac{\partial^2}{\partial \theta^2} \left(\frac{\partial w^{(l)}}{\partial r} - \frac{w^{(l)}}{r}  \right) \right].
\label{transverse_force__cont}
\end{align}
On $r=a_i$ we have $l=i$ and on $r=a_c$ we have $l=e$. 

\subsection{Multipole representation for a plane wave scattered by a coated circular inclusion}

Excluding the positions of point sources, the fields $w^{(i)}, w^{(c)}$ and $w^{(e)}$ can be written 
\begin{equation}
w^{(k)} (r,\theta) = \sum_{n=-\infty}^\infty \left[A_n^{(k)} J_n (\beta_k r) + E_n^{(k)} {H_n^{(1)}} (\beta_k r) + B_n^{(k)} I_n (\beta_k r) + F_n^{(k)} K_n (\beta_k r) \right] e^{in\theta}, \label{wice_rep} \\
\end{equation}
where $k$ is either $i, c$ or $e$, depending on the respective region. We note that when $k=i$, the coefficients $E_n^{(i)}$ and $F_n^{(i)}$ are identically zero  to 
ensure that there are no forces or moments  
at the origin. Using the representations (\ref{wice_rep}), equations (\ref{disp_and_normal_der_disp_cont})-(\ref{transverse_force__cont}) give us two matrix relations
\begin{equation}
{\cal A}^{(ci)} 
\left(
\begin{array}{c}
A_n^{(c)} \\ E_n^{(c)} \\ B_n^{(c)} \\ F_n^{(c)}
\end{array}
\right) = 
{\cal B}^{(ci)} 
\left(
\begin{array}{c}
A_n^{(i)} \\ B_n^{(i)}
\end{array}
\right) \quad \mbox{on}~~ r=a_i,\label{matrix_inclusion_coating}
\end{equation}
where ${\cal A}^{(ci)}$ and ${\cal B}^{(ci)} $ are $4\times4$ and $4\times2$ matrices, respectively; and
\begin{equation}
{\cal A}^{(ec)} 
\left(
\begin{array}{c}
A_n^{(e)} \\ E_n^{(e)} \\ B_n^{(e)} \\ F_n^{(e)}
\end{array}
\right) = 
{\cal B}^{(ec)} 
\left(
\begin{array}{c}
A_n^{(c)} \\ E_n^{(c)}  \\ B_n^{(c)} \\ F_n^{(c)}
\end{array}
\right) \quad \mbox{on}~~ r=a_c,\label{matrix_exterior_coating}
\end{equation}
where ${\cal A}^{(ec)}$ and ${\cal B}^{(ec)} $ are both $4\times4$ matrices. The representations of these four matrices are given in 
the Supplementary Material.
At this stage, it is 
important to emphasise that reducing the matrices ${\cal A}^{(ci)}$ and ${\cal A}^{(ec)}$ to block-diagonal structure is essential to pursue {
the derivation of an analytical formula for the scattering matrix ${\cal S}$}. We note that in what follows ${\cal A}^{(ci,*)}, {\cal A}^{(ec,*)}$ denote the required matrices in block-diagonal form and ${\cal B}^{(ci,*)}, {\cal B}^{(ec,*)}$ are the associated matrices, with all these matrices being given in 
the Supplementary Material.

We use the four matrices ${\cal A}^{(ci,*)}$, ${\cal B}^{(ci,*)}$, ${\cal A}^{(ec,*)}$ and ${\cal B}^{(ec,*)}$ to construct two further matrices ${\cal C}^{(ci,*)}$ and ${\cal C}^{(ec,*)}$ as follows:
\begin{equation}
{\cal C}^{(ci,*)} = {{\cal A}^{(ci,*)}}^{-1} {\cal B}^{(ci,*)}, \quad
{\cal C}^{(ec,*)} = {{\cal A}^{(ec,*)}}^{-1} {\cal B}^{(ec,*)},
\end{equation}
where the expressions which would represent the results of the matrix inversions in the previous equation are too cumbersome to be given explicitly. The final step in the construction is to cascade via a matrix product the progression from the interface between the inclusion and the inner coating to the interface between the outer coating and the matrix, giving the explicit transfer matrix ${\cal C}^{(ei,*)}$, which incorporates the  interface conditions at both interfaces:

\begin{equation}
{\cal C}^{(ei,*)} = {\cal C}^{(ec,*)} {\cal C}^{(ci,*)}.
\end{equation}

The transfer matrix ${\cal C}^{(ei,*)}$ can then be used to give the wave coefficients $A_n^{(e)}$, $B_n^{(e)}$, $E_n^{(e)}$ and $F_n^{(e)}$ in the exterior region in terms of the wave coefficients $A_n^{(i)}$, $B_n^{(i)}$ in the inclusion region:
\begin{equation}
\left(
\begin{array}{c}
A_n^{(e)} \\ E_n^{(e)} \\ B_n^{(e)} \\ F_n^{(e)}
\end{array}
\right) = {\cal C}^{(ei,*)}
\left(
\begin{array}{c}
A_n^{(i)} \\ B_n^{(i)}
\end{array}
\right). \label{main_matrix_eqn}
\end{equation}
 
The final step in this procedure is to eliminate the wave coefficients in the inclusion region to determine the $2\times 2$ scattering matrix ${\cal S}_n$ which relates outgoing wave coefficients in the exterior region to incoming wave coefficients in the exterior region. The relevant equations resulting from the procedure are described in 
the Supplementary Material.
and lead to the following matrix equation
$$
\left(
\begin{array}{c}
E_n^{(e)} \\ F_n^{(e)}
\end{array}
\right) = {\cal S}_n
\left(
\begin{array}{c}
A_n^{(e)} \\ B_n^{(e)}
\end{array}
\right),
$$
where on the left-hand side we have outgoing wave coefficients, and on the right-hand side incoming wave coefficients. 
If we now assume that the scattering occurs from a plane incident wave travelling along the $x_1$-axis, the coefficients $A_n^{(i)}, B_n^{(i)}$ associated with the inclusion, $A_n^{(c)}, B_n^{(c)}, E_n^{(c)}, F_n^{(c)}$ associated with the coating and  $E_n^{(e)}, F_n^{(e)}$ associated with the exterior, respectively, can be readily found from (\ref{main_matrix_eqn}) (see 
Supplementary Material).
Note that in this particular case $A_n^{(e)} = i^n$ since
\begin{equation}
w_0(x_1)=e^{i\beta_e x_1} = \sum_{n=-\infty}^\infty i^n J_n(\beta_e r) e^{i n\theta},
\label{incident_plane_wave}
\end{equation}
and the $B_n^{(e)}$ are identically zero.

It can be shown that the general expression for the scattered field is (see, for example, \cite{JO_OS_RCM_ABM_NVM}, \cite{YuKK}, \cite{ANN_CV})
\begin{equation}
w_{(sc)}(r,\theta) = {\cal S}_{0}^{11}
H_0^{(1)}(\beta_e r)+{\cal S}_{0}^{21}
K_0(\beta_e r) +\sum_{n=1}^\infty 2 i^n [{\cal S}_{n}^{11}
 H_n^{(1)}(\beta_e r)+{\cal S}_{n}^{21}
 K_n(\beta_e r)] \cos  \left(n \theta\right),
\label{wsc_monopole_dipole_general_simplified}
\end{equation}
where ${\cal S}_n^{ij},\,i,j=1,2$, denote the entries of the scattering matrix.
At this point, we emphasise that the coefficient ${\cal S}_0^{11}$ in the right-hand side of the expression (\ref{wsc_monopole_dipole_general_simplified}) is in fact equal to the scattering coefficient $E_0^{(e)}$ (using $A_0^{(e)}=1$ and $B_0^{(e)}=0$). As has been claimed previously \cite{MF_PYC_HB_SE_SG_AA}, with a suitably designed coating $E_0^{(e)} \approx 0$. (In fact, $E_0^{(e)}$ can be made equal to zero exactly, with an appropriate choice of the material parameters of the coating - see (\ref{CIEnAnrel}), (\ref{scattering_coeff}) below.)

\subsection{An algebraic system for active source amplitudes in the presence of an incident plane wave}
\label{algebsys}

Now, we assume that a finite number of active sources are applied in the exterior of the coated inclusion in order to reduce the scattering of an incident plane wave by the inclusion. Here we derive formally an algebraic
system for the active source amplitudes, which are computed to make the required number of multipole coefficients vanish in the representation of the outgoing scattered field. 

\begin{figure}[H]
\begin{center}
\includegraphics[scale=1.1]{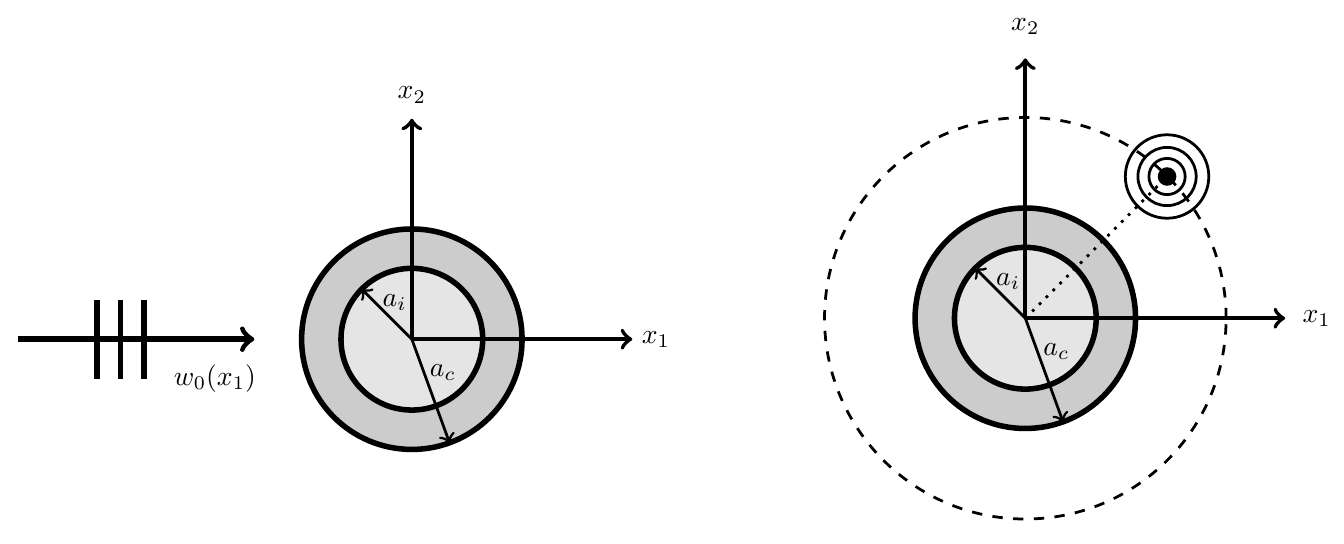}
\put(-280,-10) {(a)}
\put(-100,-10) {(b)}
\caption{Model problems: (a) plane wave incident on a coated inclusion; (b) point source near the inclusion.
}
\label{modpr}
\end{center}
\end{figure}

In order to find the total wave scattered from the coated inclusion,
we identify  two model problems illustrated in Fig. \ref{modpr}. These problems deal with scattering from the coated circular inclusion, with the centre at the origin.  
In the first model problem, the incident field is a plane wave  advancing along the $x_1$-axis, whereas in the second problem, the incident field is generated by a point source of  unit amplitude placed outside the coated inclusion.

{\bf The plane wave problem.} The first model problem involves scattering of a plane wave from the coated inclusion.
The displacement field $w^{(p)}$ satisfies the equation
\begin{equation}
\begin{split}
\Delta^2 w^{(p)}  - \beta^4 w^{(p)} & =0 \quad \mbox{in} \,\, \mathbb{R}^2 \setminus \bar{{\cal D}}, 
\end{split}
\end{equation}
supplemented  by interface conditions (\ref{disp_and_normal_der_disp_cont})--(\ref{transverse_force__cont}) on the interior and exterior interfaces of the coated inclusion.
Here ${\cal D}$ is the region occupied by the coated inclusion. The total field $w^{(p)}$ can be written as the sum of the incident field $w_{(i)}$ and the scattered field $w_{(sc)}^{(p)}$, that is
\begin{equation}\label{plane_wave_field}
w^{(p)} ({\bf x}) = w_{(i)} ({\bf x}) +w_{(sc)}^{(p)} ({\bf x}),
\end{equation}
where $w_{(i)}$ is the same as $w_0$ defined in (\ref{incident_plane_wave}).

Outside the coated inclusion, the outgoing scattered field  $w_{(sc)}^{(p)}$
has the representation
\begin{equation}\label{plane_wave_scattered_field}
w_{(sc)}^{(p)}  = \sum_{n=-\infty}^\infty \left[ {E}_n^{(p)} H_{n}^{(1)}(\beta_e r) + {F}_n^{(p)} K_{n}(\beta_e r) \right]
\, e^{i n \theta}.
\end{equation}
For $\beta_e r \gg 1$, the evanescent term in the scattered field can be neglected since $K_n(\beta_e r) = {\it O} (\exp(-\beta_e r)/(\beta_e r))$ (see (9.7.2) in \cite{MA_IAS}), and hence
\begin{equation}
\label{asymp}
w_{(sc)}^{(p)} \sim
\sum_{n=-\infty}^\infty {E}_n^{(p)} H_{n}^{(1)}(\beta_e r) e^{in \theta } ~ \mbox{for} ~ \beta_e r  \gg 1.
\end{equation}

 {\bf The single source model problem.} The second model problem involves scattering of a radial wave from the coated inclusion (see Fig. \ref{modpr}(b)).  
The displacement field, solving this model problem, is denoted by $w^{(s,j)}$ and satisfies the equation
\begin{equation}
\begin{split}
\Delta^2 w^{(s,j)}  - \beta_e^4 w^{(s,j)} + \delta({\bf x} - {\bf a}^{(j)})& =0 \quad \mbox{in} \,\,\mathbb{R}^2 \setminus \bar{{\cal D}}, 
\end{split}
\end{equation}
where $\delta({\bf x} - {\bf a}^{(j)})$ denotes the Dirac delta function, centred at ${\bf a}^{(j)}$. It is supplemented by the transmission conditions (\ref{disp_and_normal_der_disp_cont})--(\ref{transverse_force__cont}).  The solution $w^{(s,j)}$admits the representation 
\begin{equation}\label{arbitrary_source_field}
w^{(s,j)} ({\bf x}) = w_{(i)}^{(s,j)} ({\bf x}) + w_{(sc)}^{(s,j)} ({\bf x})  = G({\bf x}- {\bf a}^{(j)} ) + w_{(sc)}^{(s,j)} ({\bf x}),
\end{equation}
and hence,
\begin{equation}\label{arbitrary_source_field_asymp}
w^{(s,j)} ({\bf x}) \sim \sum_{n=-\infty}^\infty {A}_n^{(s,j)} H_{n}^{(1)}(\beta_e r) e^{in\theta},  \quad j=\,1, \dots,N, \quad \text{as} ~~r \to \infty,
\end{equation}
where exponentially small terms satisfying the modified Helmholtz equation, are not shown.
Here $G({\bf x} - {\bf a}^{(j)} )$ denotes the Green's function for the biharmonic operator, $w_{(sc)}^{(s,j)}$ is the scattered field due to the unit source at ${\bf a}^{(j)}$, and ${A}_n^{(s,j)}$ are constant coefficients.

Assuming that the coefficients ${E}_n^{(p)}$ and ${A}_n^{(s,j)}$ are given,  active cloaking is achieved by introducing a set of $N$ control sources of complex amplitudes ${\cal B}_j $ placed at the points ${\bf a}^{(j)}$ in the exterior of the scatterer ${\cal D}$.
After the truncation to order $L$ in the expansions (\ref{plane_wave_scattered_field}), (\ref{arbitrary_source_field}), we choose $N=2 L+1$, so that    the total displacement field $w_{(total)}$ is approximately equal to the incident field $w_{(i)}$, that is
\begin{equation}\label{total_field}
w_{(total)} = w^{(p)} + \sum_{j=1}^N {\cal B}_j w^{(s,j)} \approx w_{(i)}.
\end{equation}
To find ${\cal B}_j$, we substitute (\ref{plane_wave_scattered_field})  and (\ref{arbitrary_source_field}) into (\ref{total_field}), and obtain the following system of linear algebraic equations
\begin{equation}\label{fourier_coeff_eqn}
{E}_k^{(p)}  + \sum_{j=1}^{2L+1} {\cal B}_j {A}_k^{(s,j)} = 0,
 \quad k = -L,\dots, L.
 \end{equation}

This algebraic system holds not only for the flexural waves in Kirchhoff plates, but it is also valid for the membrane waves governed by the Helmholtz equation. Of course, in the latter case the coefficients
${E}_k^{(p)}$ and ${A}_k^{(s,j)}$ will change accordingly. It is assumed that the positions of active sources are chosen so that the system (\ref{fourier_coeff_eqn}) is non-degenerate. 

In the text below, this algebraic system is implemented for the case of 4, 8 and 12 sources positioned on a circular contour of a sufficiently large radius $a_s$, and with its centre  coinciding with the centre of the coated inclusion. This leads to an approximate active cloaking, which is efficient in the required frequency range.
The numerical examples, based on our semi-analytic method, are also presented and demonstrate clearly the effectiveness of the method even in the resonant regimes.

\section{Resonant scattering of membrane waves by a coated inclusion}
\label{membrane}

The ultimate aim of this paper is to cloak an inclusion with a suitably designed coating from flexural waves in resonant regimes using active control sources. In such regimes,  a small change in frequency results in abrupt variations in scattering and thus it becomes important to eliminate frequency sensitivity of active sources. 
With this in mind, as a preliminary problem, it is 
instructive to study the interaction of flexural waves in a membrane containing a coated inclusion. This is because solutions to the biharmonic equation are superpositions of solutions to the Helmholtz (a mixture of propagating and evanescent waves) and modified Helmholtz equations (all of which are evanescent waves).  

In this section, we give an illustrative example for active cloaking in resonant regimes in a membrane, where waves are governed by the Helmholtz equation.
The out-of-plane displacement of the membrane satisfies the following equations and transmission conditions in three regions, with  $u^{(e)}$, $u^{(c)}$ and $u^{(i)}$, denoting the displacements in the exterior, coating and inclusion respectively:

\begin{equation}
\label{CIdisp}
\Delta u^{(k)}({\bf x}) + \beta_k^2 u^{(k)} ({\bf x})  + f ({\bf x})= 0, \quad {\bf x} \in {\Omega}_k, 
\end{equation}
where $\beta_k$ is defined as $\beta_k = \omega \sqrt{\rho_k/\mu_k}$, $k$ runs over  $e,$ $c$ or $i$, and the domains $\Omega_k$ are as in (\ref{domains});  $\mu_k,\,k=i,c,e,$ are the shear moduli in respective regions, and $f({\bf x})$ is the force term representing the active sources.

\begin{figure}[H]
\begin{center}
\includegraphics[scale=1.3]{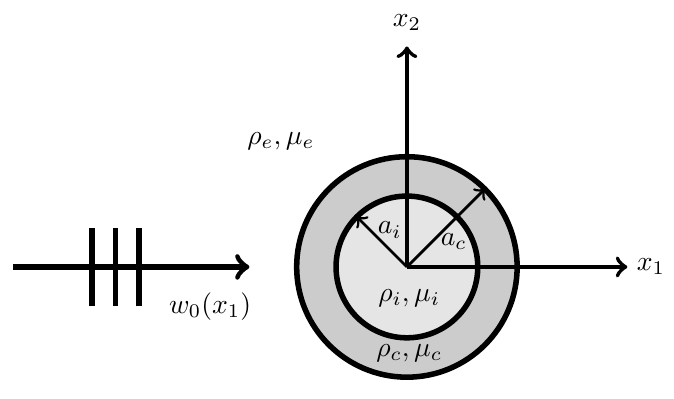}
\caption{A plane wave $w_0(x_1)$ is incident on the coated inclusion along the $x_1$-axis. The inclusion,  coating and  exterior are associated with the subscripts $i$, $c$ and $e$, respectively. 
}
\label{setting_membrane}
\end{center}
\end{figure}

We denote, as before, the radii of the inclusion and the exterior boundary of the coating by $a_i$ and $a_c$, respectively (see Fig. \ref{setting_membrane}).
Thus, the interface conditions can be written as
\begin{equation}
\label{interface_cond_ic}
\ u^{(i)}=u^{(c)}, \quad \mu_i { {\partial {u^{(i)}}} \over {\partial r } }= \mu_c {{\partial {u^{(c)}}} \over {\partial r}} \quad 
\mbox{on}~~r=a_i,
\end{equation}
and
\begin{equation}
\label{interface_cond_ce}
u^{(c)}=u^{(e)}, \quad  \mu_c { {\partial {u^{(c)}}} \over {\partial r } }= \mu_e {{\partial {u^{(e)}}} \over {\partial r}} \quad 
\mbox{on}~~r=a_c.
\end{equation}

Excluding the positions of point sources,
the associated solutions of equations (\ref{CIdisp}) are
\begin{eqnarray}
\ u^{(k)} (r,\theta) &=& \sum_{n=-\infty} ^{\infty} {\left[A_n^{(k)} J_n(\beta_e r) + E_n^{(k)} H_n^{(1)} (\beta_er)\right]e^{in\theta}}, \quad k=i, c, e,\label{ksol}
\end{eqnarray}
where $E_n^{(i)}$ are identically zero.

\subsection{Derivation of the relation between coefficients $E_n^{(e)}$ and $A_n^{(e)}$}
Using the representations (\ref{ksol}) with the interface conditions (\ref{interface_cond_ic}) and (\ref{interface_cond_ce}), it is straightforward to derive the outgoing wave coefficient $E_n^{(e)}$ in terms of the coefficient $A_n^{(e)}$ (characterising the incoming wave) as
\begin{equation}
\label{CIEnAnrel}
\ E_n^{(e)}= {\cal T}_n A_n^{(e)},
\end{equation}
where 
\begin{equation}
{\cal T}_n = -\frac{{\cal N}_n {\cal M}_n^{11} -{\cal M}_n^{21}} {{\cal N}_n {\cal M}_n^{12}- {\cal M}_n^{22}}.
\label{scattering_coeff}
\end{equation}
Here ${\cal T}_n$ stands for the scattering coefficient, and the representations for ${\cal M}_n^{kl} $ ($k,l=1,2$) and ${\cal N}_n$ are given 
in the Supplementary Material.

We note that once the relation (\ref{CIEnAnrel}) is established, the coefficients $A_n^{(c)}, E_n^{(c)}$ associated with the coating, as well as the coefficient $A_n^{(i)}$ associated with the inclusion can be readily found (see 
 Supplementary Material).

Also note that the scattered field in this particular case can be represented as follows (compare with (\ref{wsc_monopole_dipole_general_simplified}))
\begin{eqnarray}
u_{(sc)}&=& 
{\cal T}_0 H_0^{(1)}(\beta_e r)
+ \sum_{n=1}^\infty 2i^n {\cal T}_n H_n^{(1)} (\beta_e r)\, \cos (n\theta).
\end{eqnarray}

\subsection{Resonant scattering by a high-contrast inclusion with no coating}
\label{NoCoatingMembrane}

\begin{figure}[H]
\begin{tikzpicture}
\node[inner sep=0pt] (figleft) at (0.0,0.0)
{\includegraphics[width=.50\textwidth]{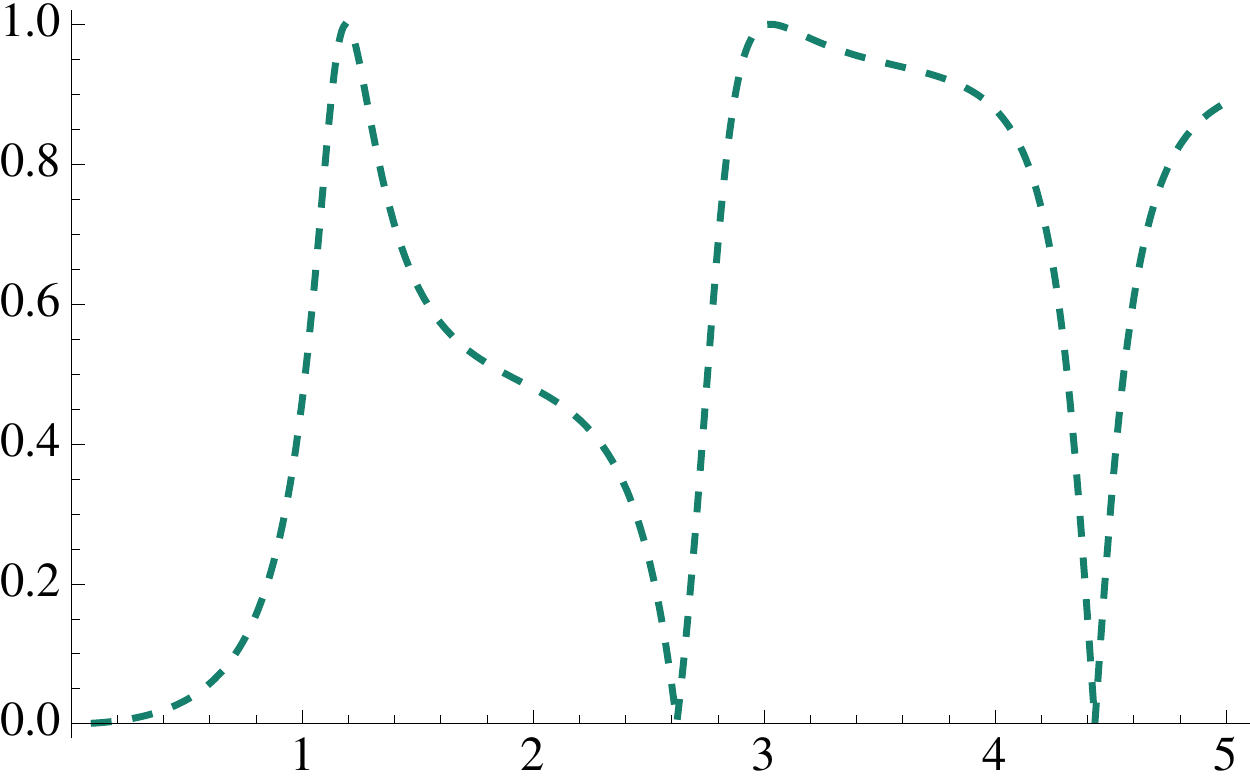}};
\node at (4.5,-2.1) {$\omega$};
\node at (-4.0,3.0) {$|E_0^{(e)}|$};
\draw [blue, thick] (-3.63,-2.0) rectangle (-3.5,-2.4);
\node at (1.0,-3.0) {(a)};
\node[inner sep=0pt] (figright) at (8.0,0.0)
{\includegraphics[width=.40\textwidth]{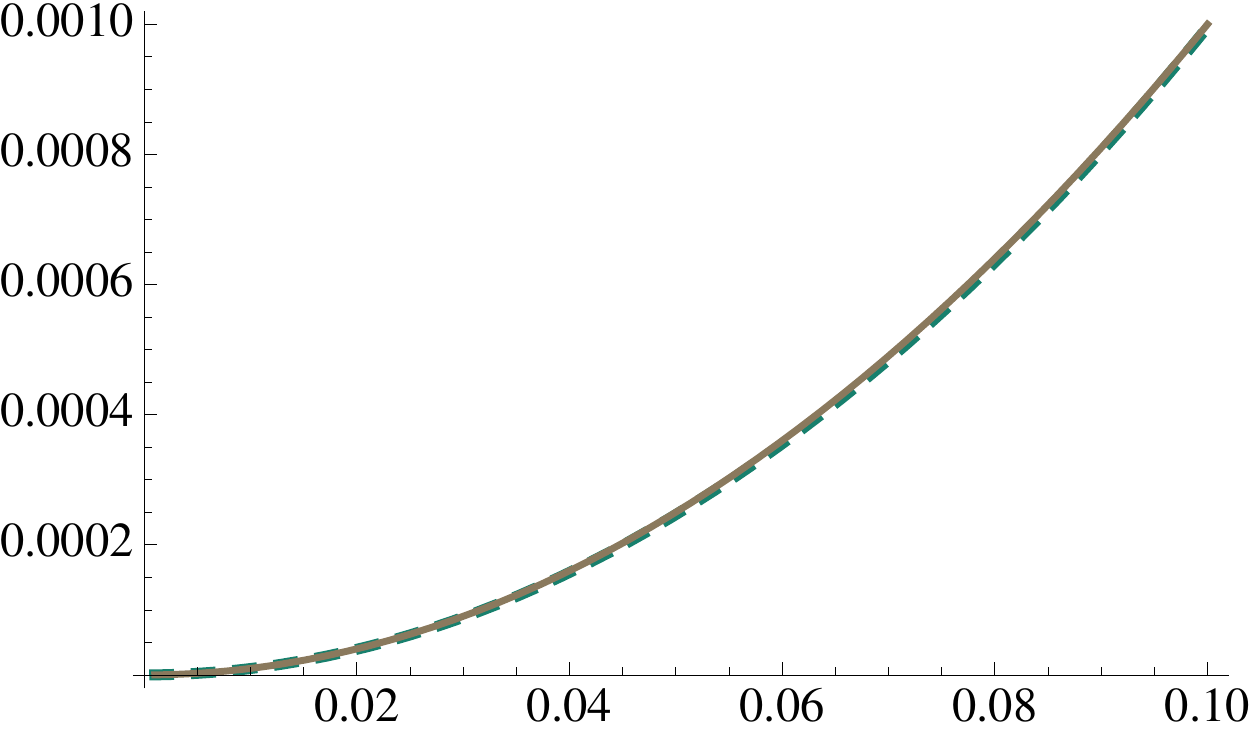}};
\node at (11.7,-1.55) {$\omega$};
\node at (5.0,2.5) {$|E_0^{(e)}|$};
\node at (9.0,-3.0) {(b)};
\end{tikzpicture}
\caption{Monopole amplitude for membrane wave scattering by an uncoated inclusion with the parameters  $\rho_i=1.5$, $\rho_e=1.0$, $\mu_i=0.1$, $\mu_e=1.0$, $a_i=0.50$: (a) $|E_0^{(e)}|$ as a function of $\omega$, (b) Detail from (a) (solid blue rectangle) showing  $|E_0^{(e)}|$ (dashed green) and the empirical fit $0.10 \,\omega^{2}$ (solid grey).}
\label{Acoust_scat_uncoated_inclusion}
\end{figure}
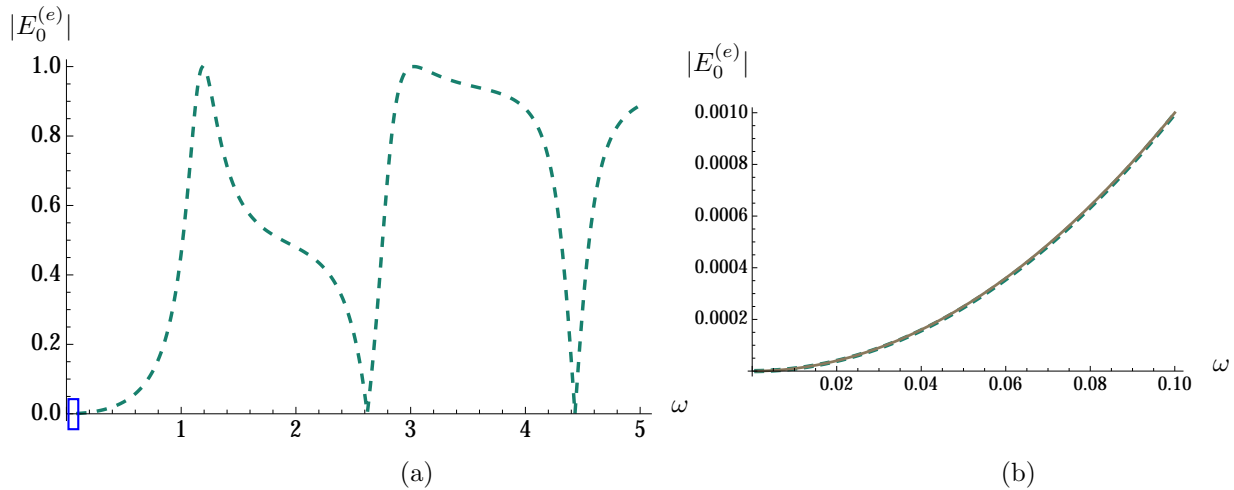

In the case of an inclusion with no coating it can be shown that formula (\ref{CIEnAnrel}) reduces to
\begin{equation}
E_n^{(e)} = \frac{J_n (\beta_i a_i) {J_n}' (\beta_e a_i) - \sqrt{(\mu_i \rho_i)/(\mu_e \rho_e)} {J_n}'(\beta_i a_i) J_n (\beta_e a_i)}{\sqrt{(\mu_i \rho_i)/(\mu_e \rho_e)} H_n^{(1)} (\beta_e a_i) {J_n}' (\beta_i a_i) - {H_n^{(1)}}' (\beta_e a_i) J_n (\beta_i a_i)}\, A_n^{(e)}. \label{EfromA_no_coating}
\end{equation} 
In particular, for the monopole term $(n=0)$
\begin{equation}
|E_0^{(e)}| \sim \frac{\pi a_i^2}{4} \frac{|\rho_i-\rho_e|}{\mu_e}\, \omega^2 \quad \text{as} ~~ \beta_i a_i, \beta_e a_i \to 0, \label{E0e_asy}
\end{equation}
where $\beta_i a_i, \beta_e a_i$ are non-dimensional quantities (considering a plane wave travelling in the $x_1$ direction as before, $A_0^{(e)} = 1$). We also note that the monopole term for static problems corresponds to a net force produced by an elastic inclusion, and is consequently zero; this differs significantly from the dynamic case, when the inclusion possesses a non-zero mass density. In the dynamic case, the coefficient $E_0^{(e)}$ of the monopole term is defined by the formula (\ref{EfromA_no_coating}), and plays an important role in scattering since it determines the scattering cross-section (see Chapter 13 of Born and Wolf \cite{BandW}). For low frequencies, as evident from equation (\ref{E0e_asy}), the coefficient $|E_0^{(e)}|$ is small, but, in general, it is a rapidly varying function of frequency, and  may indeed show strongly resonant behaviour for high-contrast inclusions (here caused by the ratio of $\mu_e/\mu_i=10$, see Fig. \ref{Acoust_scat_uncoated_inclusion}). 

Figure \ref{Acoust_scat_uncoated_inclusion}(a) shows the monopole coefficient $|E_0^{(e)}|$ as a function of $\omega$ for a specific choice of high-contrast inclusion; whilst the densities of the exterior and the inclusion are comparable ($\rho_i=1.5$, $\rho_e=1.0$), the shear modulus of the exterior is tenfold that of the inclusion ($\mu_i=0.1$, $\mu_e=1.0$). A prominent feature of the  low frequency region (enclosed by a solid blue rectangle), where $|E_0^{(e)}|$  is small, 
is that both the function $|E_0^{(e)}|$ itself and its derivative tend to zero with $\omega$ (for which a diagram of the scattered field is given in Fig. \ref{Acoust_scat_uncoated_inclusion}(b)).
Sufficiently away from the low frequency region, the major features of Fig. \ref{Acoust_scat_uncoated_inclusion} include a strong peak near
$\omega=1.2$, a narrow zero near $\omega=2.62$ and a rapid rise to a maximum near $\omega=3.0$. The region of rapid variation of $|E_0^{(e)}|$ with $\omega$ between 2.62 and 3.0 would
give rise to difficulties for active cloaking of a frequency-swept incident wave, and thus will be discussed in detail below. 

\begin{figure}[H]
\begin{center}
\subfigure[]{\label{omega_030_no_coating}\includegraphics[width=0.33\textwidth]{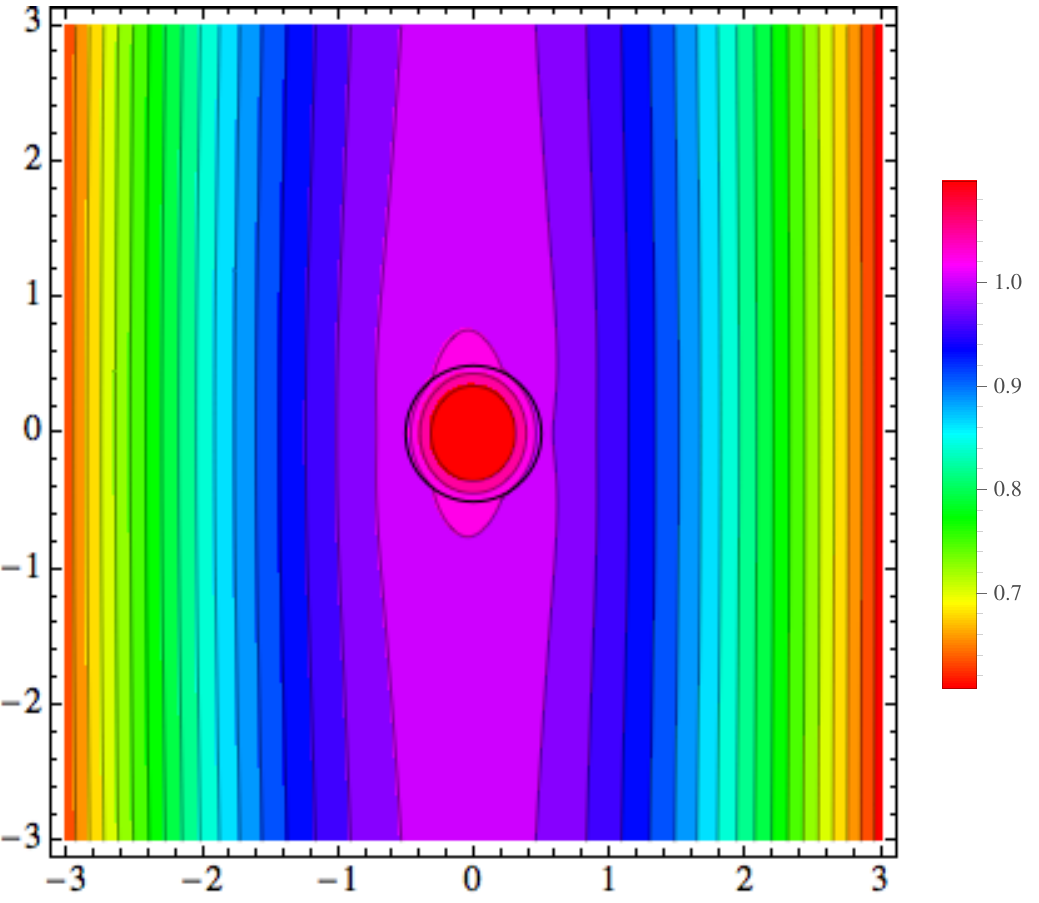} ~~
\includegraphics[width=0.33\textwidth]{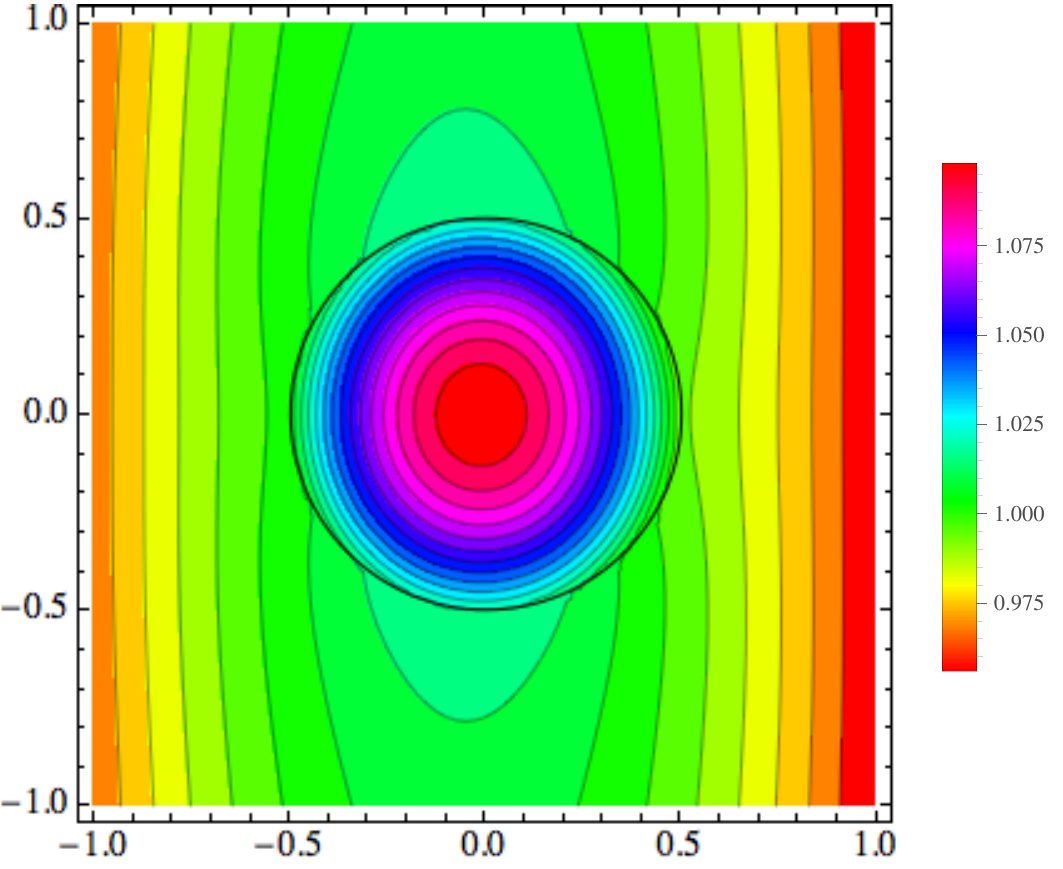}}
\\
\subfigure[]{\label{omega_262_no_coating}\includegraphics[width=0.33\textwidth]{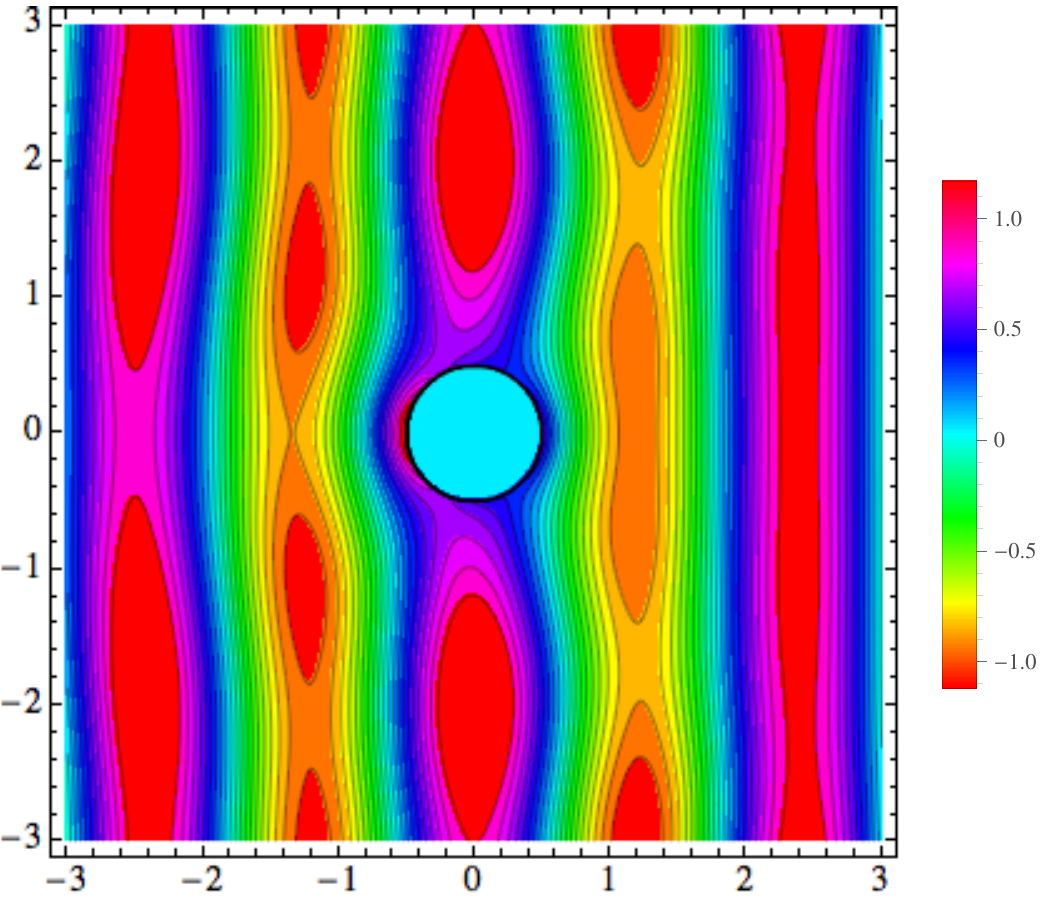}~~
\includegraphics[width=0.33\textwidth]{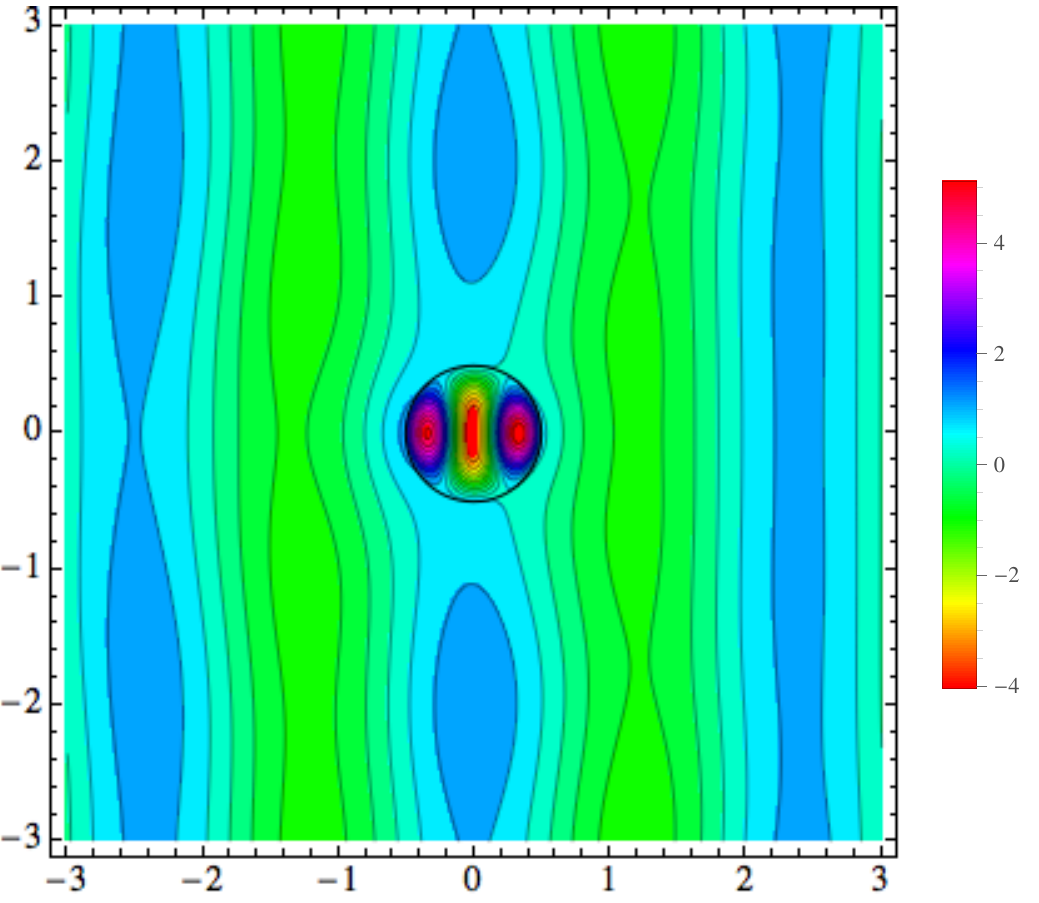}~~
\includegraphics[width=0.33\textwidth]{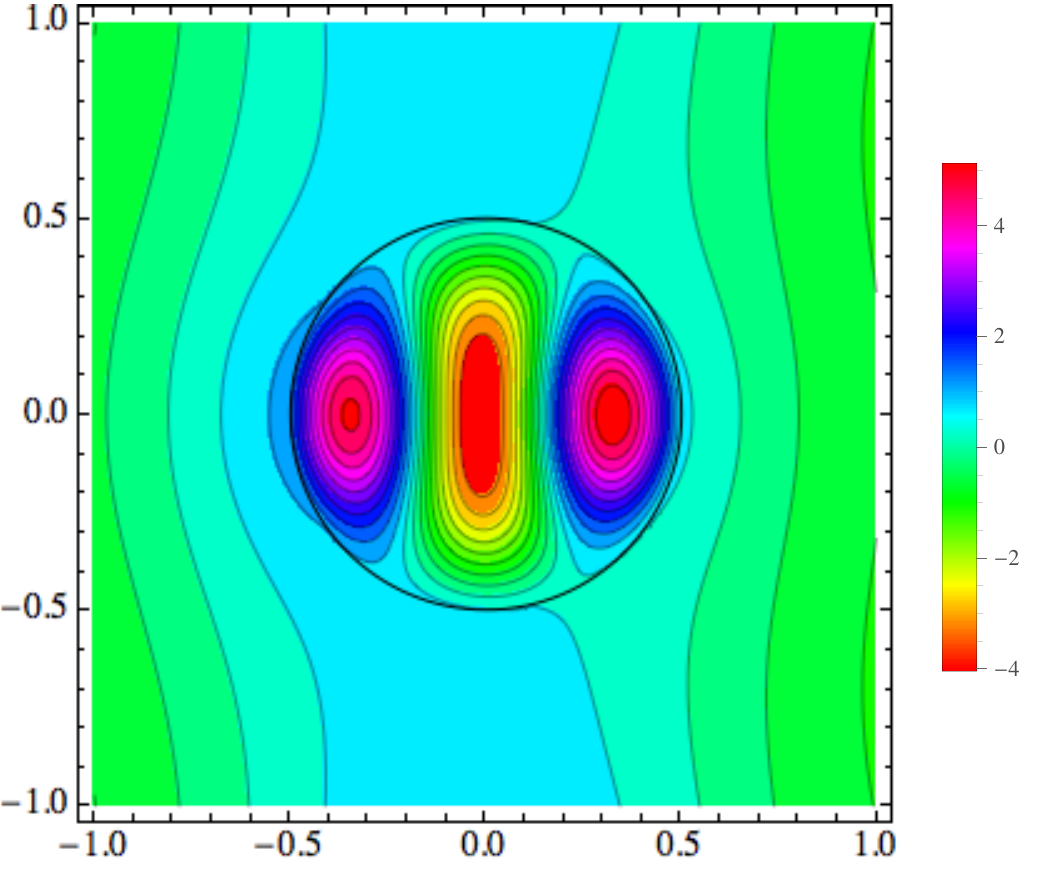}}
\\
\subfigure[]{\label{omega_300_no_coating}\includegraphics[width=0.33\textwidth]{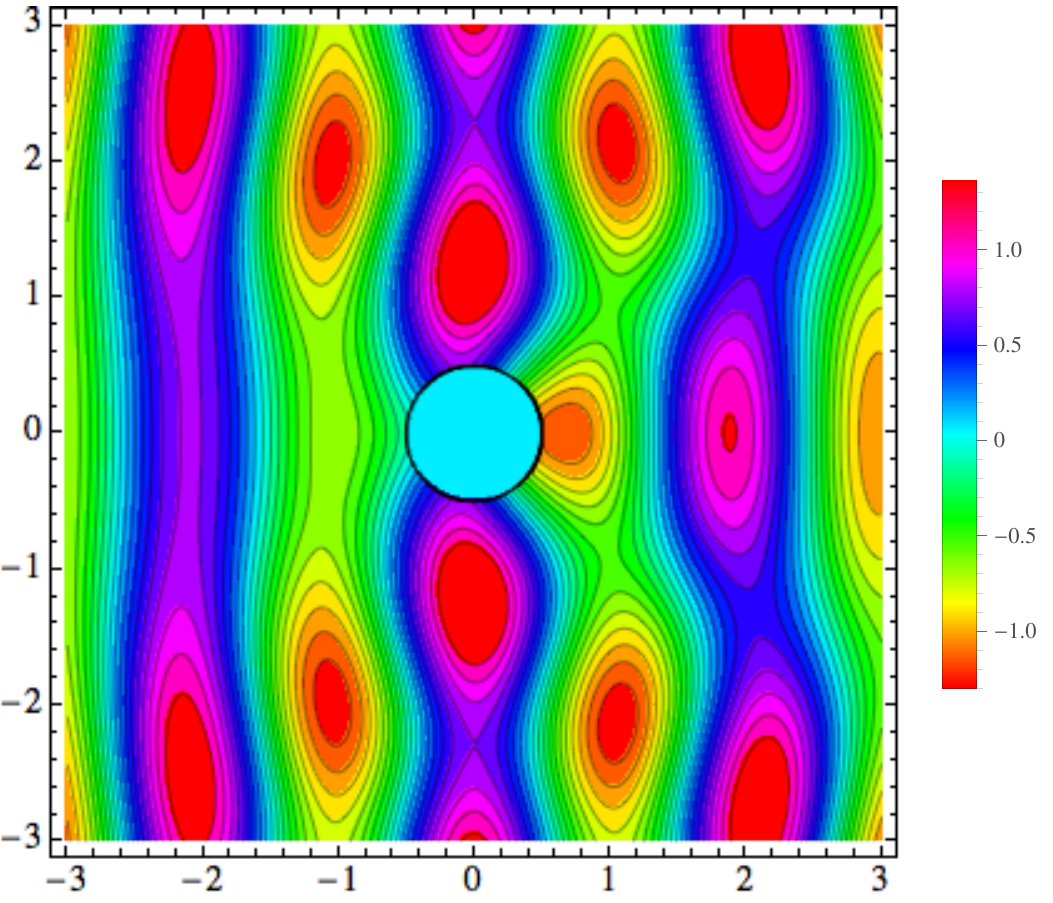}~~
\includegraphics[width=0.33\textwidth]{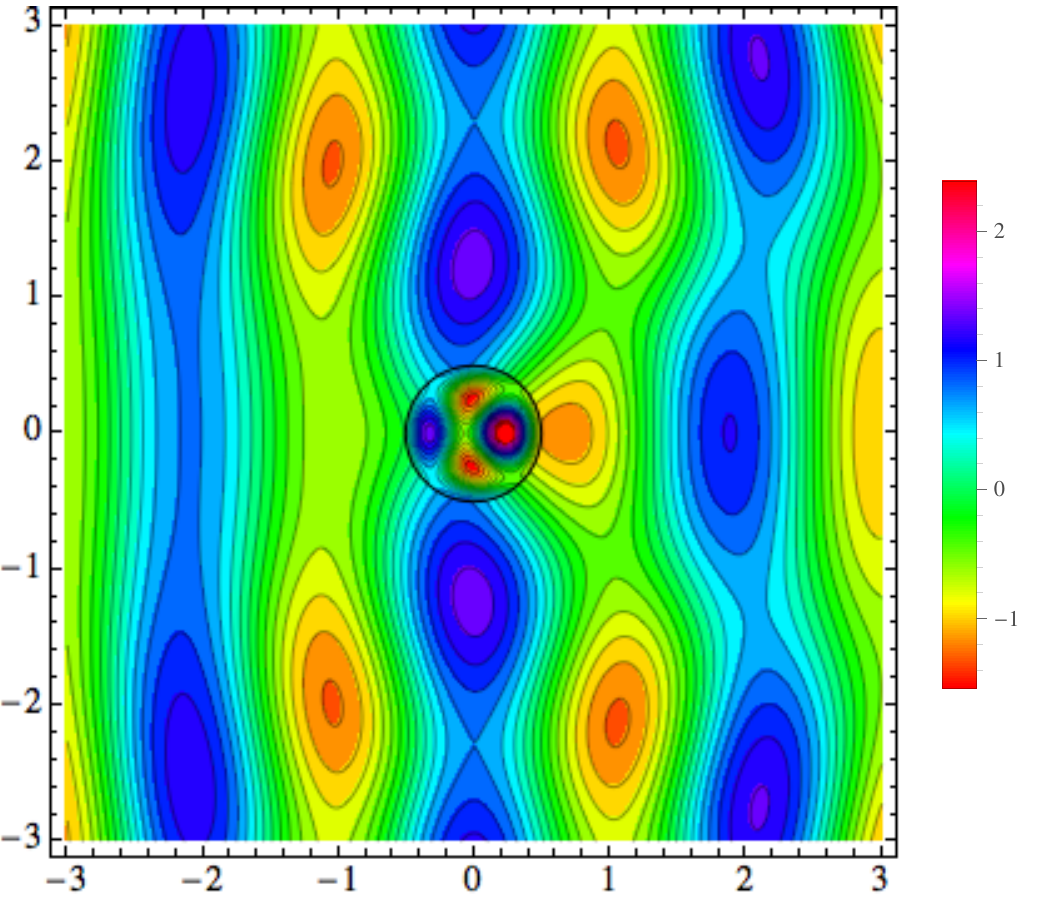} ~~
\includegraphics[width=0.33\textwidth]{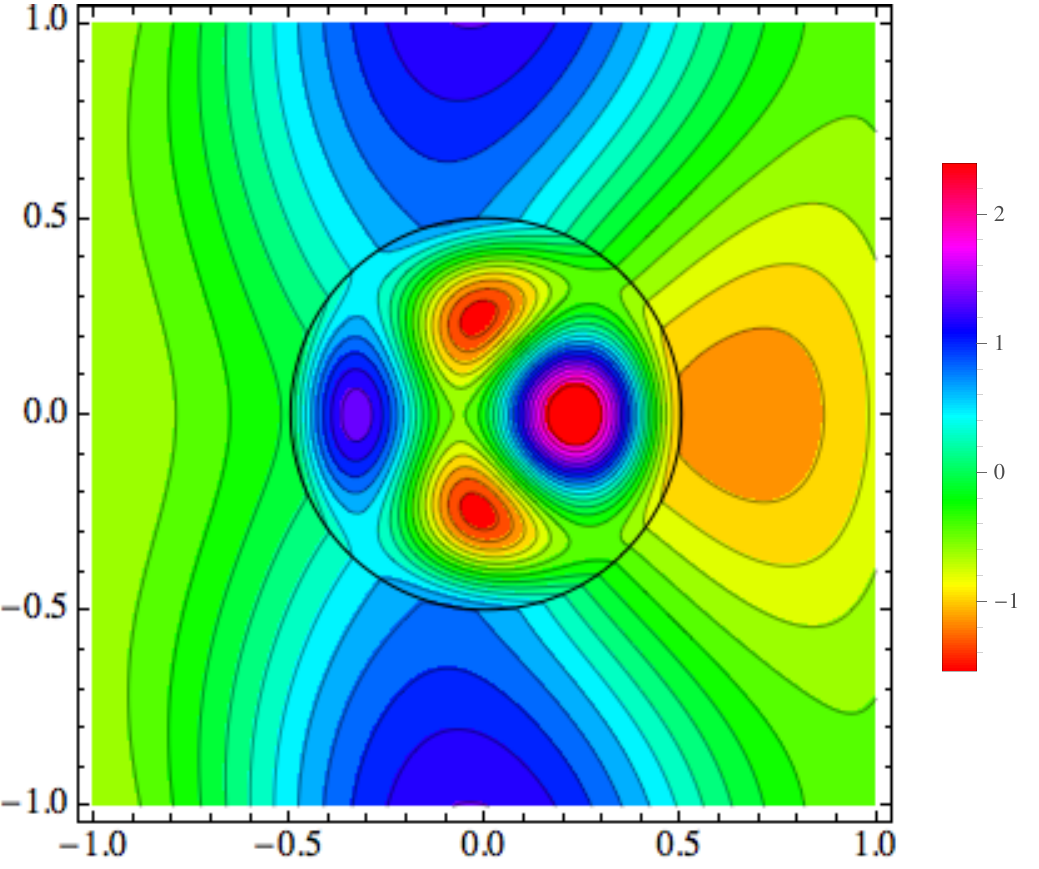}}
\caption{Membrane wave scattering by an uncoated inclusion for the parameters  $\rho_i=1.5$, $\rho_e=1.0$, $\mu_i=0.1$, $\mu_e=1.0$, $a_i=0.50$.  (a)  Total field for $\omega=0.3$, (b)  Total field for $\omega=2.62$ where $E_0^{(e)}$ vanishes, (c)  Total field for $\omega=3.0$ where $|E_0^{(e)}|$ has nearly a maximum. 
In the last frame of each row, we present a zoomed-in version of the displacement inside and in the close vicinity of the inclusion. 
The first two frames in the inset (b) represent the same field outside the inclusion, but with different colour maps, where the first frame excludes the internal field, whereas the second frame includes it. The same comment applies to the first two frames of the inset (c).  In all contour plots horizontal and vertical axes are $x_1$- and $x_2$-axes, respectively.}
\end{center}
\label{membrane_nocoating_contours}
\end{figure}

The influence of $|E_0^{(e)}|$ on the scattered field is illustrated in Figs. \ref{omega_030_no_coating}, \ref{omega_262_no_coating} and \ref{omega_300_no_coating}: 
Fig. \ref{omega_030_no_coating} shows the total displacement for a small frequency value of $\omega=0.3$, which clearly shows a low scattering pattern. Figs. \ref{omega_262_no_coating} (with $\omega=2.62$) and \ref{omega_300_no_coating} (with $\omega=3.0$) contrast the  total displacement field patterns around the inclusion
for the frequencies corresponding respectively to the first non-trivial zero of $|E_0^{(e)}|$ and its subsequent maximum. In Fig. \ref{omega_262_no_coating} scattering is primarily due to terms of dipole order, but the wave fronts behind the inclusion are
practically straight. In Fig. \ref{omega_300_no_coating}, both monopole and dipole terms are contributing to the distortion of the wavefronts behind the inclusion. The scattering patterns change dramatically even though the frequency difference is small.

\newpage

\subsection{Perturbation of a plane wave by a coated inclusion}

We now discuss two ways of scattering control by an elastic coating placed around an inclusion.  First, we consider the very low frequency regime dominated by the 
monopole term, leading to the mass-compensation criterion for the choice of the parameters of the coating around the inclusion, as  mentioned in the Introduction (see formula (\ref{mass_balance})). Second, we analyse the higher frequency regime, which may include resonance modes.

\subsubsection{Reduction of the monopole term for small frequencies}

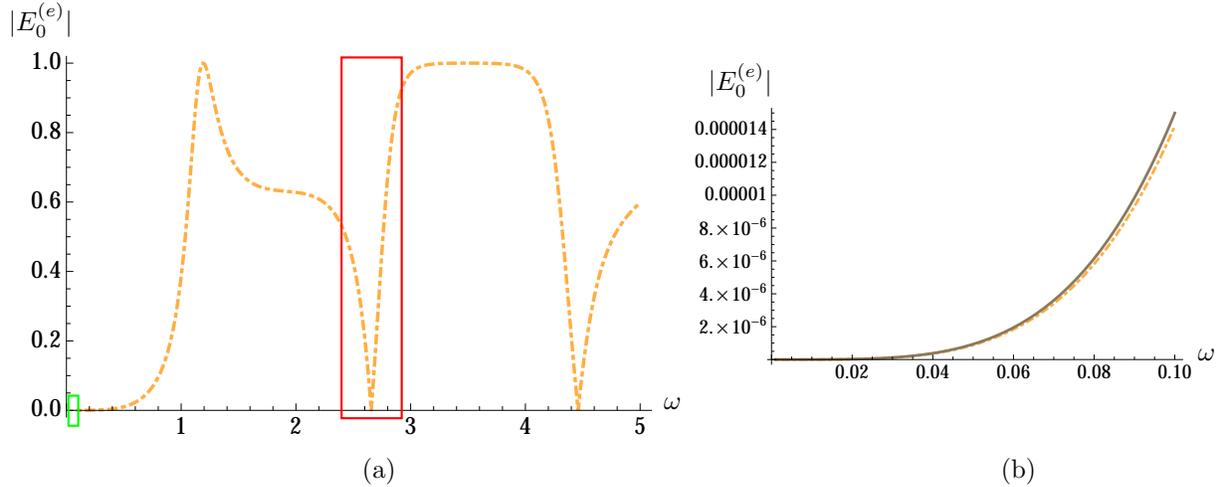
\begin{figure}[H]
\begin{tikzpicture}
\node[inner sep=0pt] (figleft) at (0.0,0.0)
{\includegraphics[width=.50\textwidth]{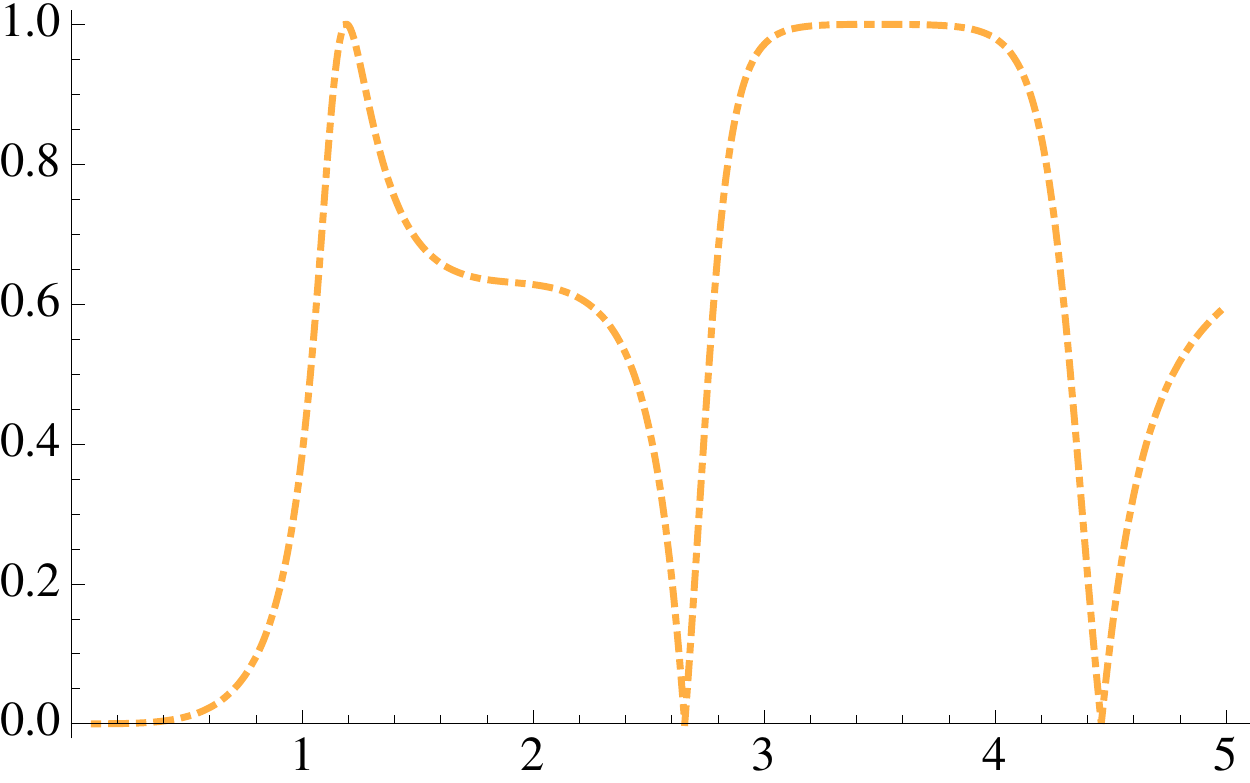}};
\node at (4.4,-2.1) {$\omega$};
\node at (-4.0,3.0) {$|E_0^{(e)}|$};
\draw [green, thick] (-3.63,-2.0) rectangle (-3.5,-2.4);
\node at (0.5,-3.0) {(a)};
\node[inner sep=0pt] (figright) at (8.0,0.0)
{\includegraphics[width=.40\textwidth]{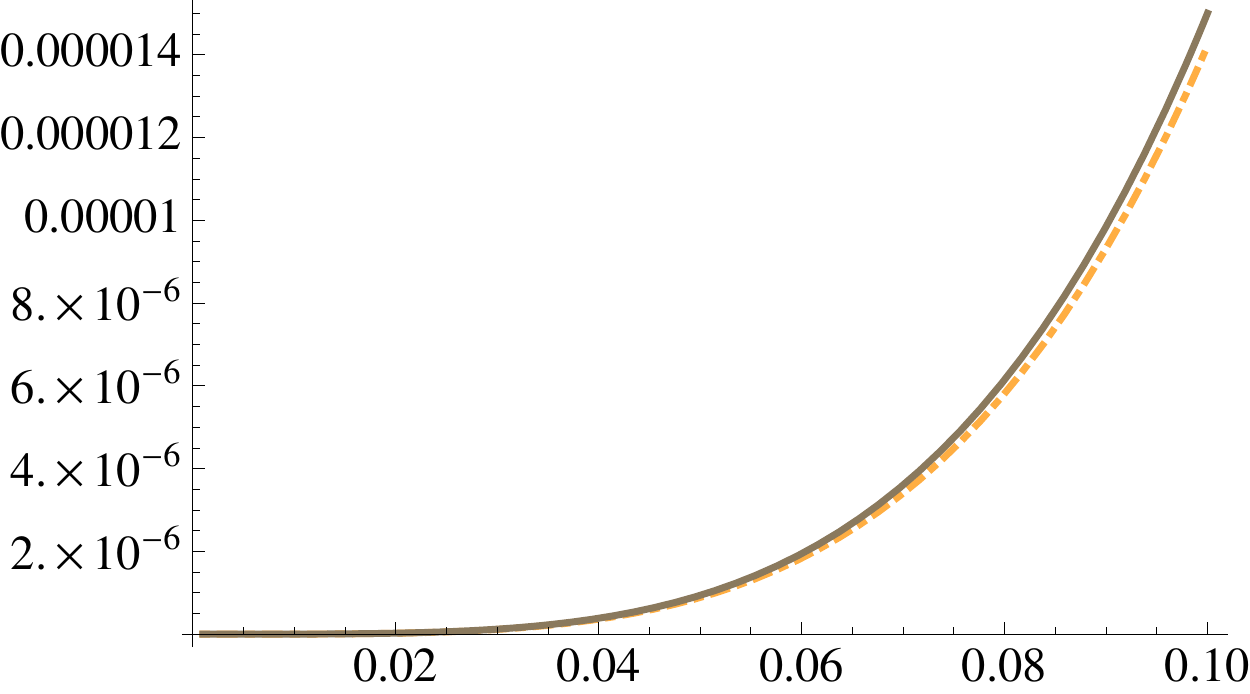}}; 
\node at (11.5,-1.45) {$\omega$};
\node at (5.3,2.2) {$|E_0^{(e)}|$};
\node at (9.0,-3.0) {(b)};
\draw [red, thick] (-0.0,-2.3) rectangle (0.8,2.5);
\end{tikzpicture}
\caption{Monopole amplitude for membrane wave scattering by a coated inclusion with parameter values:
$\rho_i=1.5$,  $\rho_e=1.0$, $\rho_c= (\rho_e - \rho_i(a_i/a_c)^2)/(1 - (a_i/a_c)^2) \approx 0.635462$, $\mu_i=0.1$,  $\mu_c=1.0$, $\mu_e=1.0$,  
$a_i=0.50$, $a_c=0.77$: 
$|E_0^{(e)}| $ as a function of  $\omega$. Note, the curve flattens approximately for 
$\omega \in  [3,4.2]$. 
(b) Detail from (a) (green rectangle) showing $|E_0^{(e)}| $ (orange/dot-dashed) and 
the empirical fit $0.15 \omega^{4}$ (grey/solid).} 
\label{E0_coated_inclusion1}
\end{figure}

Formula (\ref{CIEnAnrel}) leads to
\begin{equation}
|E_0^{(e)}| = \frac{\pi}{4 \mu_e} \left|\rho_i a_i^2 + \rho_c (a_c^2-a_i^2)-\rho_e a_c^2\right|\omega^2 + \textit{O} (\omega^4) \quad \text{as} ~~ \beta_i a_i, \beta_c a_i, \beta_c a_c, \beta_e a_c \to 0.
\label{E0e_with_coating}
\end{equation}
This suggests the design (\ref{mass_balance}) for a coating which compensates the density  difference between the inclusion and exterior, so that the scattering cross-section at low frequencies is minimised accordingly.
In particular $|E_0^{(e)}|  \approx 0$ in the low frequency regime implies that (\ref{mass_balance}) holds,
that is, the average mass density of the inclusion combined with the coating equals the mass density of the exterior material.  
When $\rho_c$ is chosen according to the formula (\ref{mass_balance}), whilst the shear moduli of the coating $\mu_c$ and the exterior $\mu_e$ are chosen to be equal to each other, the graph of $|E_0^{(e)}|$ versus $\omega$ is given in Fig. \ref{E0_coated_inclusion1}(a). Comparing this with Fig. \ref{Acoust_scat_uncoated_inclusion}(a) and their respective detailed views in Figs \ref{Acoust_scat_uncoated_inclusion}(b) and \ref{E0_coated_inclusion1}(b), it is evident that in the low frequency regime, the scattering cross-section is substantially reduced. 
However, it is also shown in Fig. \ref{E0_coated_inclusion1}(a) (see for example, the region enclosed by the red rectangle) that for higher frequencies we observe resonance scattering associated with rapid variation and increase of $|E_0^{(e)}|$. Consequently, the formula (\ref{mass_balance}) is of no use in such regimes.

\subsubsection{
Passive control of scattering by an appropriate coating }

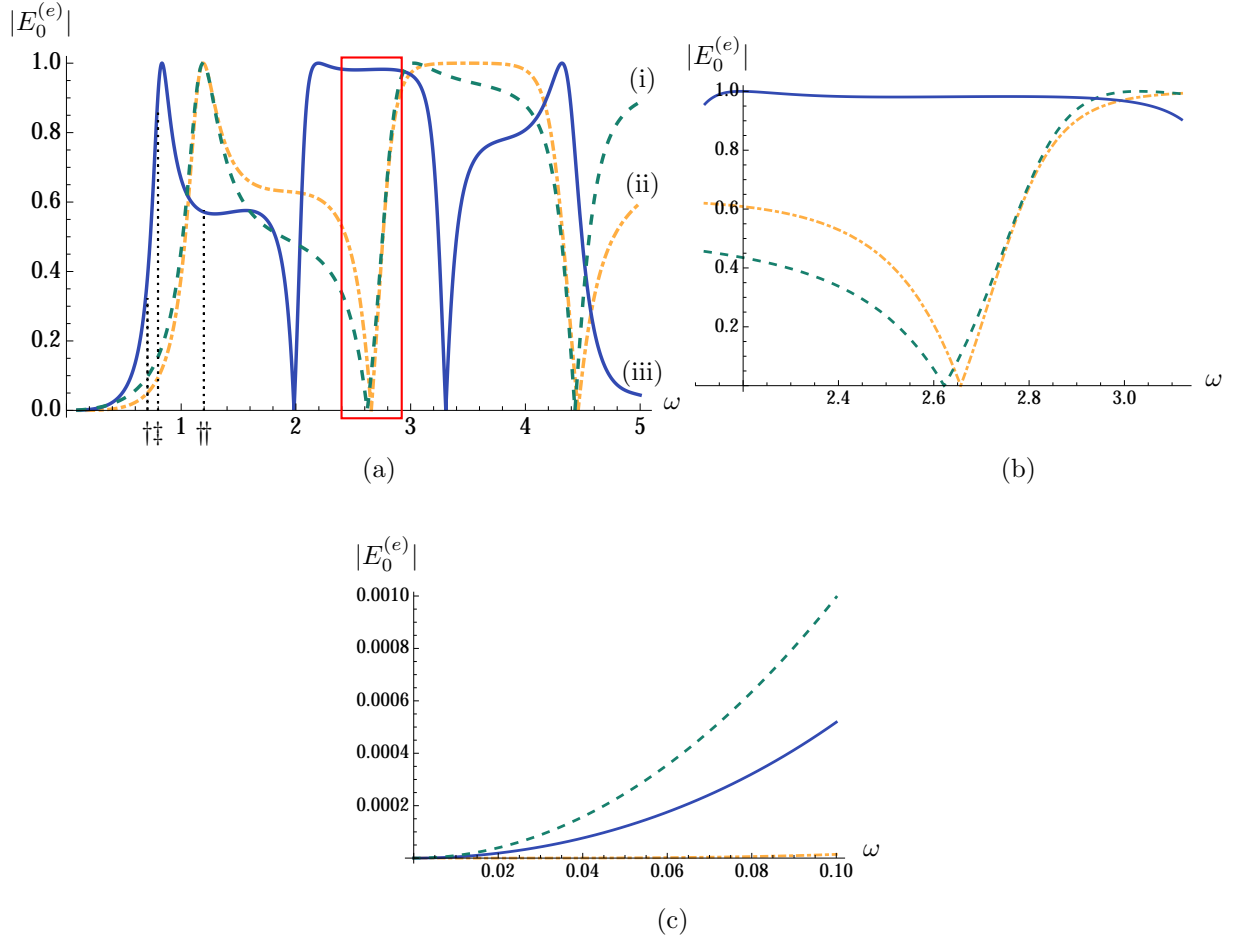
\begin{figure}[H]
\begin{tikzpicture}
\node[inner sep=0pt] (figleft) at (0.0,0.0)
{\includegraphics[width=.50\textwidth]{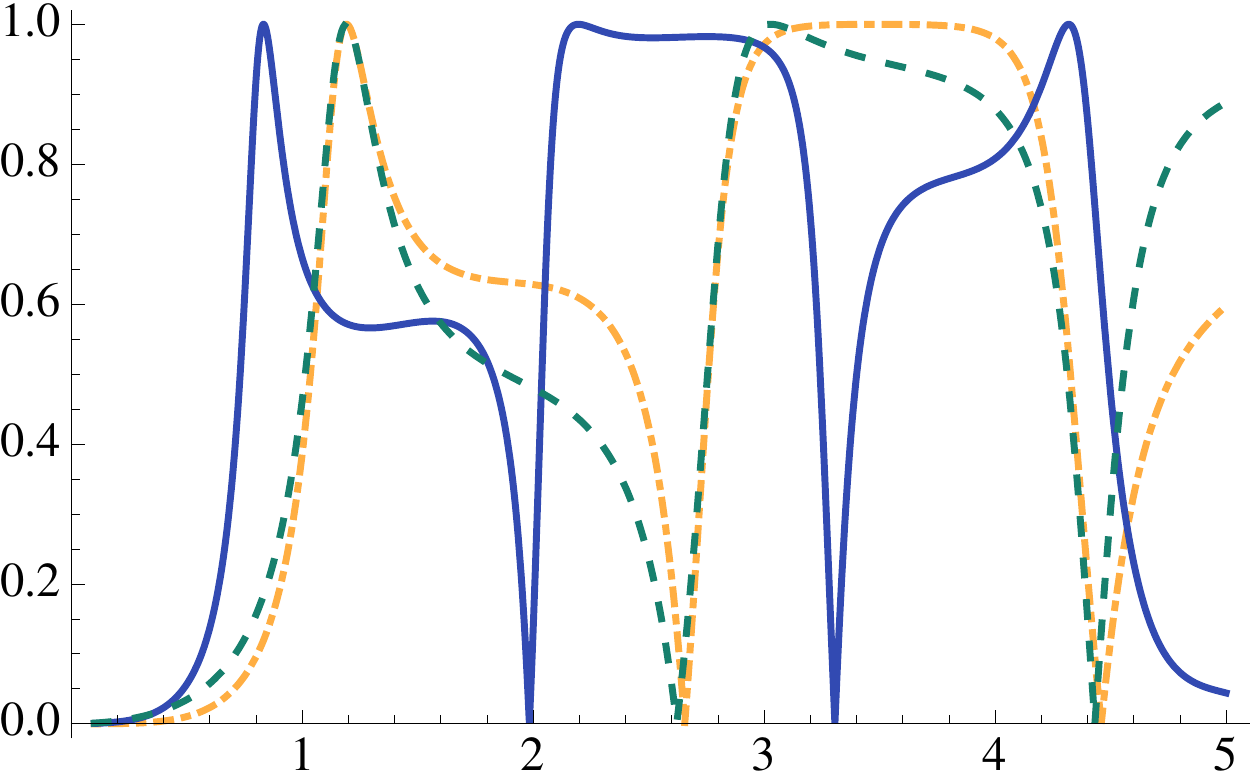}};
\node at (4.4,-2.1) {$\omega$};
\node at (-4.0,3.0) {$|E_0^{(e)}|$};
\node at (4.0,2.2) {(i)};
\node at (4.0,0.8) {(ii)};
\node at (4.0,-1.7) {(iii)};
\node at (0.5,-3.0) {(a)};
\draw [black, thick, dotted] (-2.58,-2.2) rectangle (-2.58,-0.70);
\node at (-2.58,-2.5) {$\tiny{\dagger}$};
\draw [black, thick, dotted] (-2.44,-2.2) rectangle (-2.44,1.8);
\node at (-2.44,-2.5) {$\tiny{\ddagger}$};
\draw [black, thick, dotted] (-1.83,-2.2) rectangle (-1.83,0.47);
\node at (-1.87,-2.5) {$\tiny{\dagger}$};
\node at (-1.79,-2.5) {$\tiny{\dagger}$};
\node[inner sep=0pt] (figright) at (8.0,0.0)
{\includegraphics[width=.40\textwidth]{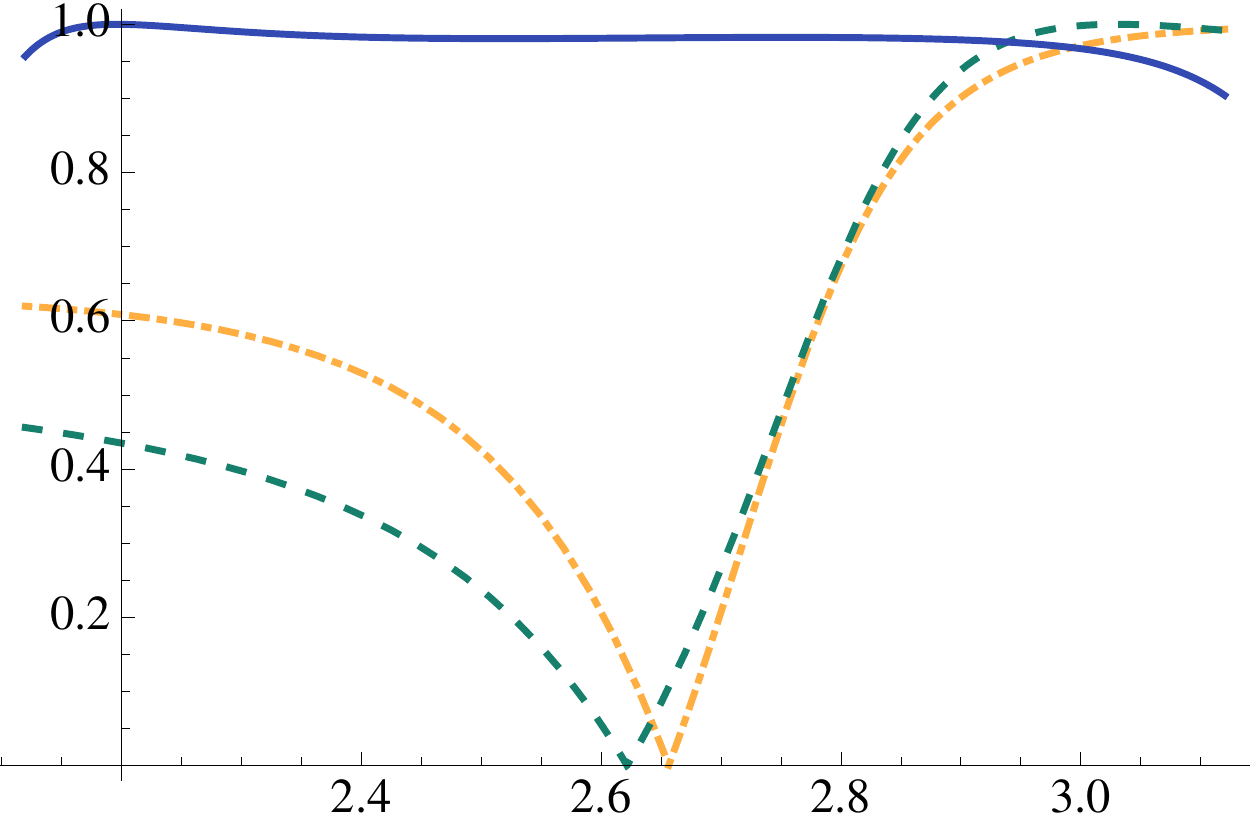}};
\node at (11.60,-1.80) {$\omega$};
\node at (5.0,2.55) {$|E_0^{(e)}|$};
\node at (9.0,-3.0) {(b)};
\draw [red, thick] (-0.0,-2.3) rectangle (0.8,2.5);
\node[inner sep=0pt] (figdown) at (3.5,-6.5)
{\includegraphics[width=.40\textwidth]{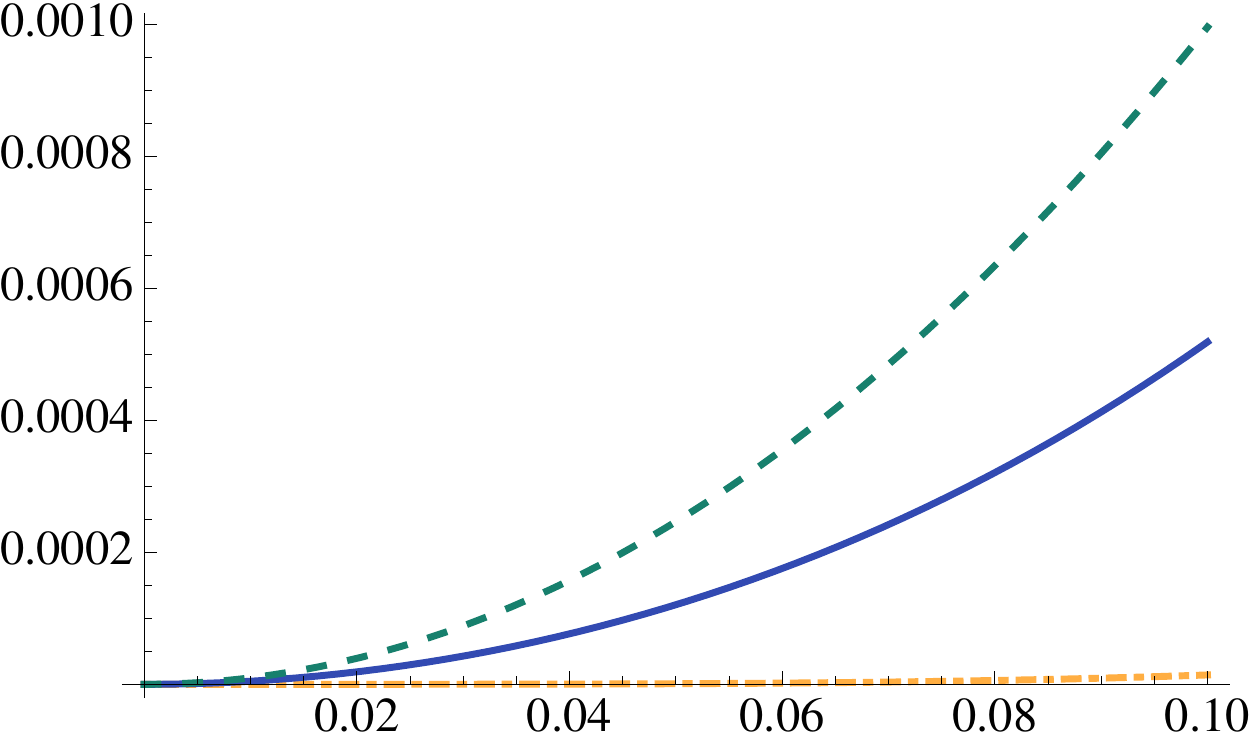}};
\node at (7.05,-8.05) {$\omega$};
\node at (0.6,-4.1) {$|E_0^{(e)}|$};
\node at (4.40,-9.0) {(c)};
\end{tikzpicture}
\caption{
(a) Monopole amplitude for membrane wave scattering by (i) an uncoated inclusion with parameter values: $\rho_i=1.5$, $\rho_e=1.0$, $\mu_i=0.1$, $\mu_e=1.0$, $a_i=0.50$ (green/dashed), (ii) a coated inclusion with parameter values:
$\rho_i=1.5$,  $\rho_e=1.0$,  $\mu_i=0.1$,  $\mu_c=1.0$, $\mu_e=1.0$,  
$a_i=0.50$, $a_c=0.77$  and 
$\rho_c= (\rho_e - \rho_i(a_i/a_c)^2)/(1 - (a_i/a_c)^2) \approx 0.635462$ (orange/dot-dashed) and (iii) a coated inclusion with parameter values:
$\rho_i=1.5$,  $\rho_c= 0.81$,   $\rho_e=1.0$,  $\mu_i=0.1$,  $\mu_c=0.11$, $\mu_e=1.0$,  $a_i=0.5$, $a_c=0.77$ (blue/solid). 
Note, the red box highlights where the blue/solid curve flattens approximately for $\omega \in [2.4,2.9]$, a lower frequency range than that seen in the green/dashed and orange/dot-dashed curves. 
(b) Monopole amplitude for membrane wave scattering for the parameters of (i)-(iii) in (a), but for a range of $\omega$ highlighted by the red box in (a).
We see strong variation in $|E_0^{(e)}|$ for an uncoated inclusion in this frequency range, but it is clear that with an appropriate choice of parameter values for the coating, we can create a region where $|E_0^{(e)}|$ remains fairly flat.
(c)  Monopole amplitude for membrane wave scattering for the parameters of (i)-(iii) in (a) where $\omega$ is small. The upper two curves (green/dashed and blue/solid) in (c) correspond to second order variation of $|E_0^{(e)}|$ with $\omega$ while the third (orange/dot-dashed) corresponds to a fourth order variation.}
\label{E0_coated_inclusion2}
\end{figure}

The method of active cloaking that will be discussed in the next section, has the advantage of being efficient even at higher frequency regimes. 
We illustrate here the limitations of passive design, where the parameters of the coating are chosen to control the monopole term in the scattered field.
At a frequency when the monopole coefficient $|E_0^{(e)}|$ is small but rapidly varying, active cloaking requires a rapid adjustment of the  source amplitudes (see the red box enclosing $\omega \approx 2.62$ in Fig. \ref{E0_coated_inclusion1}(a)). In these particular regions the scattered field increases significantly and thus we use the term ``resonant scattering".
To overcome this difficulty, the design of the coating can be revised to reduce the variation of the coefficient $|E_0^{(e)}|$ for this  frequency range. In Fig. \ref{E0_coated_inclusion2}(a), we combine data from Figs \ref{Acoust_scat_uncoated_inclusion}(a), \ref{E0_coated_inclusion1}(a) together with $|E_0^{(e)}|$ resulting from the revised design. The revised choice of parameters gives a ``flattened" curve near the resonant scattering frequency $\omega \approx 2.62$ (see the region enclosed by the red box). 
A magnified view is given in Fig. \ref{E0_coated_inclusion2}(b).
Figure \ref{E0_coated_inclusion2}(c) shows the behaviour of $|E_0^{(e)}|$ in the low frequency regime.  It is noted that the mass balance criterion for the coating of the inclusion clearly reduces the monopole coefficient from being \textit{O}$(\omega^2)$  to \textit{O}$(\omega^4)$. However, this is of little import since the monopole amplitude is already very small even for an uncoated inclusion.

\begin{figure}[H]
\begin{center}
\subfigure[]{\label{E0_flat_tot_ex}\includegraphics[width=.33\textwidth]{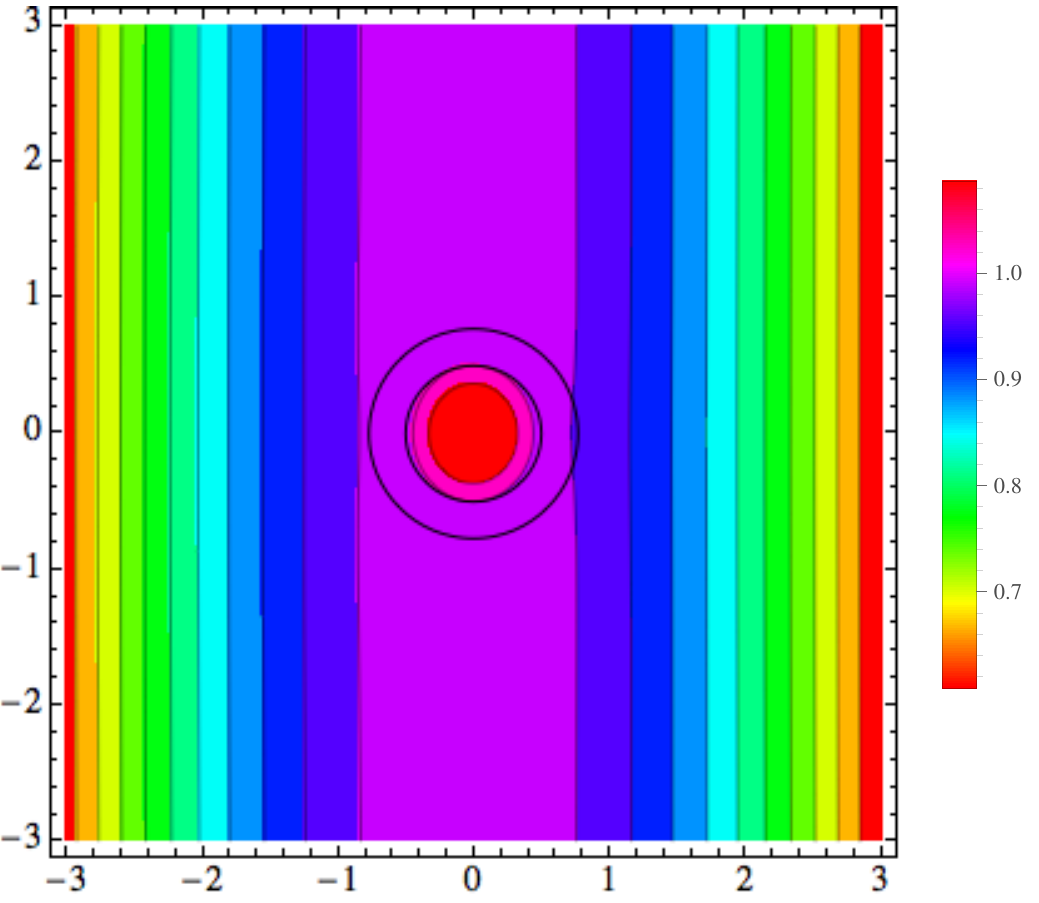}}~~
\subfigure[]{\label{E0_flat_tot_ex}\includegraphics[width=.33\textwidth]{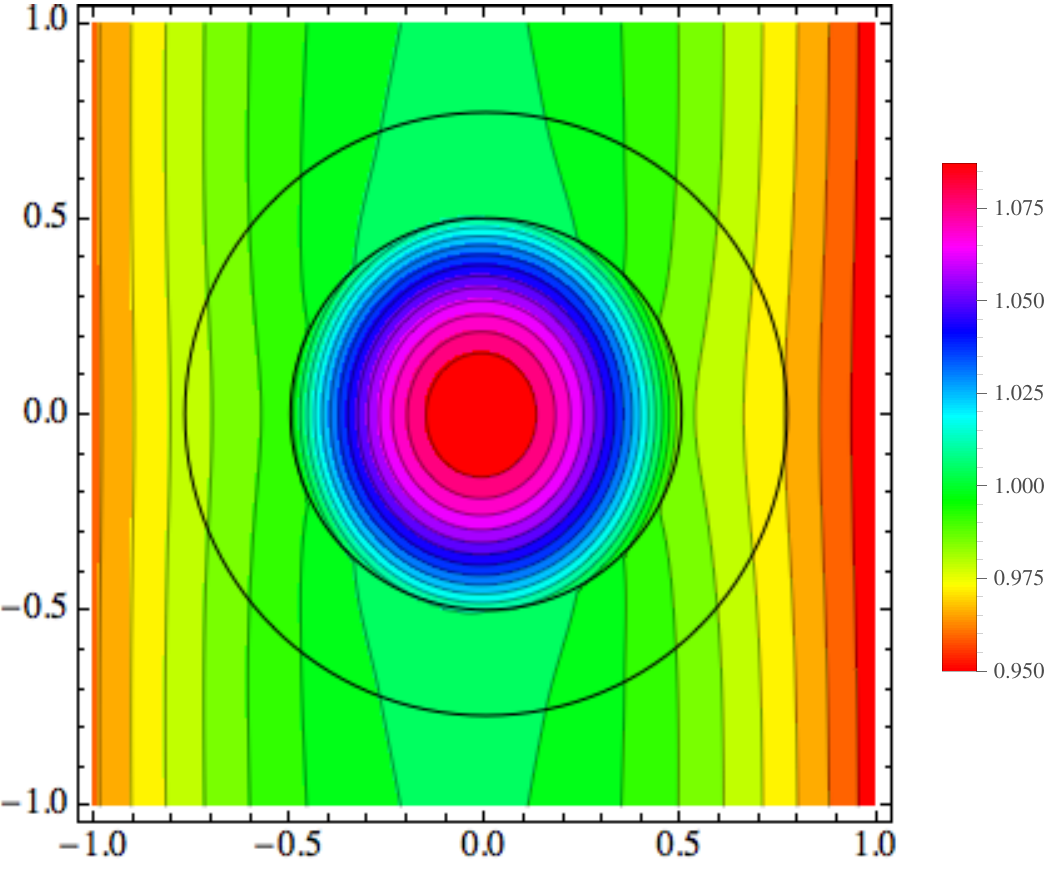}}~~
\caption{(a) Total displacement field for membrane wave scattering by a coated inclusion with parameter values: $\rho_i=1.5$,  $\rho_e=1.0$, $\rho_c= (\rho_e - \rho_i(a_i/a_c)^2)/(1 - (a_i/a_c)^2) \approx 0.635462$, $\mu_i=0.1$,  $\mu_c=1.0$, $\mu_e=1.0$,  $a_i=0.50$, $a_c=0.77$ and $\omega = 0.3$. (b) A zoomed in version of the displacement inside the inclusion, the coating and in the close vicinity of the coating. 
}
\label{useless_coating}
\end{center}
\end{figure}

\begin{figure}[H]
\begin{center}
\begin{tikzpicture}
\node[inner sep=0pt] (figleft) at (0.0,0.0)
{\includegraphics[width=.50\textwidth]{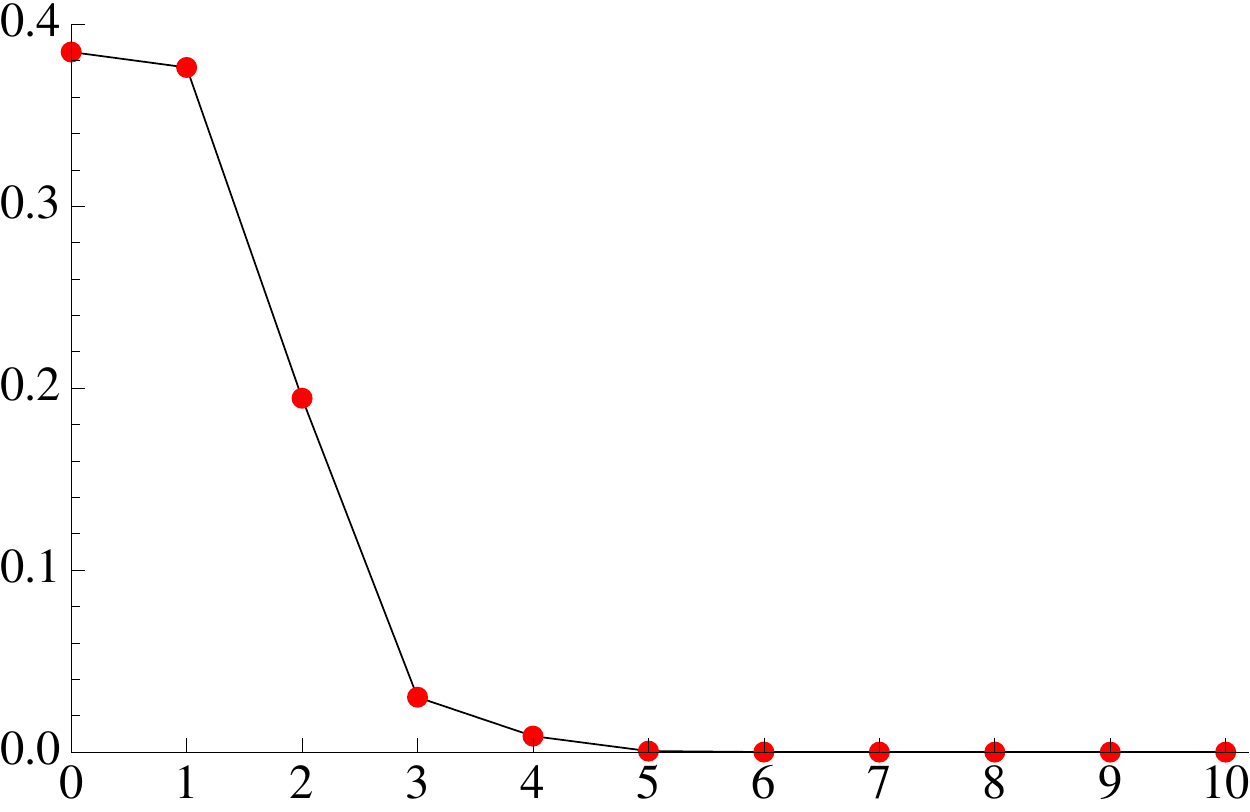}};
\node at (4.4,-2.2) {$n$};
\node at (-4.0,3.0) {$|E_n^{(e)} H_n^{(1)}(\beta_e a_s)|$};
\end{tikzpicture}
\end{center}
\caption{Membrane wave scattering: absolute value of the component $n$,  $E_n^{(e)} H_n^{(1)}(\beta_e a_s)$, of the displacement field, where $a_s$ is the radius of circle on which the control sources are placed, as a function of $n$ for $\omega=2.62$. Parameter values as for (blue/solid) curve in Fig. \ref{E0_coated_inclusion2}: $\rho_i=1.5$,  $\rho_c= 0.81$,   $\rho_e=1.0$,  $\mu_i=0.1$,  $\mu_c=0.11$, $\mu_e=1.0$,  $a_i=0.5$, $a_c=0.77$, $a_s=1.57$.}
\label{EneVersusnPlate_a}
\end{figure}

Next, in Fig. \ref{useless_coating}, we present  contour plots of the membrane wave amplitude for a small frequency value; the inclusion has a coating whose density is chosen according to the formula (\ref{mass_balance}). 
The  parameter values, with the exception of the mass density and elastic modulus of the coating, are chosen exactly as in Fig. \ref{omega_030_no_coating}; we choose the elastic modulus of the coating equal to that of the exterior and the frequency value as $\omega = 0.3$. As in Fig. \ref{E0_coated_inclusion2}(c), it is apparent that the scattering has been reduced 
compared with Fig. \ref{omega_030_no_coating}, however, the benefit of such a reduction is very low, as even without the coating there is no significant scattering.

For the higher frequency near $\omega = 2.62$, we have chosen the coating to ``flatten" the monopole term $E_0^{(e)}$  curve  over the resonant range of $\omega$. We then observe in Fig. \ref{EneVersusnPlate_a} that the multipole
field components of higher order become non-negligible, and hence passive reduction of scattering would not be appropriate any longer. Active cloaking by point sources, as described in \cite{JO_OS_RCM_ABM_NVM},  enables us to make the multipole coefficients vanish to the required order. 

\subsection{Active sources: cloaking of membrane waves in resonant regimes}
\label{active_section}

As in \cite{JO_OS_RCM_ABM_NVM}, we add active sources of membrane waves to annul a number of multipole coefficients of the propagating scattered field. Our study in \cite{JO_OS_RCM_ABM_NVM} confirmed that for an effective cloaking of flexural waves in a Kirchhoff plate at least four sources were required. Thus we first attempt to create a cloak for the coated inclusion with four active sources. These sources are symmetrically located on the $x_1$- and $x_2$-axes, all at a distance $a_s>a_c$ away from the origin. Note that the symmetry of the inclusion dictates that we set the amplitudes of the sources on the $x_2$-axis to be the same. Since the Green's function for the Helmholtz operator is given by 
\begin{equation*}
G(x_1- u, x_2 - v)=\frac{1}{4 i} H_0^{(1)}(\beta_e\sqrt{(x_1- u)^2+(x_2 - v)^2}),
\end{equation*}
where $(u,v)$ is the position of the point source, using Graf's addition theorem
for the sources located at $(-a_s,0), (a_s,0), (0,-a_s)$ and $(0,a_s)$ we obtain
\begin{eqnarray*}
Q_+\, G(x_1 - a_s,x_2) &=& \frac{Q_+}{4 i}\sum_{l=-\infty}^\infty  e^{i l\theta} H_l^{(1)}(\beta_e a_s)J_l(\beta_e r), \\
Q_- \,G(x_1 + a_s,x_2) &=& \frac{Q_-}{4 i}\sum_{l=-\infty}^\infty  e^{i l (\pi - \theta)} H_l^{(1)}(\beta_e a_s)J_l(\beta_e r), \\
P \,G(x_1,x_2 \mp a_s) &=& \frac{P}{4i} \sum_{l=-\infty}^\infty  e^{i l(\theta\mp \pi/2)} H_l^{(1)}(\beta_e a_s)J_l(\beta_e r).
\end{eqnarray*}
Here $Q_\pm$ are the amplitudes of the two sources located on the $x_1$-axis at $(\mp a_s,0)$, whereas $P$ is the amplitude of both sources located on the $x_2$-axis at $(0,\mp a_s)$; note that these amplitudes are to be found.

The $n$-th order coefficient for the total wave incident on the coated inclusion is thus
$$
A_n^{(e)} = i^n + \frac{Q_+}{4i} H_n^{(1)} (\beta_e a_s) + \frac{Q_-}{4i} (-1)^n H_n^{(1)} (\beta_e a_s) + \frac{P}{2i} H_n^{(1)} (\beta_e a_s) \cos(n\pi/2).
$$
As in \cite{JO_OS_RCM_ABM_NVM} we set the $n$-th order coefficient of the $H_n^{(1)}$ term equal to zero when $r>a_s$ and obtain
\begin{eqnarray*}
  4i{\cal T}_0 &=& -(Q_++Q_-+2P) [{\cal T}_0 H_0^{(1)}(\beta_e a_s) + J_0(\beta_e a_s)],\\
  4 {\cal T}_1 &=&  (Q_+-Q_-)  [{\cal T}_1 H_1^{(1)}(\beta_e a_s) + J_1 (\beta_e a_s)],\\
   4i {\cal T}_2 &=& 
( Q_+ + Q_- + 2P)  [{\cal T}_2 H_2^{(1)} (\beta_e a_s) + J_2(\beta_e a_s)],
\end{eqnarray*}
which we solve for $Q_\pm$ and $P$ for given $\beta_e, a_s$. Finally, we write the displacement outside the coated inclusion as
\begin{eqnarray*}
w(r,\theta) &\approx& w_0 (x_1) + Q_- \,G(r\cos \theta+a_s, r\sin \theta) + Q_+ \,G(r\cos \theta-a_s, r\sin \theta) \\
&+& P \,G(r\cos\theta, r \sin\theta-a_s) + P \,G(r\cos\theta, r \sin\theta+a_s) + \sum_{n=-N}^N E_n^{(e)} H_n^{(1)} (\beta_e r) \,e^{in\theta}, 
\end{eqnarray*}
where $N$ needs to be chosen sufficiently large to ensure wave amplitude accuracy ($N \leq 8$ for all the examples considered in this paper).

\begin{figure}[H]
\begin{center}
\subfigure[]{\label{E0_flat_tot_ex_large_scale}\includegraphics[width=.33\textwidth]{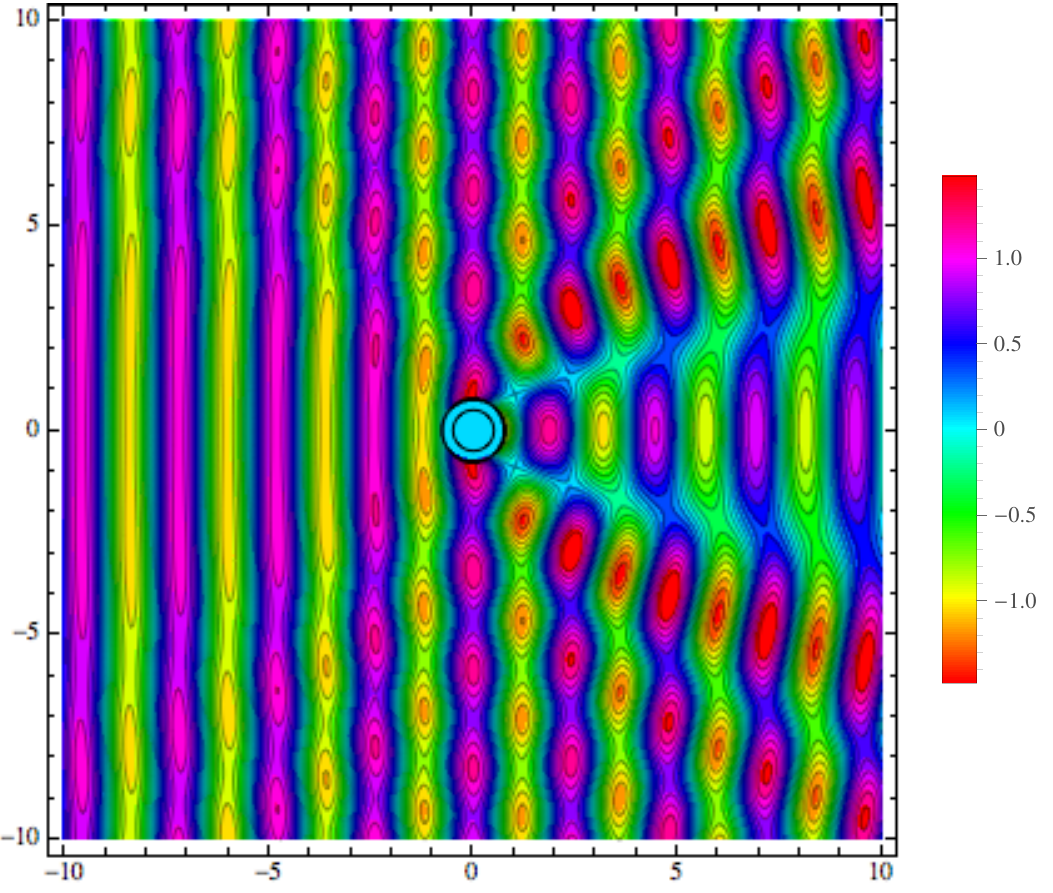}}~~
\subfigure[]{\label{E0_flat_tot_ex_zoom_withsources157}\includegraphics[width=.33\textwidth]{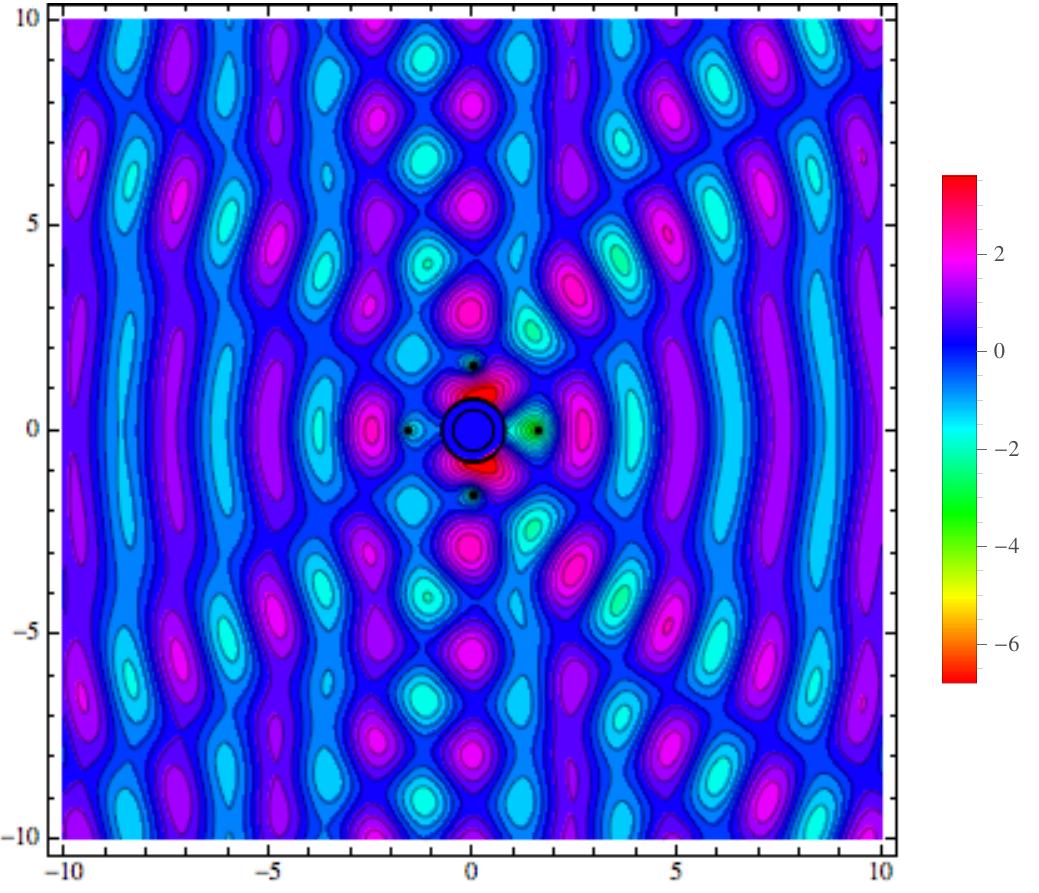}}~~
\subfigure[]{\label{E0_flat_tot_ex_zoom_witheightsources157}\includegraphics[width=.33\textwidth]{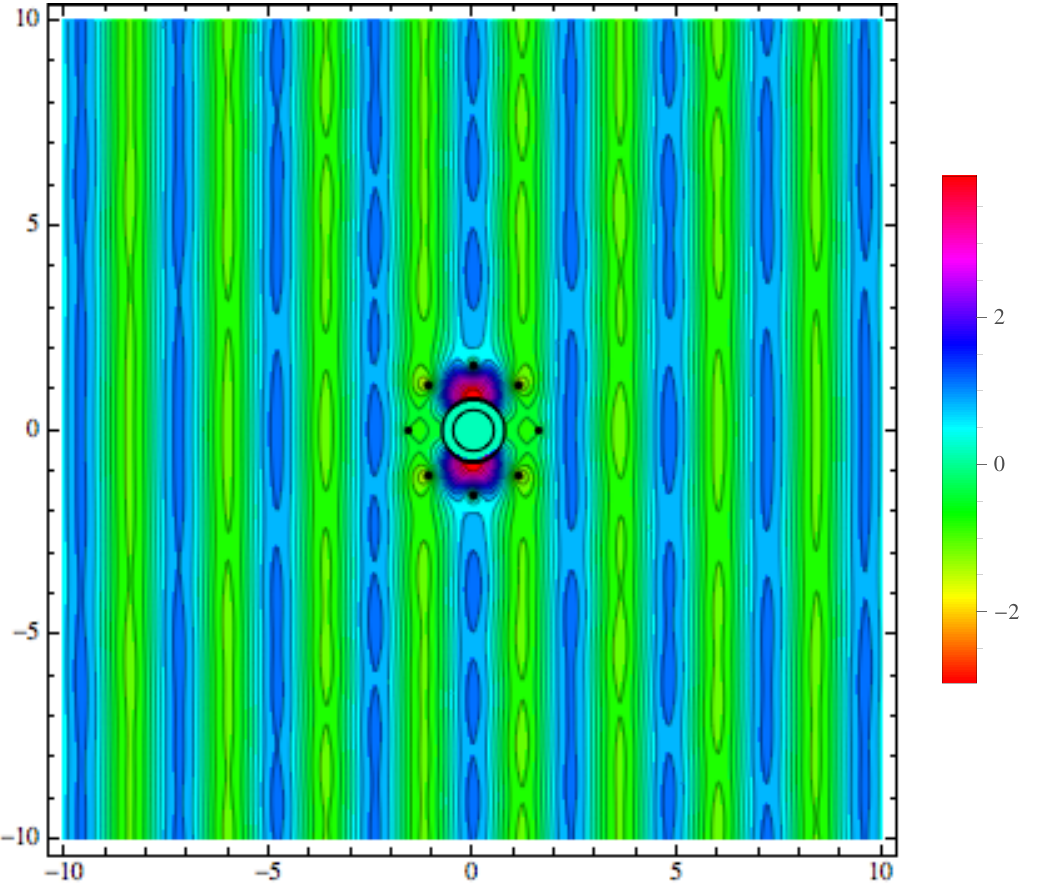}}~~
\caption{Membrane wave scattering: (a) Total displacement field exterior to a coated inclusion with parameter values $\omega=2.62$,  $\rho_i=1.5$,  $\rho_c= 0.81$,   $\rho_e=1.0$,  $\mu_i=0.1$,  $\mu_c=0.11$, $\mu_e=1.0$,  $a_c=0.77$,  $a_i=0.5$ over a square region of side length 20. (b) Total displacement field exterior to a coated inclusion in the presence of 4 active sources located on the axes (black dots), all 1.57 units away from the origin. Parameter values as in (a). (c) Total displacement field exterior to a coated inclusion in the presence of 8 active sources (black dots), all 1.57 units away from the origin; sources off the $x_1$- and $x_2$-axes lie on $x_2 = \pm x_1$. Parameter values as in (a). 
}
\label{E0eFlatMembraneWithSources}
\end{center}
\end{figure}

In Fig. \ref{E0eFlatMembraneWithSources}(a), we present the membrane wave amplitude for the frequency $\omega=2.62$ for an inclusion with a suitably designed coating chosen to give the flattened variation of $|E_0^{(e)}|$ in Fig. \ref{E0_coated_inclusion2}(a); this is the frequency value at which a small deviation creates large fluctuations in the monopole term. Comparing Fig. \ref{E0eFlatMembraneWithSources}(a) with Fig. \ref{omega_262_no_coating} (left) or even with Fig. \ref{omega_300_no_coating} (left), it is clear that the increase in strength of the monopole term results in higher scattering. Figures \ref{E0eFlatMembraneWithSources}(b) and (c) illustrate two attempts, with four and eight active sources located at 1.57 units away from the origin, to cloak the coated inclusion. The equations to be solved to obtain the amplitudes of eight and twelve sources can be found in 
the Supplementary Material.

\begin{figure}[H]
\begin{center}
\subfigure[]{\label{E0_flat_cloak_0p70}\includegraphics[width=.30\textwidth]{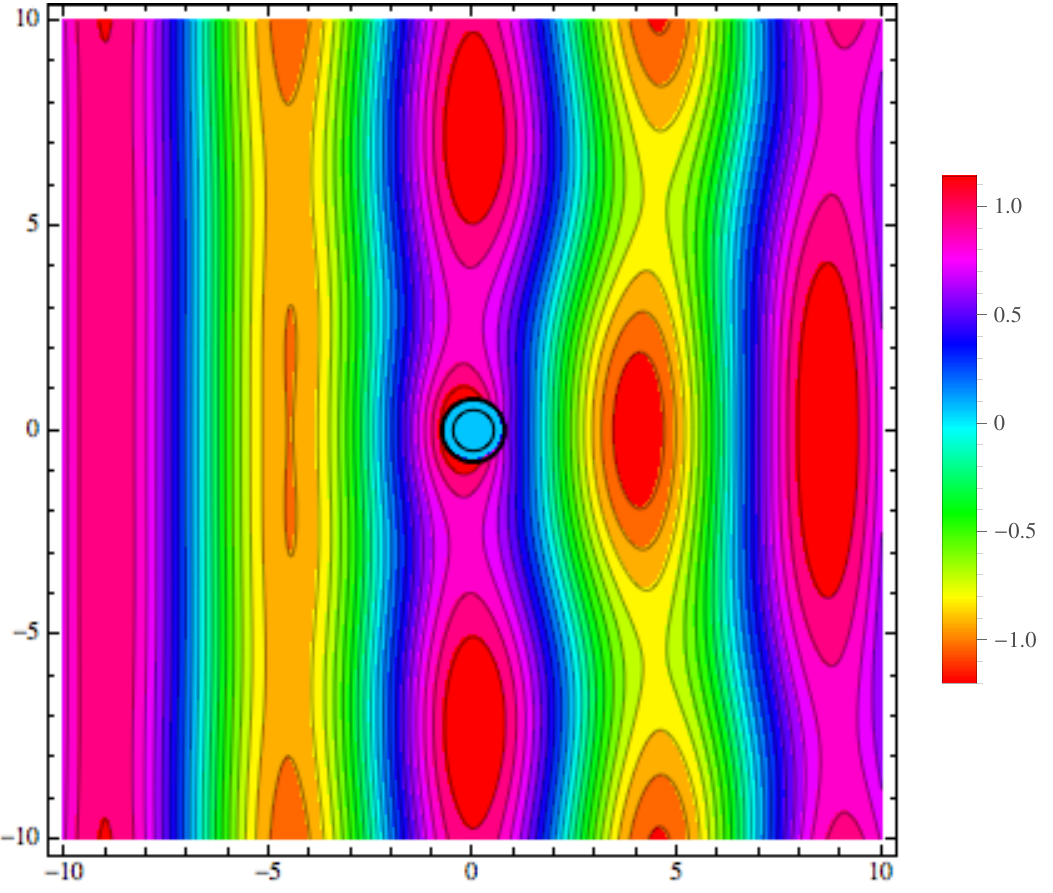} ~~
\includegraphics[width=.30\textwidth]{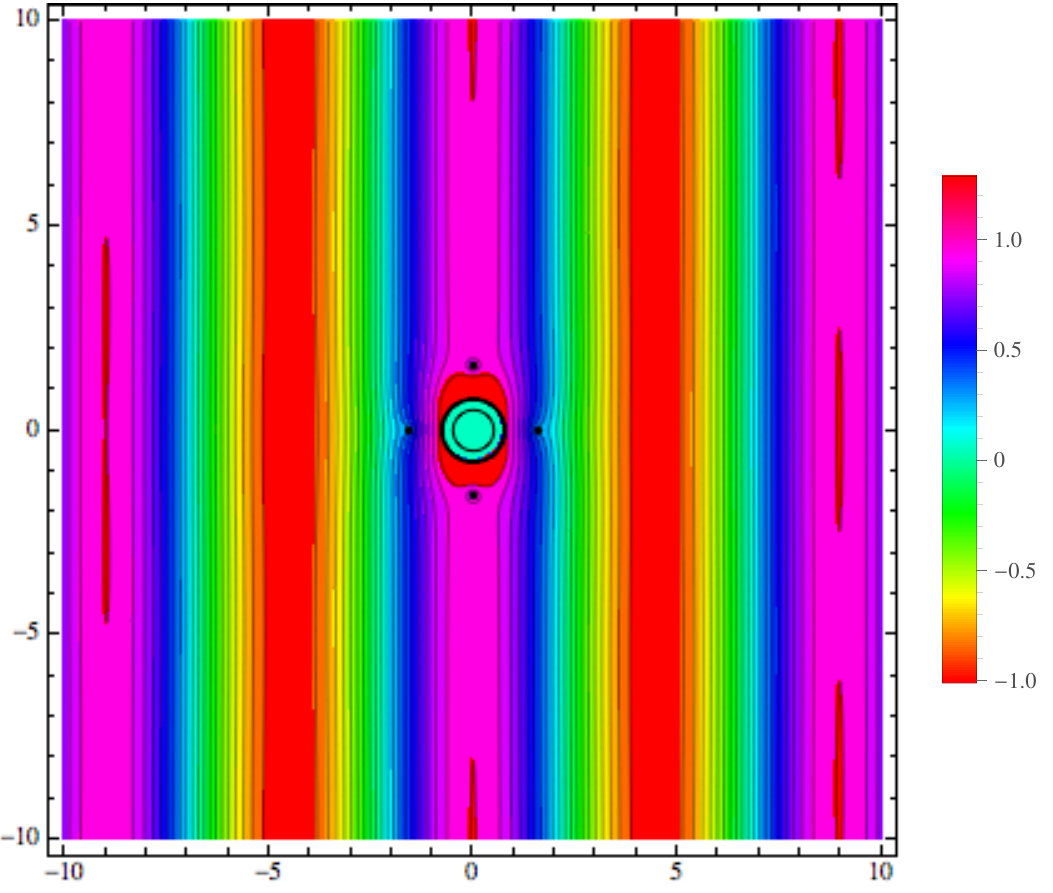}~~
\includegraphics[width=.30\textwidth]{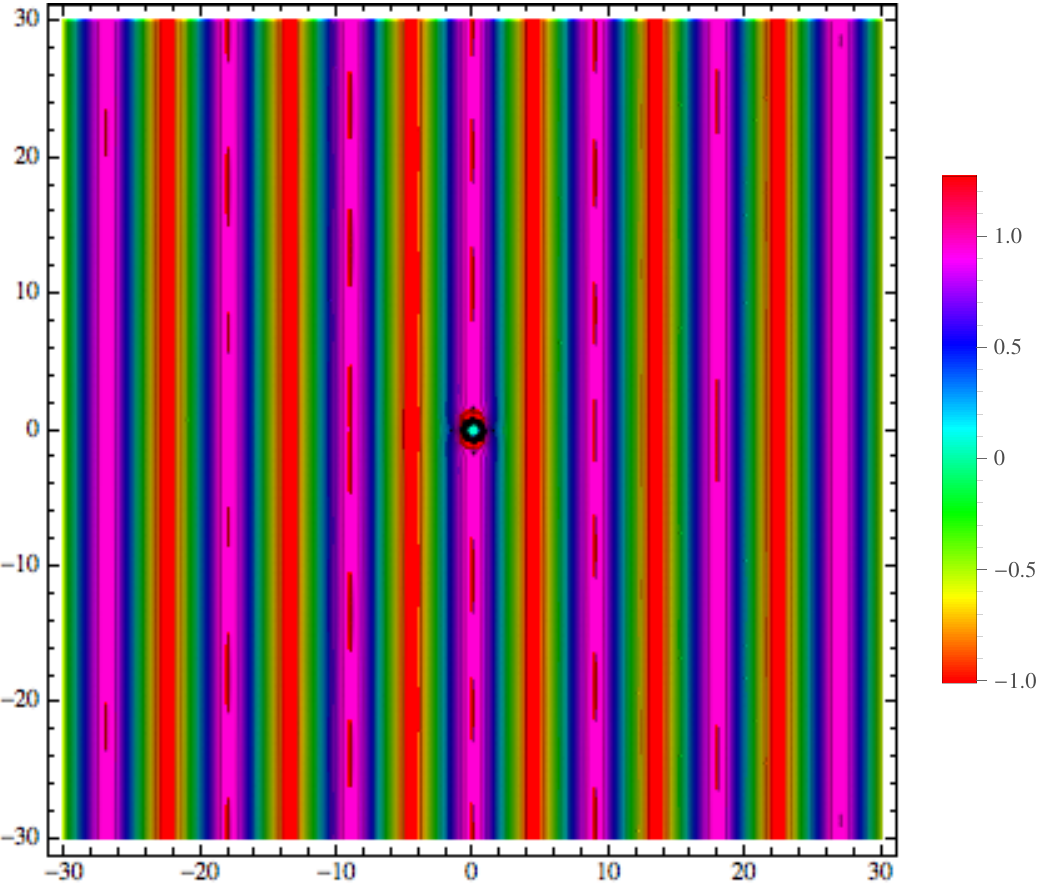}}
\subfigure[]{\label{E0_flat_cloak_0p80}\includegraphics[width=.30\textwidth]{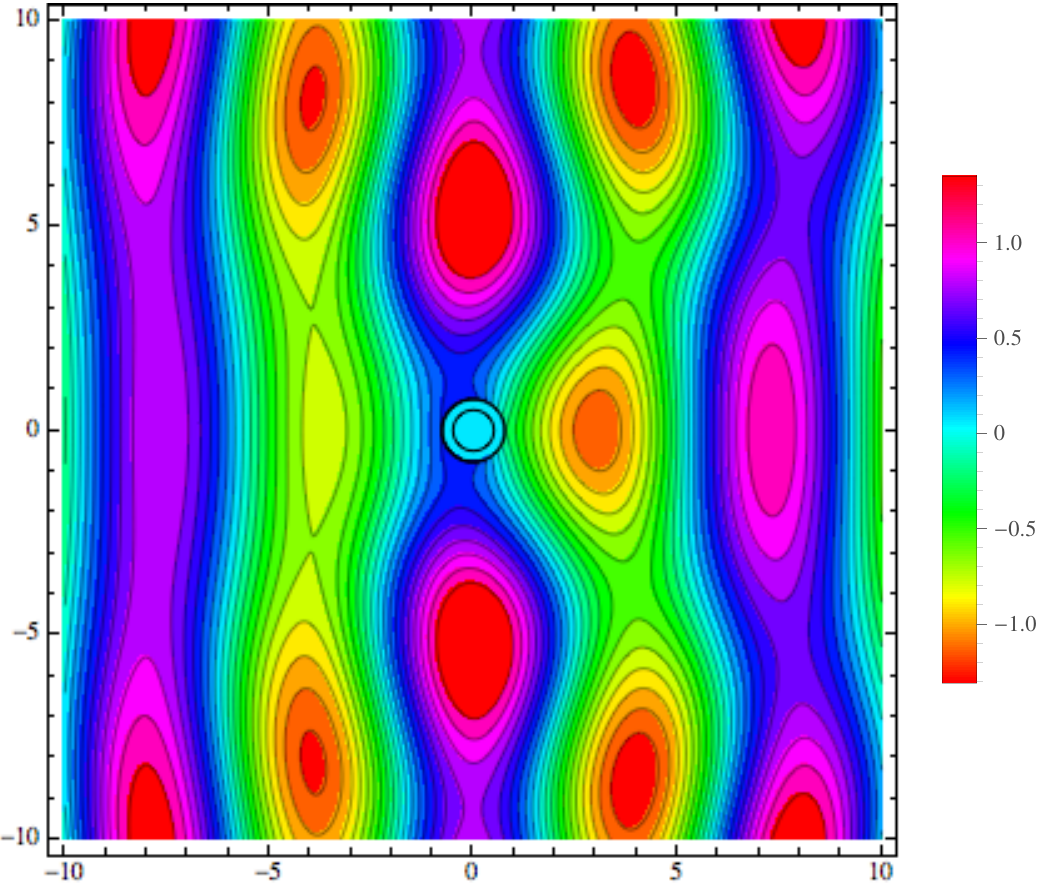} ~~
\includegraphics[width=.30\textwidth]{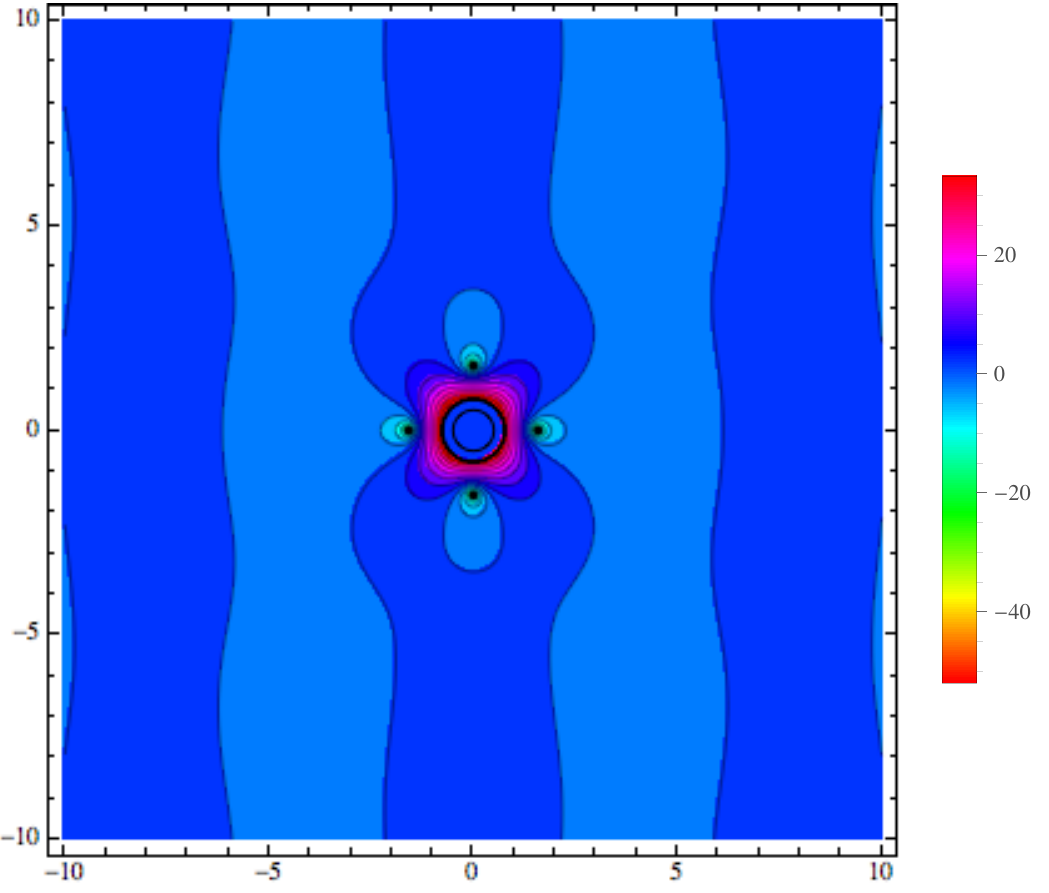}~~
\includegraphics[width=.30\textwidth]{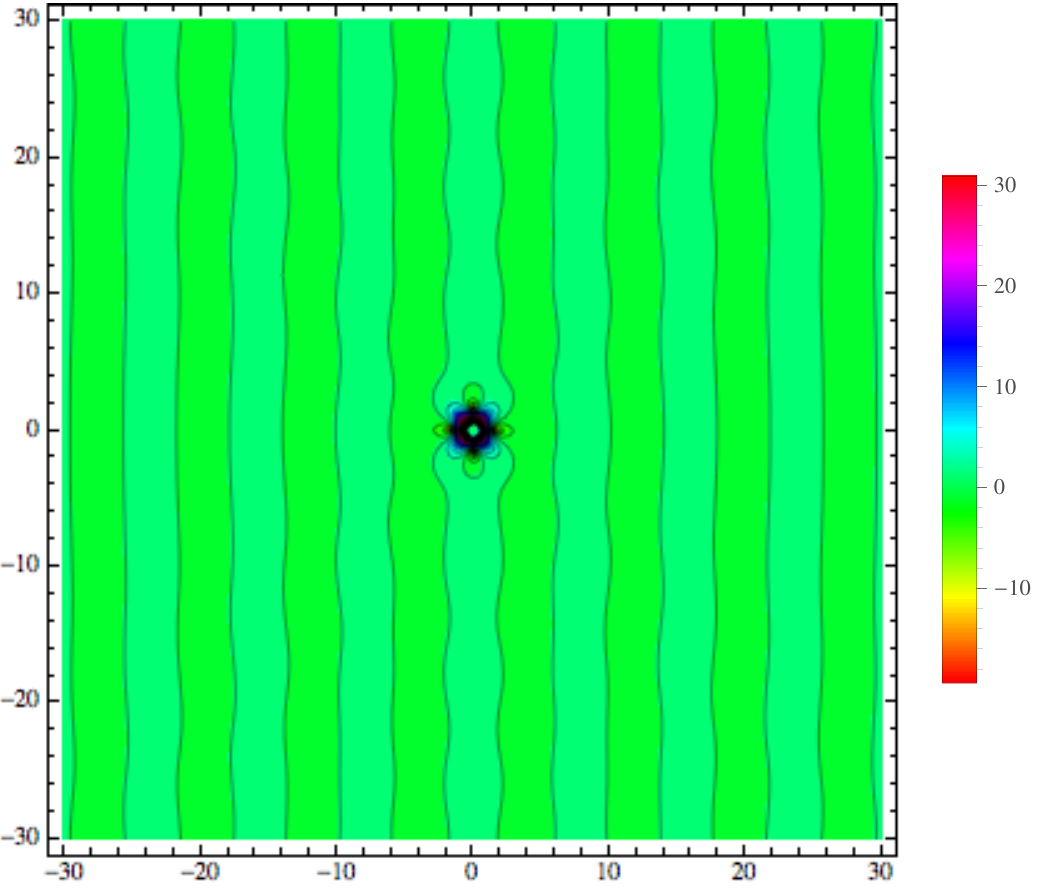}} \\
\subfigure[]{\label{E0_flat_cloak_1p20}\includegraphics[width=.30\textwidth]{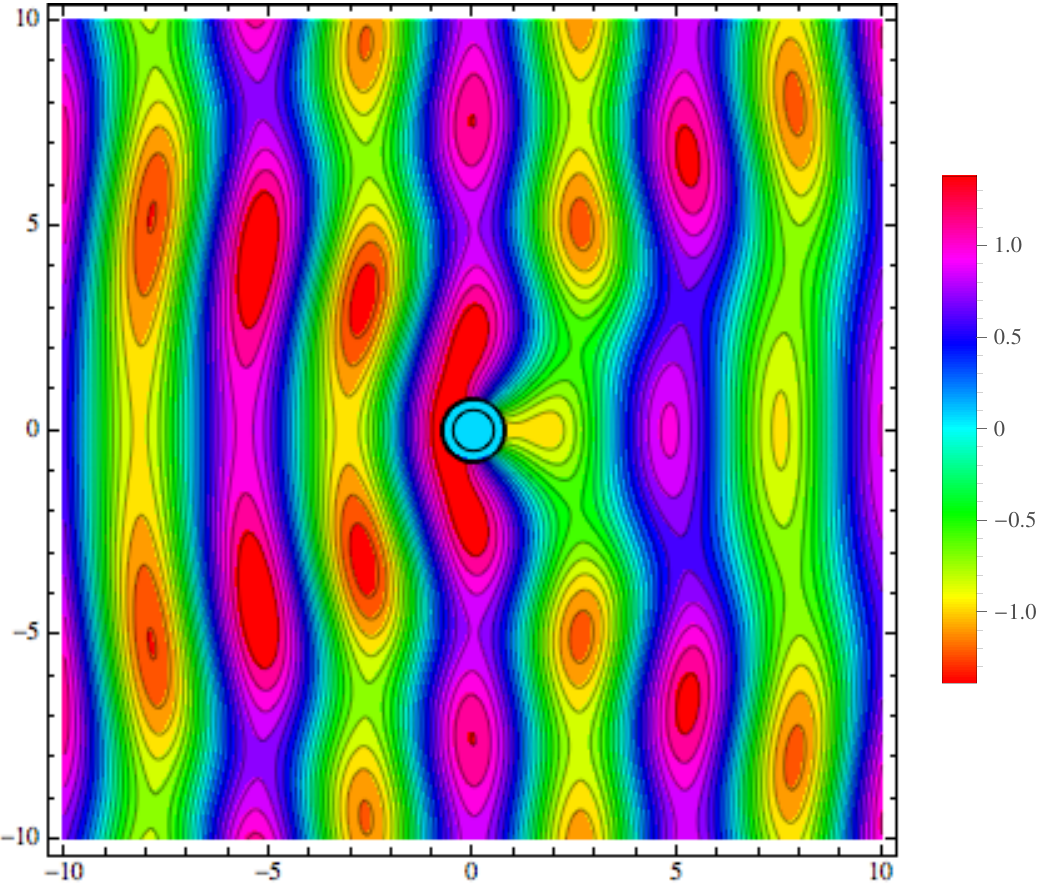} ~~
\includegraphics[width=.30\textwidth]{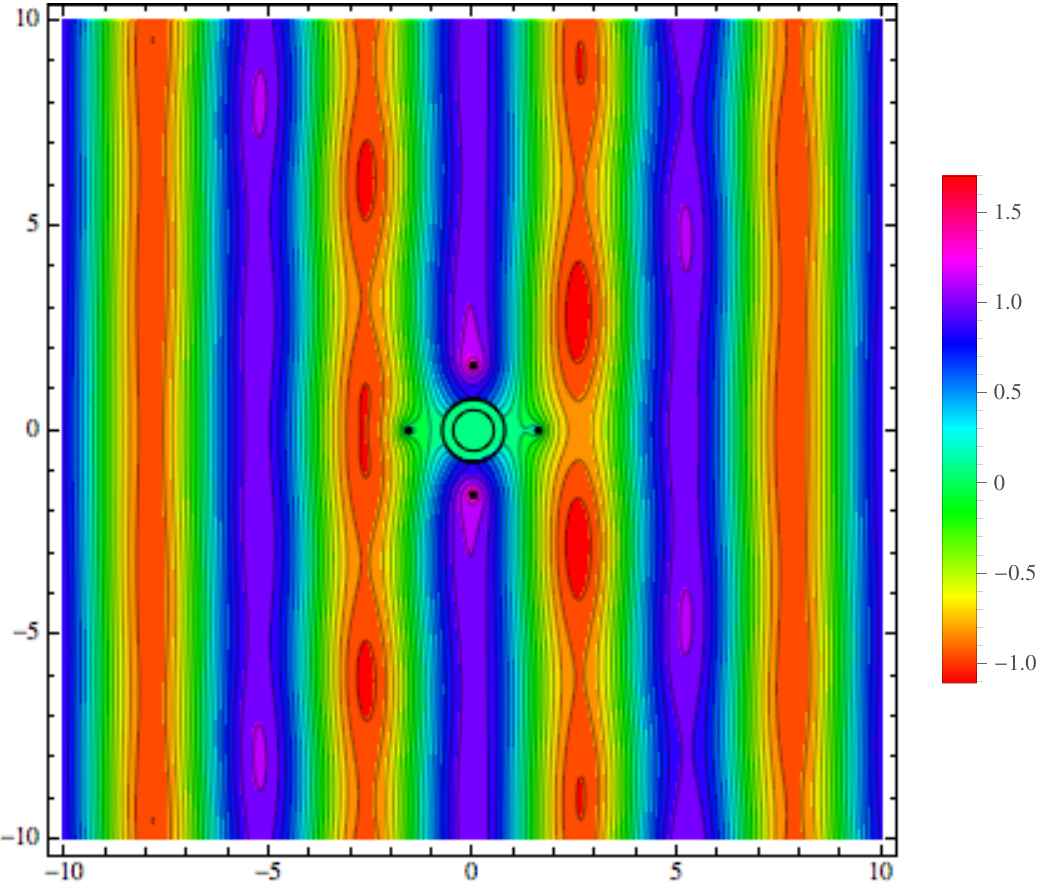} ~~
\includegraphics[width=.30\textwidth]{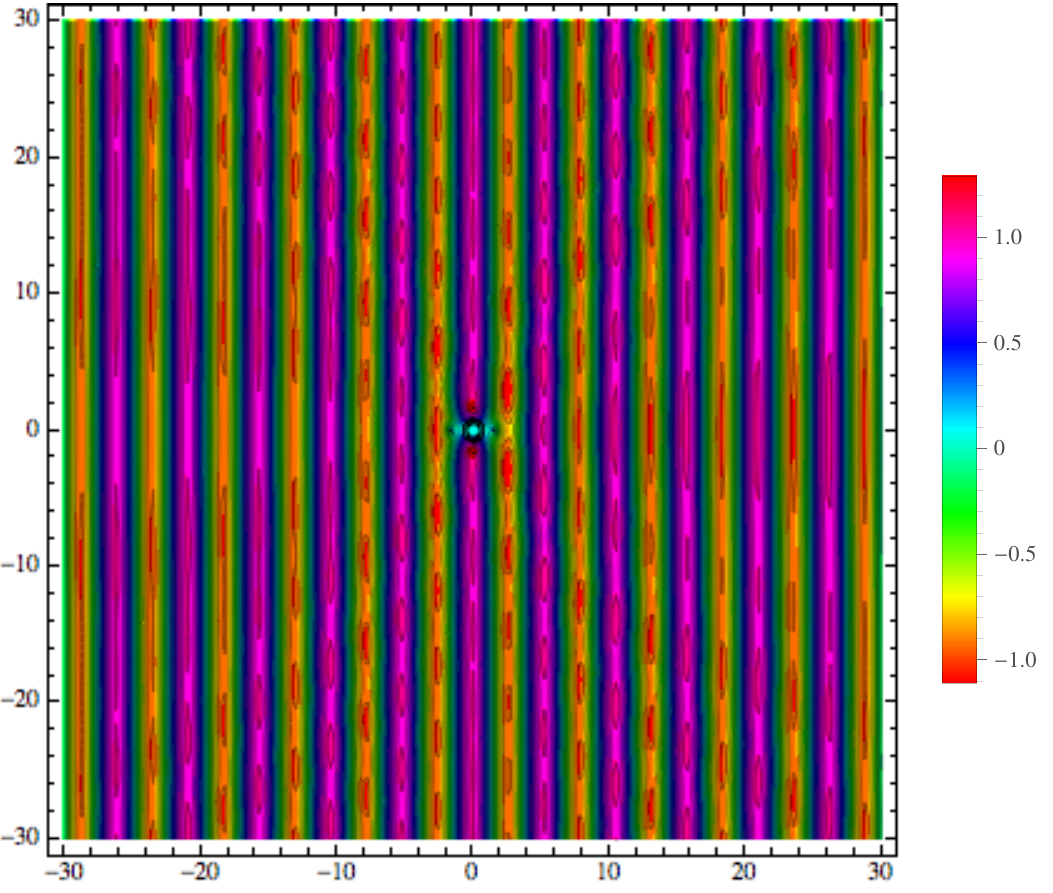}}
\caption{Membrane wave scattering: total displacement field exterior to a coated inclusion, where material properties are as in Fig. \ref{E0eFlatMembraneWithSources}, for
(a) $\omega=0.7$, (b) $\omega=0.8$, (c) $\omega=1.2$. Left: uncloaked coated inclusion; centre: coated inclusion  cloaked  using four active control sources (small black dots surrounding the coated inclusion); right: same as the central images but over a wider range of $x_1$, $x_2$ values.
}
\label{SourcesFreq0p70_0p80_1p20Membrane}
\end{center}
\end{figure}

We also remark that employing four active sources is perfectly adequate for smaller frequencies. To illustrate this, we present cloaking attempts at frequency values $\omega=0.7, 0.8, 1.2$ denoted by $\dagger, \ddagger, \dagger \hspace{-0.8px}\dagger$ in Fig. 
\ref{E0_coated_inclusion2}(a), respectively. These frequency values respectively correspond to small, large and intermediate values of the monopole term, as shown in the same figure using dotted lines. The contour plots for the membrane wave amplitudes when the coated inclusion is surrounded by four active control sources are presented in Fig. \ref{SourcesFreq0p70_0p80_1p20Membrane}: (a) corresponds to $\omega=0.7$, (b) to $\omega=0.8$ and (c) to $\omega=1.2$. In each of these plots, that on the left is with no sources, whereas the middle and the right ones have four sources, the last showing the membrane wave amplitude over a wider range of $x_1, x_2$ values. 
These computations confirm that the values of frequencies in the neighbourhood of the points of maximum for $|E_0^{(e)}|$ 
(see Fig. \ref{E0_coated_inclusion2}) correspond to highly localised vibration modes, as seen in Fig. \ref{E0_flat_cloak_0p80}. 

\begin{figure}[H]
\begin{center}
\subfigure[]{\label{}\includegraphics[width=.31\textwidth]{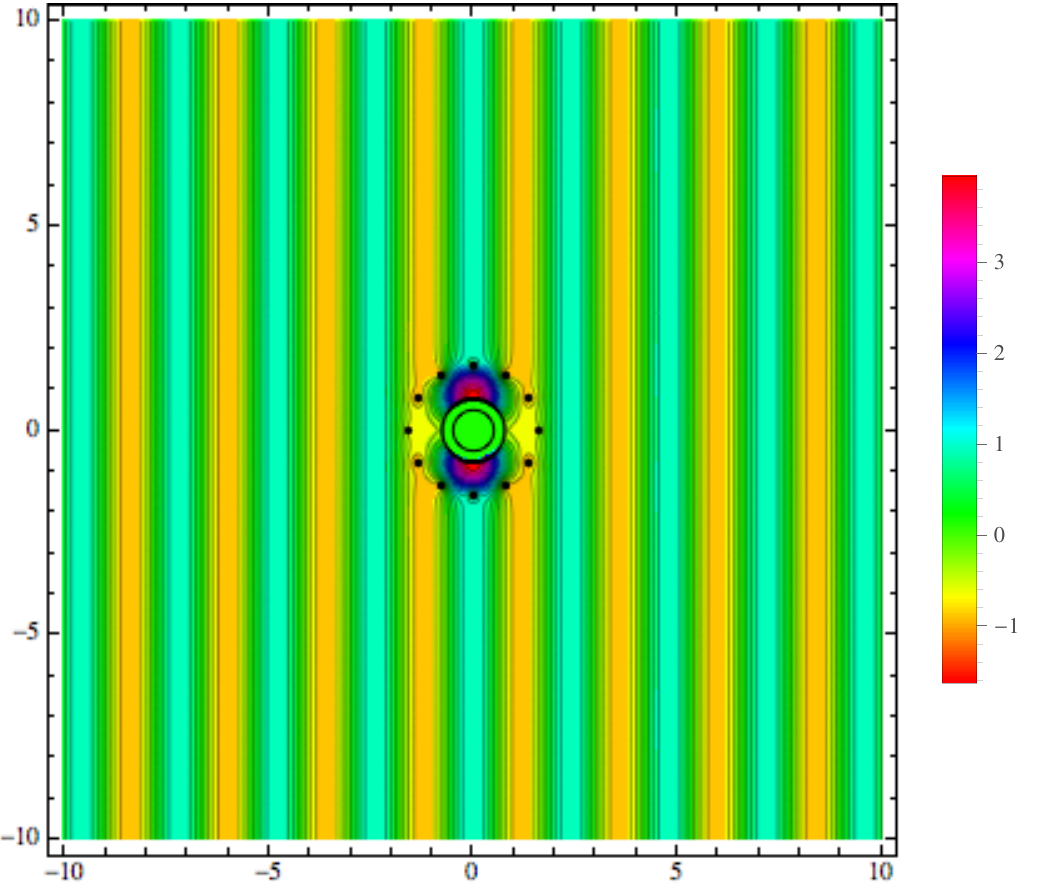}}~~
\subfigure[]{\label{}\includegraphics[width=.31\textwidth]{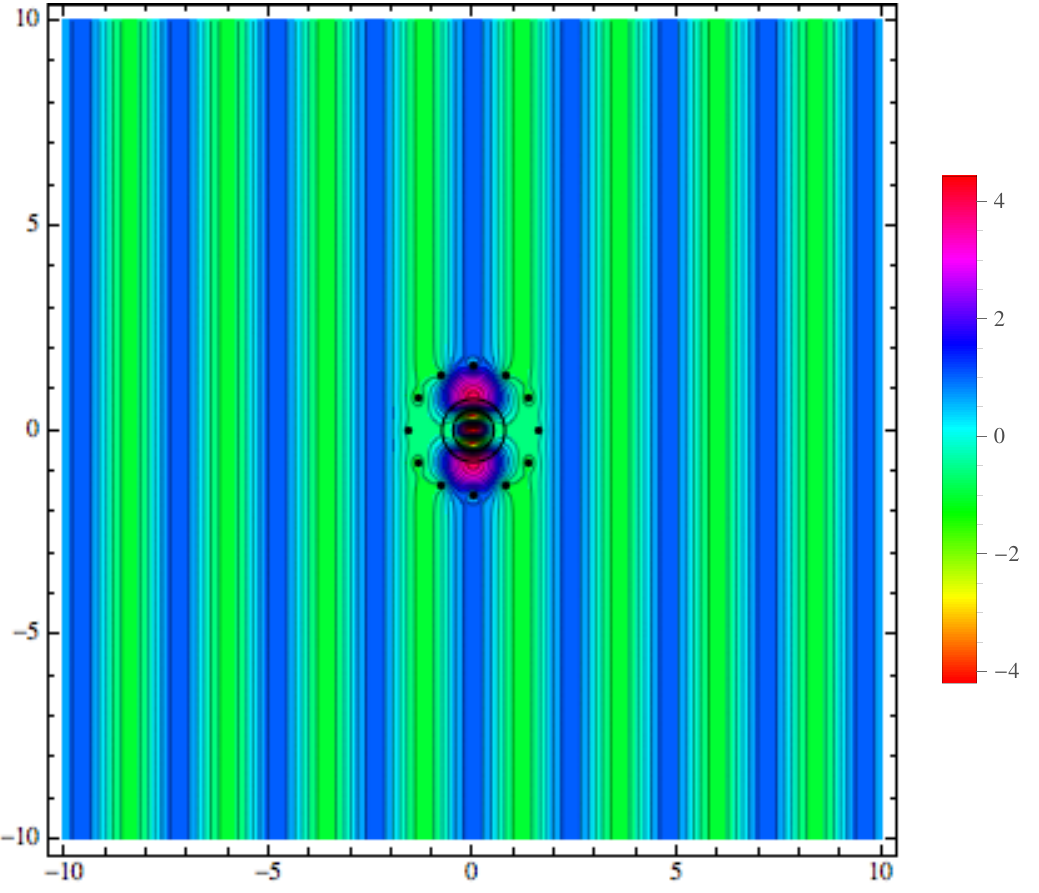}} ~~
\subfigure[]{\label{}\includegraphics[width=.31\textwidth]{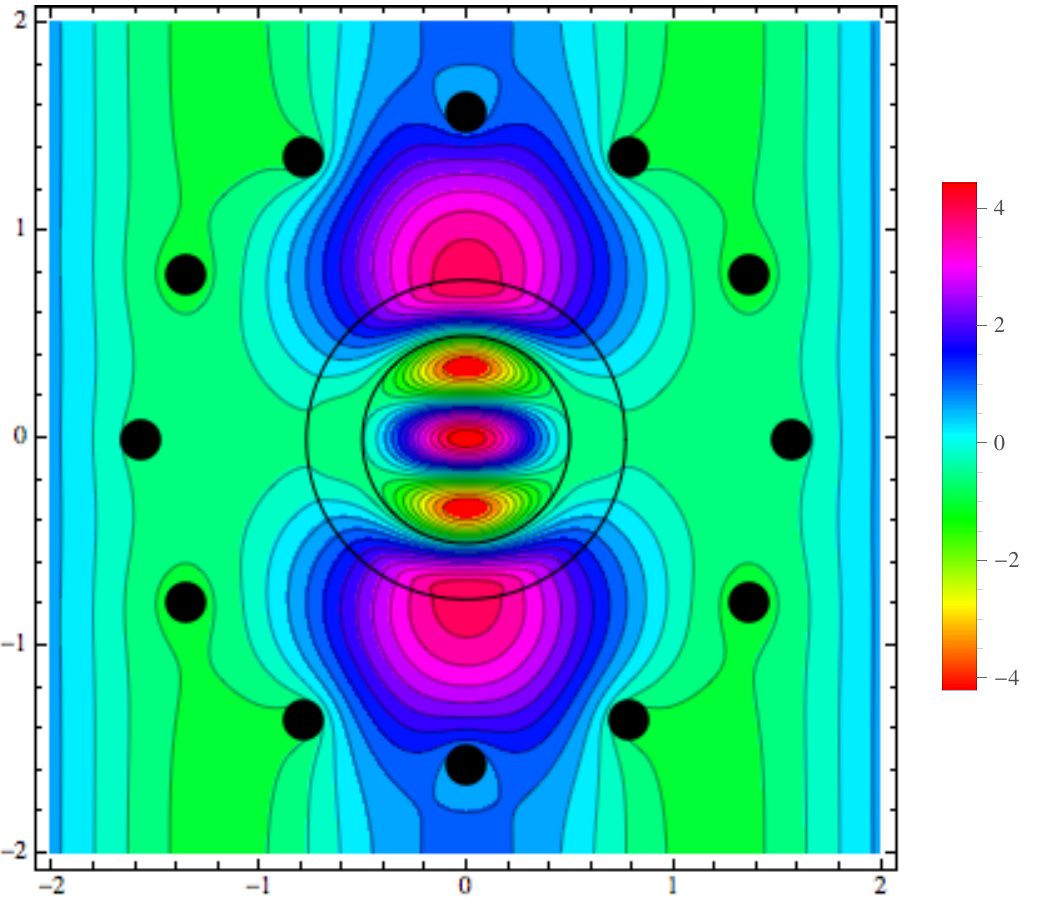}}
\caption{Total displacement field for membrane wave scattering in the presence of 12 active sources (small black dots surrounding the inclusion) all 1.57 units away from the origin: (a) exterior to the coated inclusion, (b) the entire plate including the inclusion and the coating, (c) as in (b) but a detailed view.  The material properties and frequency value are the same as in Fig. \ref{E0eFlatMembraneWithSources}.}
\label{E0eflatwithsources}
\end{center}
\end{figure}

For the higher frequencies of Fig. \ref{E0eFlatMembraneWithSources}, it is necessary to use far more than four sources to achieve good quality cloaking. This is demonstrated in Fig. \ref{E0eflatwithsources}, where cloaking of very good accuracy is achieved using twelve sources. This observation is in good accord with the plot given in Fig. \ref{EneVersusnPlate_a} of the magnitudes of multipole field components. Bearing in mind that the number of active sources is twice the value of the order showing on the horizontal axis of Fig. \ref{EneVersusnPlate_a}, it is evident that good quality cloaking requires ten or more active sources.

Table \ref{OWG_coeffs_membrane} shows the total outgoing wave coefficients propagating away from the configuration of active sources and coated inclusion.  For the case where there are no active sources, we see that there is a core set of  multipole orders in $n \in [-2,2] $ with magnitudes of order unity. When active sources are introduced, it is clear that the method reduces to zero the targeted coefficients (see shaded cells). We see an increase in magnitude for non-targeted coefficents which is generally significant for the two orders adjacent to the core. Noting for example, the magnitude $0.15$ for $n = \pm 6$ with eight sources we see why the cloaking in Fig. \ref{E0_flat_tot_ex_zoom_witheightsources157} is still not adequate, but with twelve sources, we see from Table \ref{OWG_coeffs_membrane} that it is clearly adequate.

\begin{table}[H]
\resizebox{6.5in}{!}{
\footnotesize
\begin{tabular}{|c||r|c|c|c|c|} \hline
& $n$ & {\bf no sources} & 4 sources & 8 sources & 12 sources \\ \hline
\multirow{17}{*}{$\tilde{E}_n^{(e)}$}
& $-8$
&
$-1.34\times10^{-16} + 1.16\times10^{-8} \, i$
& 
$4.14\times10^{-10} - 0.036 \, i$
& 
$4.14\times10^{-10} - 0.036 \, i$
& 
$3.01\times10^{-12} - 0.00026 \, i$
\\ \cline{2-6}
& $-7$
&
$-6.25\times10^{-7} - 3.91\times10^{-13} \, i$
& 
$0.043 + 2.67\times10^-8 \, i$
& 
$ 0.043 + 2.67\times10^{-8} \, i$
& 
$ -0.000092 - 5.72\times10^{-11} \, i$
\\ \cline{2-6}
& $-6$
&
$6.41\times10^{-10} - 0.000025 \, i$
& 
$-3.68\times10^{-6} +  0.15 \, i$
& 
$-3.68\times10^{-6} + 0.15 \, i$
& 
{\cellcolor{gray!25}$-9.72\times10^{-17} + 3.52\times10^{-17}  i$}
\\ \cline{2-6}
& $-5$
&
$0.00074 + 5.50\times10^{-7} \, i$
& 
$0.35 + 0.00026 \, i$
& 
$0.023 + 0.000017 \, i$
& 
{\cellcolor{gray!25}$1.53\times10^{-17} + 3.13\times10^{-17} \,i$}
\\ \cline{2-6}
& $-4$
&
$-0.00026 + 0.016 \, i$
&
$0.036 - 2.21 \, i$
&
{\cellcolor{gray!25}$-8.45\times10^{-17} - 5.90\times10^{-17} \, i$}
& 
{\cellcolor{gray!25}$ -8.38\times10^{-17} + 1.08\times10^{-16} \, i$}
\\ \cline{2-6}
& $-3$
& 
$-0.066 - 0.0044 \, i$
& 
$0.99 + 0.065 \, i$
&
{\cellcolor{gray!25}$8.3\times10^{-17} - 4.94\times10^{-17} \, i$}
&
{\cellcolor{gray!25}$-1.11\times10^{-16} + 1.47\times10^{-16} \, i$}
\\ \cline{2-6}
& $-2$
&
$0.218 - 0.413 i$
&  
{\cellcolor{gray!25}$0$}
& 
{\cellcolor{gray!25}$3.04\times10^{-16} + 3.33\times10^{-16} \, i$}
& 
{\cellcolor{gray!25}$1.25\times10^{-16} $}
\\ \cline{2-6}
& $-1$
& 
$  0.305 + 0.896 i$
& 
{\cellcolor{gray!25}$-1.11\times10^{-16}$}
&
{\cellcolor{gray!25}$1.11\times10^{-16} - 3.47\times10^{-17} \, i$}
& 
{\cellcolor{gray!25}$-2.22\times10^{-16} + 2.43\times10^{-17} \, i$}
\\ \cline{2-6}
& $0$
& 
$-0.963 - 0.188 i$
& 
{\cellcolor{gray!25}$-5.55\times10^{-17} - 4.44\times10^{-16} \, i$}
& 
{\cellcolor{gray!25}$-5.55\times10^{-17} + 4.44\times10^{-16} \, i$}
& 
{\cellcolor{gray!25}$-1.67\times10^{-16} + 4.44\times10^{-16} \, i$}
\\ \cline{2-6}
& $1$
& 
$ -0.305 - 0.896$
&
{\cellcolor{gray!25} $ 1.11\times10^{-16} $}
&
{\cellcolor{gray!25}$-1.11\times10^{-16} + 3.47\times10^{-17} \, i$}
& 
{\cellcolor{gray!25}$2.22\times10^{-16} - 2.43\times10^{-17} \, i$}
\\ \cline{2-6}
& $2$
& 
$0.218 - 0.413 i$
& 
{\cellcolor{gray!25}$0$}
& 
{\cellcolor{gray!25}$3.04\times10^{-16} + 3.33\times10^{-16} \, i$}
& 
{\cellcolor{gray!25}$1.25\times10^{-16} $}
\\ \cline{2-6}
& $3$
& 
$0.066 + 0.0044 \, i$
&
$-0.99 - 0.065 \, i$
& 
{\cellcolor{gray!25}$-8.3\times10^{-17} + 4.94\times10^{-17} \, i$}
& 
{\cellcolor{gray!25}$1.11\times10^{-16} - 1.47\times10^{-16} \, i$}
\\ \cline{2-6}
& $4$
&
$-0.00026 + 0.016 i$
& 
$0.036 - 2.21 \, i$
& 
{\cellcolor{gray!25}$-8.45\times10^{-17} - 5.90\times10^{-17} \, i$}
&
{\cellcolor{gray!25}$ -8.38\times10^{-17} + 1.08\times10^{-16} \, i$}
\\ \cline{2-6}
& $5$
& 
$-0.00074 - 5.50\times10^{-7} i$
&
$-0.35 - 0.00026 \, i$
&
$-0.023 - 0.000017 \, i$
& 
{\cellcolor{gray!25}$-1.53\times10^{-17} - 3.13\times10^{-17} \,i$}
\\ \cline{2-6}
& $6$
& 
$6.41\times10^{-10} - 0.000025 \, i$
&
$-3.68\times10^{-6} +  0.15 \, i$
&
$-3.68\times10^{-6} + 0.15 \, i$
& 
{\cellcolor{gray!25}$-9.72\times10^{-17} + 3.52\times10^{-17}  i$}
\\ \cline{2-6}
& $7$
& 
$6.25\times10^{-7} + 3.91\times10^{-13} \, i$
&
$-0.043 - 2.67\times10^-8 \, i$
&
$ 0.043 + 2.6\times10^{-8} \, i$
& 
$ 0.000092 + 5.72\times10^{-11} \, i$
\\ \cline{2-6}
& $8$
& 
$-1.34\times10^{-16} + 1.16\times10^{-8} \, i$
&
$4.14\times10^{-10} - 0.036 \, i$
&
$4.14\times10^{-10} - 0.036 \, i$
& 
$3.01\times10^{-12} - 0.00026 \, i$
\\ \cline{1-6}
\end{tabular}
}
\caption{The coefficients $\tilde{E}_n^{(e)}$ of $H_n^{(1)} (\beta_e r)$ terms, for a configuration with zero, four, eight and twelve control sources positioned symmetrically on a circle of radius of 1.57 units from the origin.  The frequency of the incident wave is $\omega = 2.62$.}
\label{OWG_coeffs_membrane}
\end{table}

\section{Flexural waves for Kirchhoff plates with coated inclusions: resonant regimes}
\label{flex_res}

As discussed in Section \ref{prob_form}, the fields $w^{(k)}({\bf x}),\, k=i,c,e,$ of (\ref{gov_eqn_simp}) are represented as a sum of solutions of the Helmholtz and modified Helmholtz equations,  $w_H^{(k)}$ and $w_M^{(k)}$ respectively
(see (\ref{w_eq_wHh_wMHh}), (\ref{Hh_MHh})).
In particular, when mass densities differ from the matrix to the inclusion, but elastic properties, flexural rigidities and Poisson's ratio, of the inclusion, coating and ambient matrix are the same, the four transmission conditions (\ref{disp_and_normal_der_disp_cont})--(\ref{transverse_force__cont}) on each circular interface reduce to conditions of continuity of the displacement  $w$ and its three derivatives with respect to $r$, i.e. in terms of jump conditions
\begin{equation}\label{jump_conditions}
[w({\bf x})] =0, ~[\frac{\partial^k}{\partial r^k} w({\bf x})] = 0, ~\mbox{with} ~ k=1,2,3, ~\mbox{and } {\bf x} ~\mbox{being a point on a circular interface}. 
\end{equation}
In this case and at large wavelengths (compared with the inclusion size) the displacement $w^{(e)}$ in the matrix is dominated by the Helmholtz wave, which satisfies the first two interface conditions (\ref{disp_and_normal_der_disp_cont}) for the displacement and its normal derivative, with the discrepancy in two remaining interface conditions being asymptotically small (see 
Supplementary Material).
The mass balance formula (\ref{mass_balance}), applied for membrane waves to make the monopole term in the scattered field vanish, can then be used as an approximation.
However, in the intermediate regime when the frequency of the incident wave becomes sufficiently large, the coupling between $w_H^{(k)}$ and $w_M^{(k)}$ becomes essential through the interface conditions (\ref{moment_cont}), (\ref{transverse_force__cont}). In the intermediate regime, the formula (\ref{mass_balance}) is no longer valid, just as it is invalid when the elastic properties of the inclusion, coating and ambient matrix are different.

As shown in the previous section for membrane waves, the coating around an inclusion can be used to control the monopole term in the expansion of the outgoing wave. Note also that the desired control is not necessarily aimed at making the monopole term vanish. Near the resonant frequencies of membranes, we have shown that the active cloaking cannot be achieved using the simple mass balance argument of \cite{MF_PYC_HB_SE_SG_AA}. On the other hand, the use of the coating enables one to shift away from a resonance mode, and hence to implement high-precision active cloaking by multiple sources. 

Here, we show that Kirchhoff plates allow a similar treatment of resonant regimes. However, the coated inclusion argument requires a significant rethinking, as the mass balance argument, working very well for membrane waves, proves to be inadequate for the case of Kirchhoff plates. As expected, the additional features of flexural Kirchhoff waves, scattered by an inclusion, are related to the interaction between the Helmholtz and the modified Helmholtz waves $w^{(k)}_H$ and  $w^{(k)}_M$ defined in (\ref{Hh_MHh}).

\subsection{Monopole term due to Konenkov}
The scattering of a monochromatic plane flexural wave by a circular obstacle with various boundary conditions was intensively studied by Konenkov \cite{YuKK}. In section 4 of \cite{YuKK}, the author gives a formula for the scattered field for small values of $kR$ (in original notations $k$ stands for the wave number and $R$ for the radius of the inclusion; this corresponds to $\beta_e a_i$ in our notations) at a large distance away from the obstacle. This formula translates to
$$
w_{(sc)} \sim i H_0^{(1)} (\beta_e r) \left\{-\frac{\pi}{2} \frac{\frac{\rho_i}{\rho_e} \left[\frac{D_e}{D_i}(1-\nu_e)+(1+\nu_i)\right] + 2 \left[\frac{D_e}{D_i}\nu_e-(1+\nu_i)\right]}{\frac{D_e}{D_i} \left[(1+\nu_i)+\frac{D_e}{D_i}(1-\nu_e)\right]} (\beta_e a_i)^2 + \dots\right\} ~~ \text{as } ~\beta_e a_i \to 0.
$$
With the aim of making the monopole coefficient $E_0^{(e)}$ asymptotically zero to the leading-order as $\beta_e a_i \to 0$, 
we require 
\begin{equation}
{\cal F} (\rho_i, \rho_e, \nu_i, \nu_e, D_i, D_e) = \frac{\rho_i}{\rho_e} \left[\frac{D_e}{D_i}(1-\nu_e)+(1+\nu_i)\right] + 2 \left[\frac{D_e}{D_i}\nu_e-(1+\nu_i)\right] = 0. \label{E0ePlateNoCoatingAsympZero}
\end{equation}
This immediately highlights the difference between membrane and flexural waves in our investigation: all material properties of the inclusion and the exterior are present in equation (\ref{E0ePlateNoCoatingAsympZero}) for flexural waves, whereas for membrane waves the vanishing of $E_0^{(e)}$ to the leading-order merely involves the densities (see formula (\ref{E0e_asy})).

\begin{figure}[H]
\begin{tikzpicture}
\node[inner sep=0pt] at (0.0,0.0)
{\includegraphics[width=.45\textwidth]{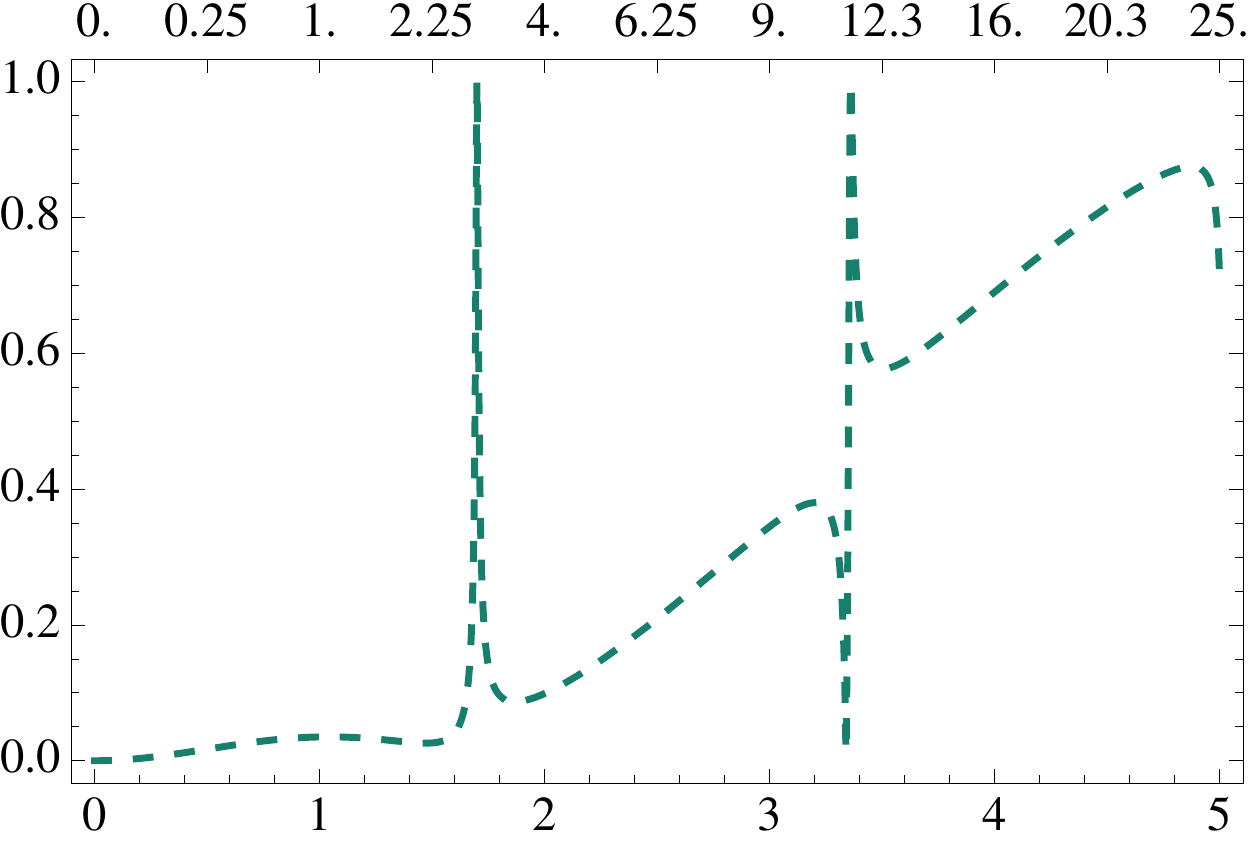}};
\node at (4.2,-2.1) {$\beta_e$};
\node at (4.2,2.2) {$\omega$};
\node at (-4.5,0.0) {$|E_0^{(e)}|$};
\draw [blue, thick] (-3.28,-1.8) rectangle (-3.15,-2.2);
\node at (0.2,-3.0) {(a)};

\node[inner sep=20pt] (figright) at (8.0,0.0)
{\includegraphics[width=.35\textwidth]{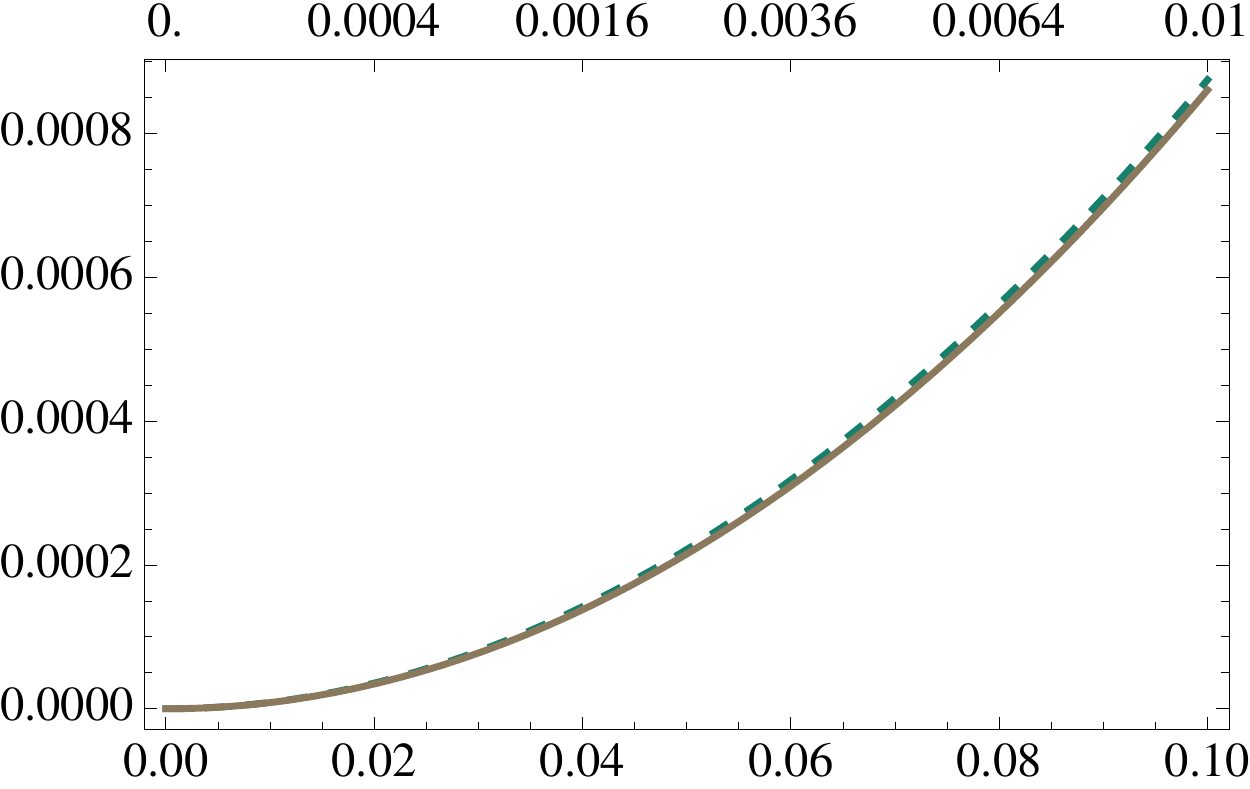}};
\node at (11.2,-1.50) {$\beta_e$};
\node at (4.5,0.0) {$|E_0^{(e)}|$};
\node at (11.2,1.6) {$\omega$};
\node at (8.5,-3.0) {(b)};

\end{tikzpicture}
\caption{Flexural wave scattering by an uncoated inclusion with the parameters  $\rho_i=0.05$, $\rho_e=1.0$, $D_i=2.5\times10^{-4}$, $D_e=1.0$, $a_i=0.50$: (a) $|E_0^{(e)}|$ as a function of $\beta_e$ (lower horizontal axis) and of $\omega$ (upper horizontal axis), (b) Detail from (a) showing  $|E_0^{(e)}|$ (green/dashed) and the empirical fit $0.086 \,\beta_e^2$ (grey/solid). }
\label{Flexural_scat_uncoated_inclusion}
\end{figure}
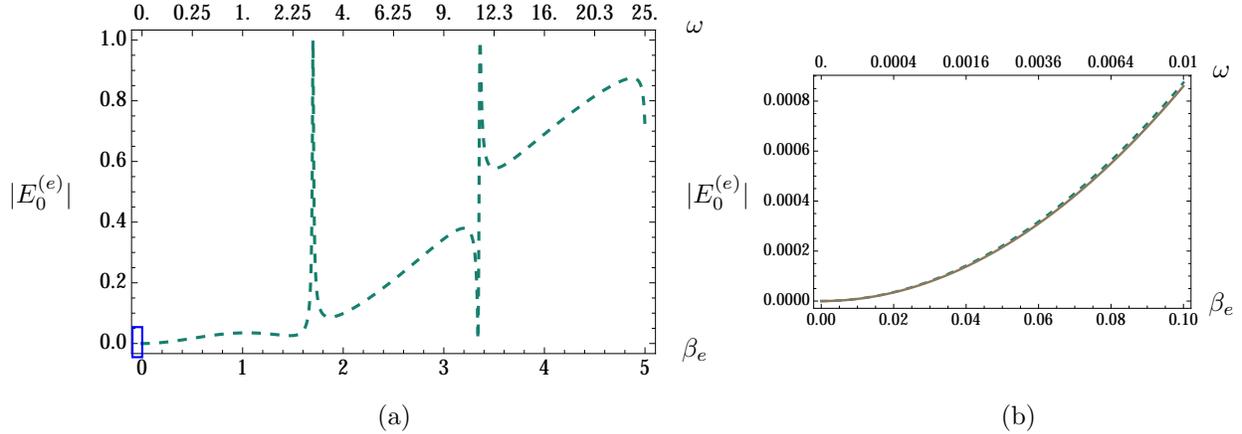

As in the case of an uncoated inclusion in the presence of membrane waves (section \ref{NoCoatingMembrane}), a strong resonant behaviour of the scattering coefficient $E_0^{(e)}$ for a high-contrast inclusion is  observed for flexural waves. Figure \ref{Flexural_scat_uncoated_inclusion}(a) shows $|E_0^{(e)}(\beta_e)|$ for a specific choice of high-contrast material properties $\rho_i=0.05$, $\rho_e=1.0$, $D_i=1.5\times10^{-4}$, $D_e=1.0$ and $r_i=0.50$. In Fig.  \ref{Flexural_scat_uncoated_inclusion}(b) we illustrate the behaviour of $|E_0^{(e)}(\beta_e)|$ for small $\omega$ (solid blue rectangle in Fig. \ref{Flexural_scat_uncoated_inclusion}(a)). We see the quadratic behaviour of the monopole term for small $\beta_e$; this is expected according to Konenkov's asymptotic formula (see the top right formula on p.156 of \cite{YuKK}) in which he states this  behaviour is 
$$
E_0^{(e)} \sim {\cal F} (\rho_i, \rho_e, \nu_i, \nu_e, D_i, D_e)\, (\beta_e a_i)^2 \quad~~ \text{as } ~\beta_e a_i \to 0,
$$ 
(in the original notations $J(\nu,\varepsilon,\sigma,\sigma')\,(kR)^2$, $kR \to 0$). 

To clarify, it is important to note that a comparison between Fig. \ref{Flexural_scat_uncoated_inclusion} and Fig. \ref{Acoust_scat_uncoated_inclusion} requires $|E_0^{(e)}|$ to be plotted as a function of $\beta_e$ for flexural waves; this is because the spectral parameters are different for membrane wave and flexural wave problems, that is $\beta_k = \omega \sqrt{\rho_k/\mu_k}$ and $\beta_k^2 = \omega\sqrt{\rho_k h / D_k}$, respectively. In other words, $|E_0^{(e)}|$ versus $\omega$ for membrane waves corresponds to $|E_0^{(e)}|$ versus $\beta_e$ (up to a multiplicative constant $\sqrt{\rho_k/\mu_k}$), which is directly comparable to Fig. \ref{Flexural_scat_uncoated_inclusion}.

\begin{figure}[H]
\begin{center}
\subfigure[]{\label{E0eSmallPlateNoCoating}
\includegraphics[width=.33\textwidth]{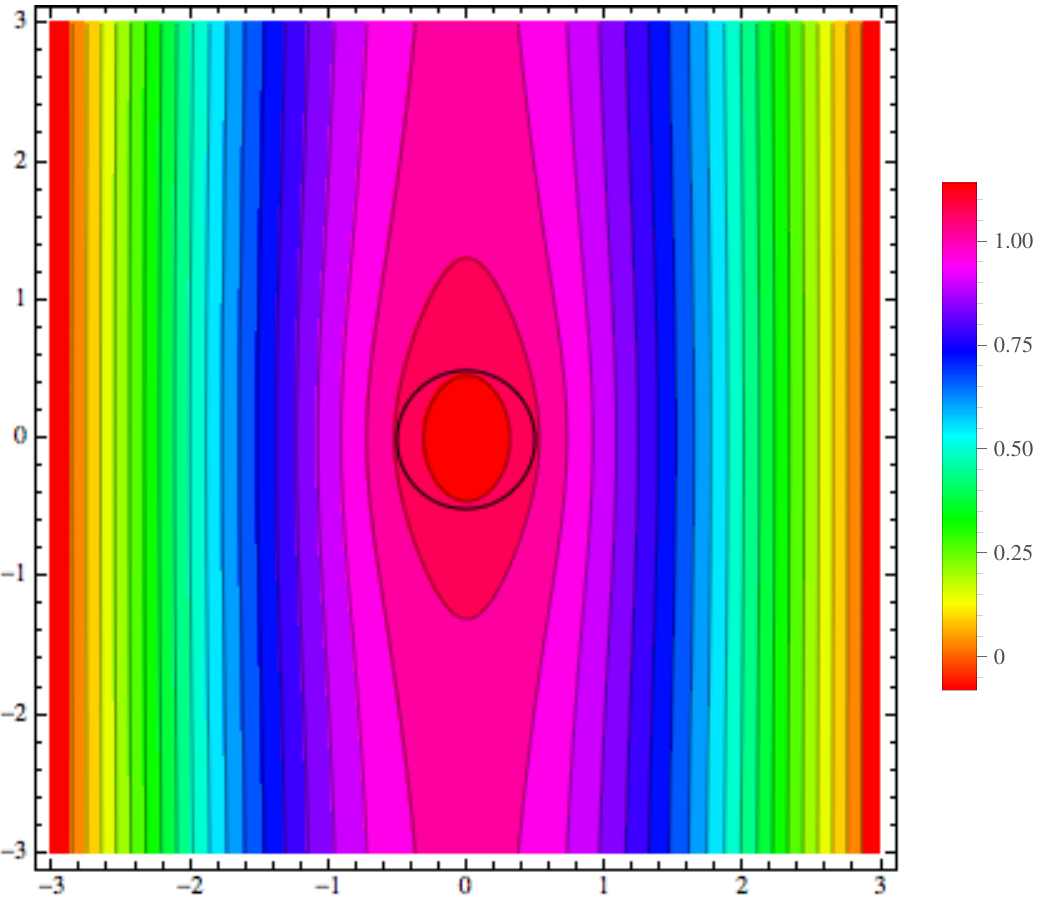} ~~
\includegraphics[width=.33\textwidth]{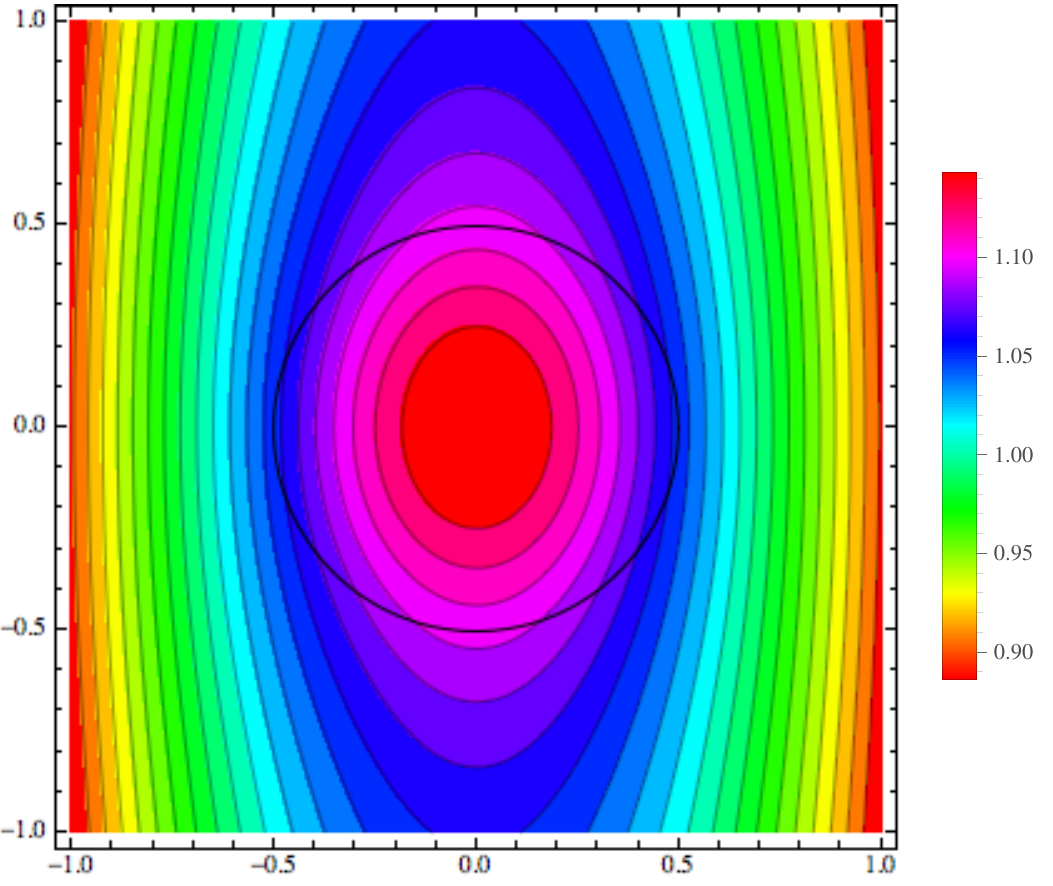}} \\
\subfigure[]{\label{E0eZeroPlateNoCoating}
\includegraphics[width=.33\textwidth]{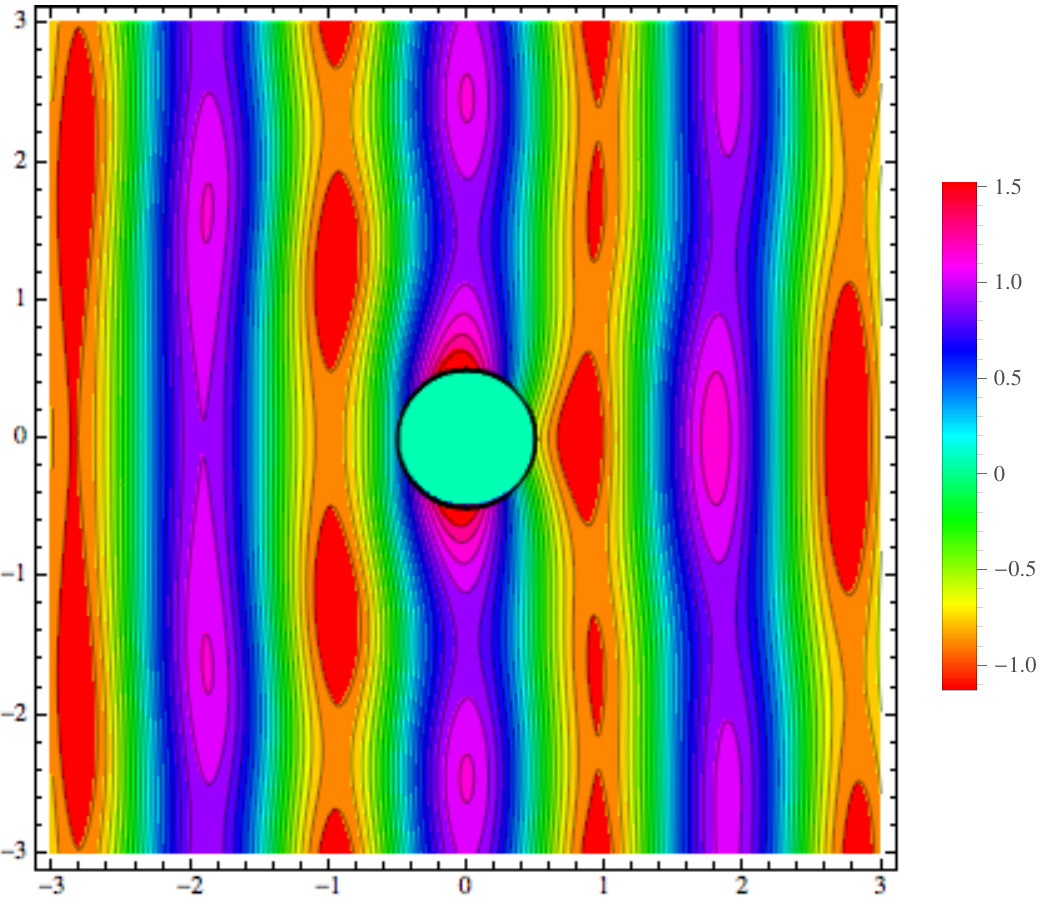}~~
\includegraphics[width=.33\textwidth]{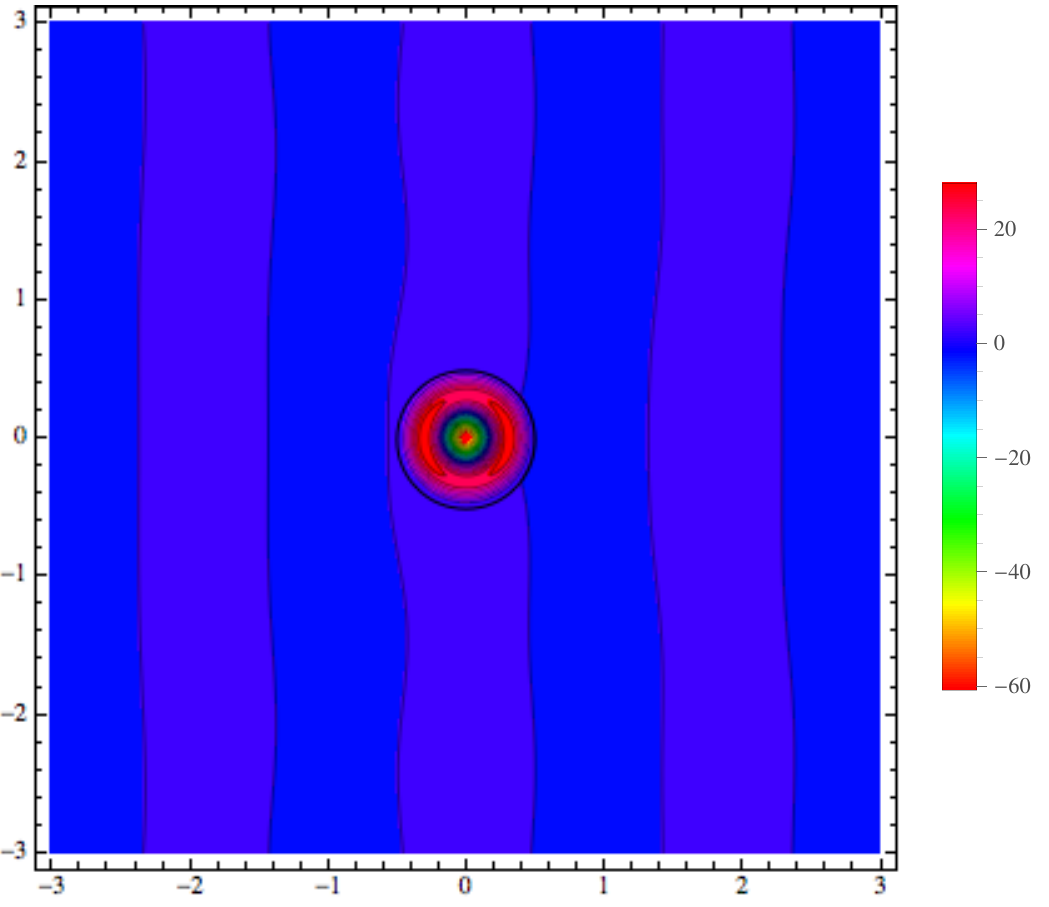}~~
\includegraphics[width=.33\textwidth]{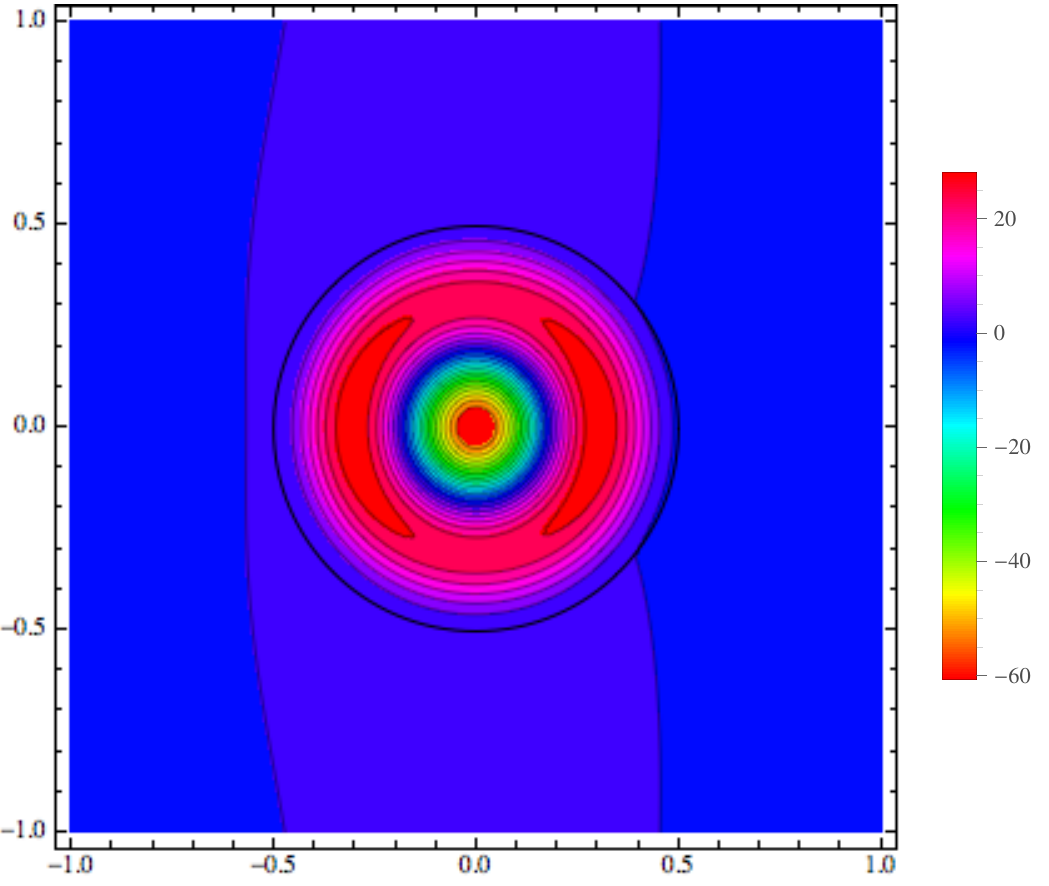}} \\
\subfigure[]{\label{E0eZeroMaxPlateNoCoating}
\includegraphics[width=.33\textwidth]{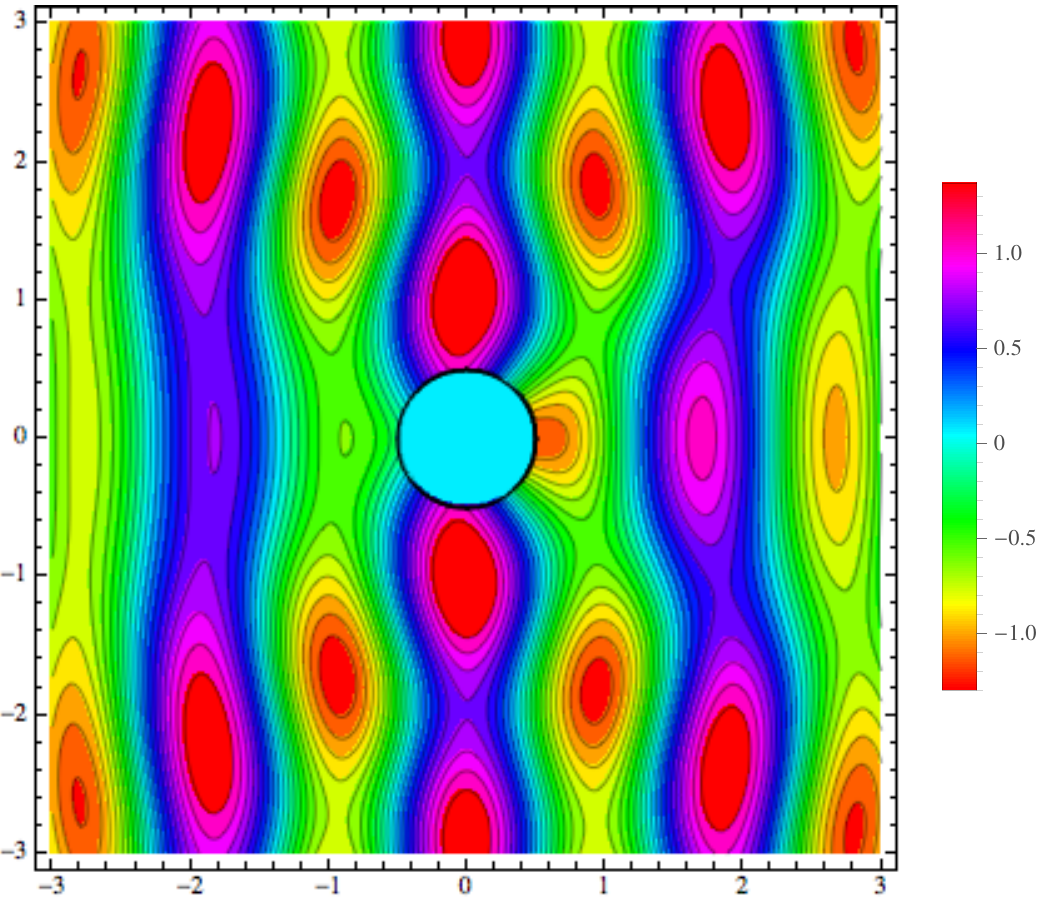}
\includegraphics[width=.33\textwidth]{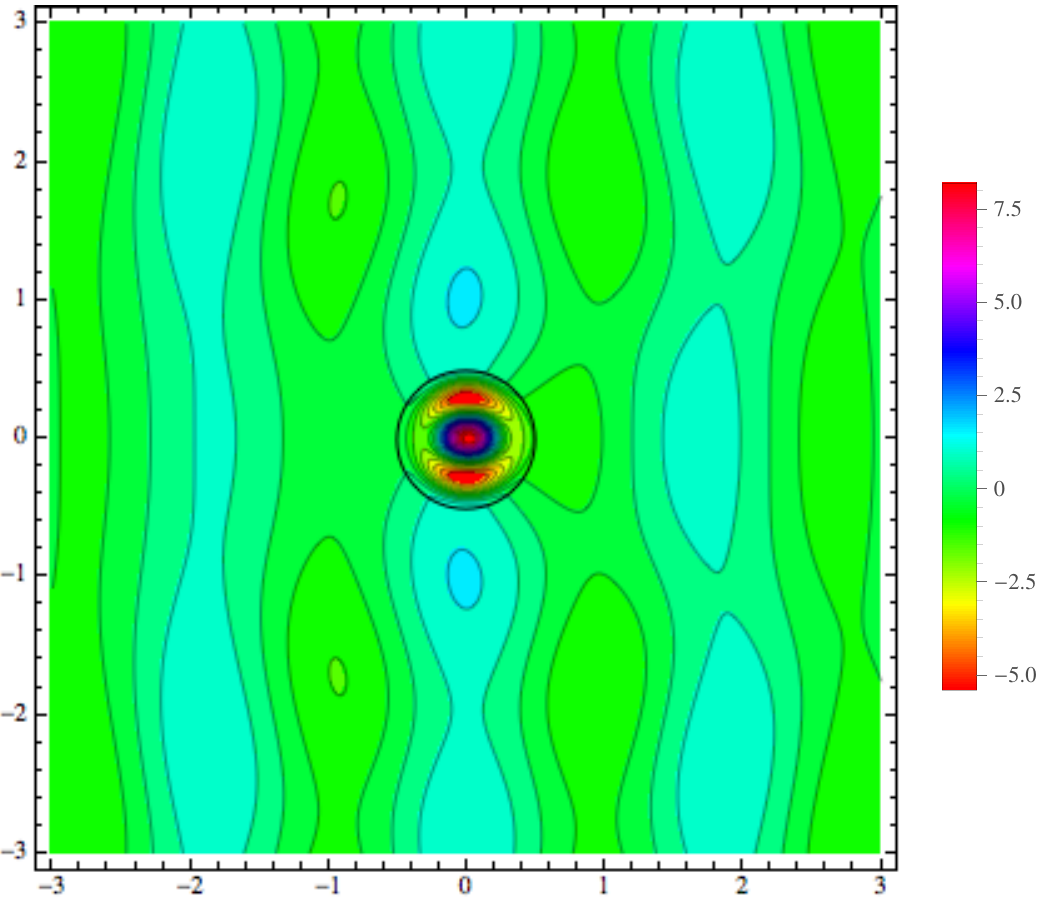} ~~
\includegraphics[width=.33\textwidth]{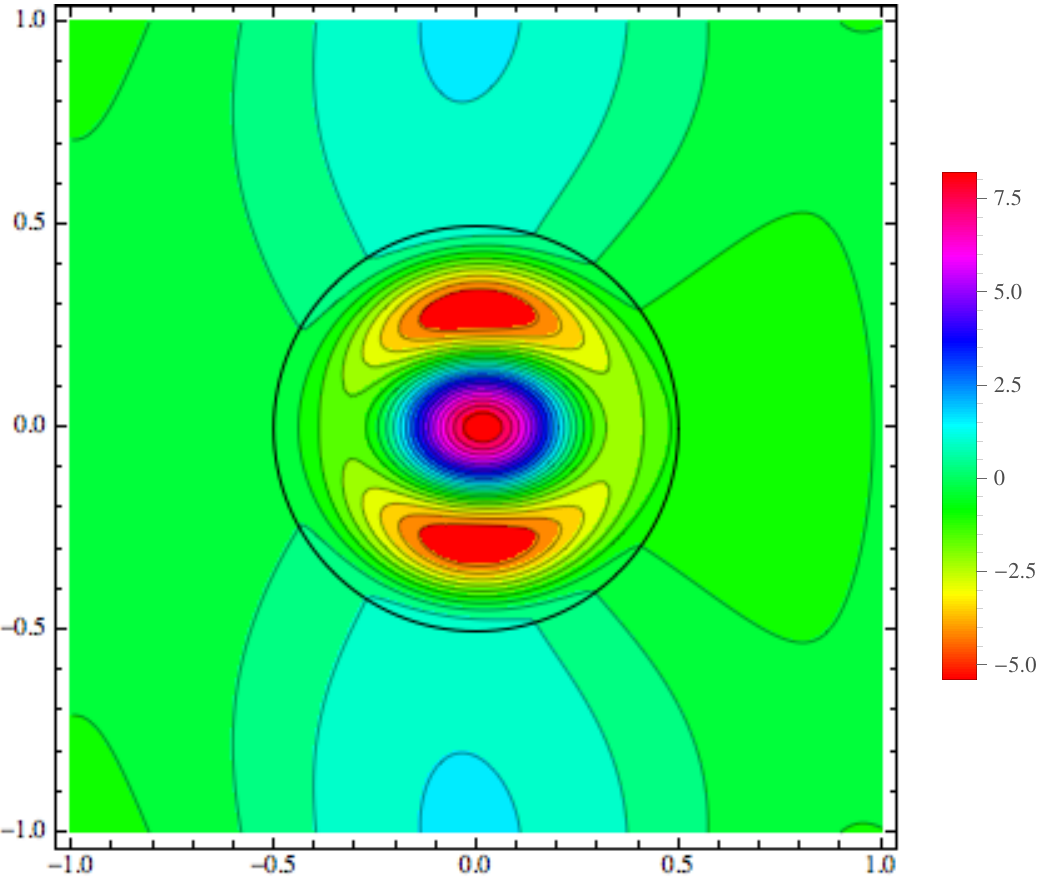}}
\caption{Flexural wave scattering by an uncoated inclusion for the parameters as in Fig. \ref{Flexural_scat_uncoated_inclusion}:
  (a)  Total field for $\omega=0.3$, (b)  Total field for $\omega=11.15$ where $E_0^{(e)}$ vanishes, (c)  Total field for $\omega=11.30$ where $|E_0^{(e)}|$ has a relatively large value. On the right, in each image we present a zoomed-in version of the displacement inside and in the close vicinity of the inclusion.   The first two frames in the inset (b) represent the same field outside the inclusion, but with different colour maps, where the first frame excludes the internal field, whereas the second frame includes it. The same comment applies to the first two frames of the inset (c).  In all contour plots horizontal and vertical axes are the $x_1$- and $x_2$-axes, respectively.}
\label{E0e_Plates_small_min_max}
\end{center}
\end{figure}

For three different frequencies we give flexural wave spatial distributions in Fig. \ref{E0e_Plates_small_min_max}. For the low frequency ($\omega=0.3$) the scattering amplitudes are small. The second frequency considered is $\omega=11.15$, the first non-trivial zero of $|E_0^{(e)}|$ where again vibrational amplitudes are small. Going from $\omega = 11.15$ to $\omega = 11.30$, causes dramatic changes in both the spatial distribution around the inclusion and in its amplitude variation.

\vspace{2cm}

\subsection{Coated inclusion: control of the monopole term in the scattered field}

To obtain a uniform asymptotic formula for $E_0^{(e)}$ in the case of a coated inclusion in a thin plate is a daunting task. Nonetheless, our aim in this section is to make the monopole coefficient $E_0^{(e)}$ asymptotically zero to the leading order, which compared to the previous is somewhat easier. The equation which must to be satisfied in order to make the scattering coefficient zero (to the leadingorder) can be derived as
\begin{eqnarray}
&& \hspace{0.28cm} [2D_e \frak{a}_1 (\frak{b}_1 + 2 D_c - 2 D_e) \beta_e^4  + \frak{a}_1 \frak{b}_1  \frak{a}_2 \beta_c^2 \beta_e^2  - 2 D_c \frak{a}_1 \frak{b}_1 \beta_c^4] \,a_c^4 +[2D_c \beta_c^4 - 2D_i \frak{a}_2 \frak{b}_2 \beta_i^4] \,a_i^4 \nonumber \\
&&+[4 D_e \frak{a}_2 (\frak{b}_2 + D_e) \beta_e^4 + \frak{a}_2 \frak{b}_2^2 \beta_c^2 \beta_e^2 -2D_i \frak{a}_1 \frak{b}_1 \beta_i^4 + 2D_c (\frak{a}_1 \frak{b}_1 - \frak{a}_2 \frak{b}_2) \beta_c^4] \, a_i^2 a_c^2
 = 0,
\label{E0ePlateAsympLeadOrder}
\end{eqnarray}
where
\begin{eqnarray*}
\frak{a}_1 &=& D_i (1+\nu_i) + D_c (1-\nu_c), \quad
\frak{a}_2 = D_i (1+\nu_i) - D_c (1+\nu_c), \\ 
\frak{b}_1 &=& D_c (1+\nu_c) + D_e (1-\nu_e), \quad
\frak{b}_2 = D_c (1-\nu_c) - D_e (1-\nu_e).
\end{eqnarray*}

This fairly unpleasant formula should of course lead us to the result obtained by Konenkov \cite{YuKK} under the assumption of identical material parameter values, for example of the exterior and the coating, hence reducing the problem to the case when there is no coating around the inclusion. Assuming $D_c=D_e, \nu_c=\nu_e, \rho_c = \rho_e$ and thus $\beta_c=\beta_e$, the left-hand side of equation (\ref{E0ePlateAsympLeadOrder}) indeed simplifies to ${\cal F} (\rho_i, \rho_e, \nu_i, \nu_e, D_i, D_e)$ in (\ref{E0ePlateNoCoatingAsympZero}).

It is also important to note that in the very special case of $D_i=D_c=D_e$, $\nu_i=\nu_c=\nu_e$ but $\rho_i \neq \rho_c \neq \rho_e$, equation (\ref{E0ePlateAsympLeadOrder}) reduces to equation (\ref{mass_balance}) which is exactly what is expected. The mass-compensation equation (\ref{mass_balance}) is, in general, inadequate to annul the leading-order part of the monopole term $E_0^{(e)}$.
If one uses the mass-compensation formula unthinkingly, the result may be beneficial or detrimental but in either case the result will, in general, still be of second order in $\beta_e$. Finally, we draw attention to equation (\ref{E0ePlateAsympLeadOrder}) noting that several parameters associated with the coating may be fine tuned to satisfy the equation. This is in stark contrast to the case of membrane waves, where density of the coating was the only parameter available for fine-tuning to minimise the monopole term.

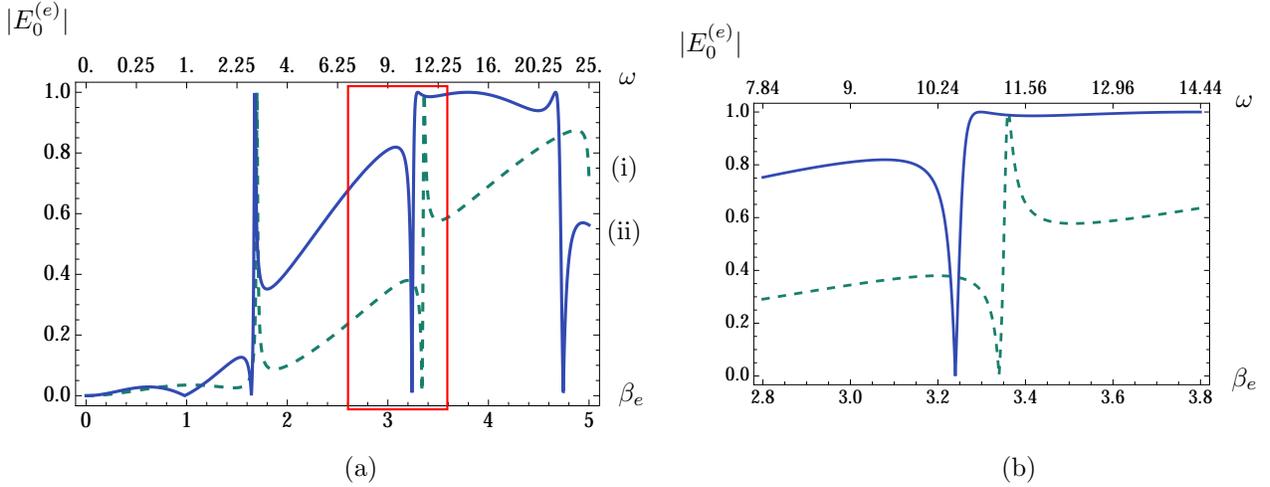
\begin{figure}[H]
\begin{tikzpicture}
\hspace{-0.15cm}
\node[inner sep=0pt] (figleft) at (0.0,0.0)
{\includegraphics[width=.45\textwidth]{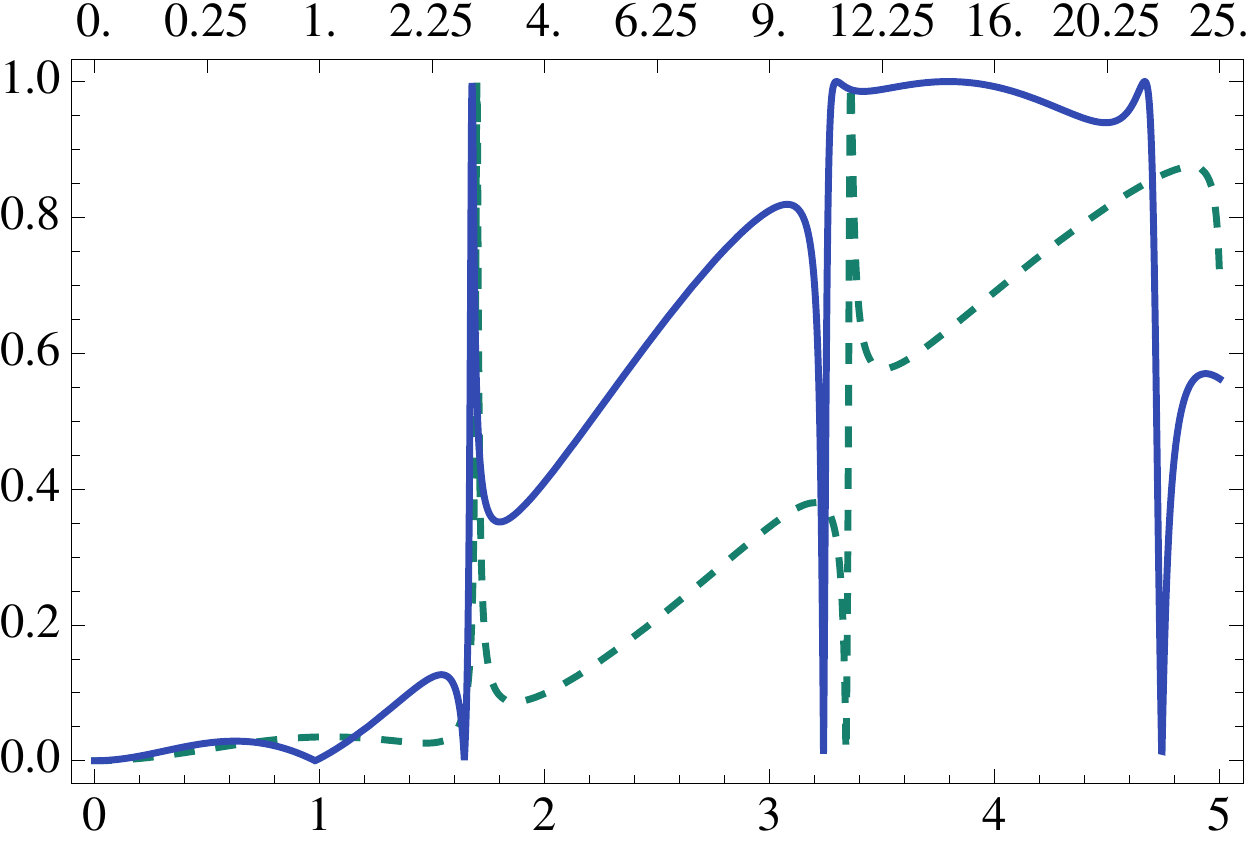}};
\node at (4.1,-2.0) {$\beta_e$};
\node at (4.05,2.2) {$\omega$};
\node at (-3.8,3.0) {$|E_0^{(e)}|$};
\node at (4.0,0.15) {(ii)};
\node at (4.0,1.0) {(i)};
\node at (0.5,-3.0) {(a)};
\hspace{0.25cm}
\node[inner sep=0pt] (figright) at (8.4,0.0)
{\includegraphics[width=.40\textwidth]{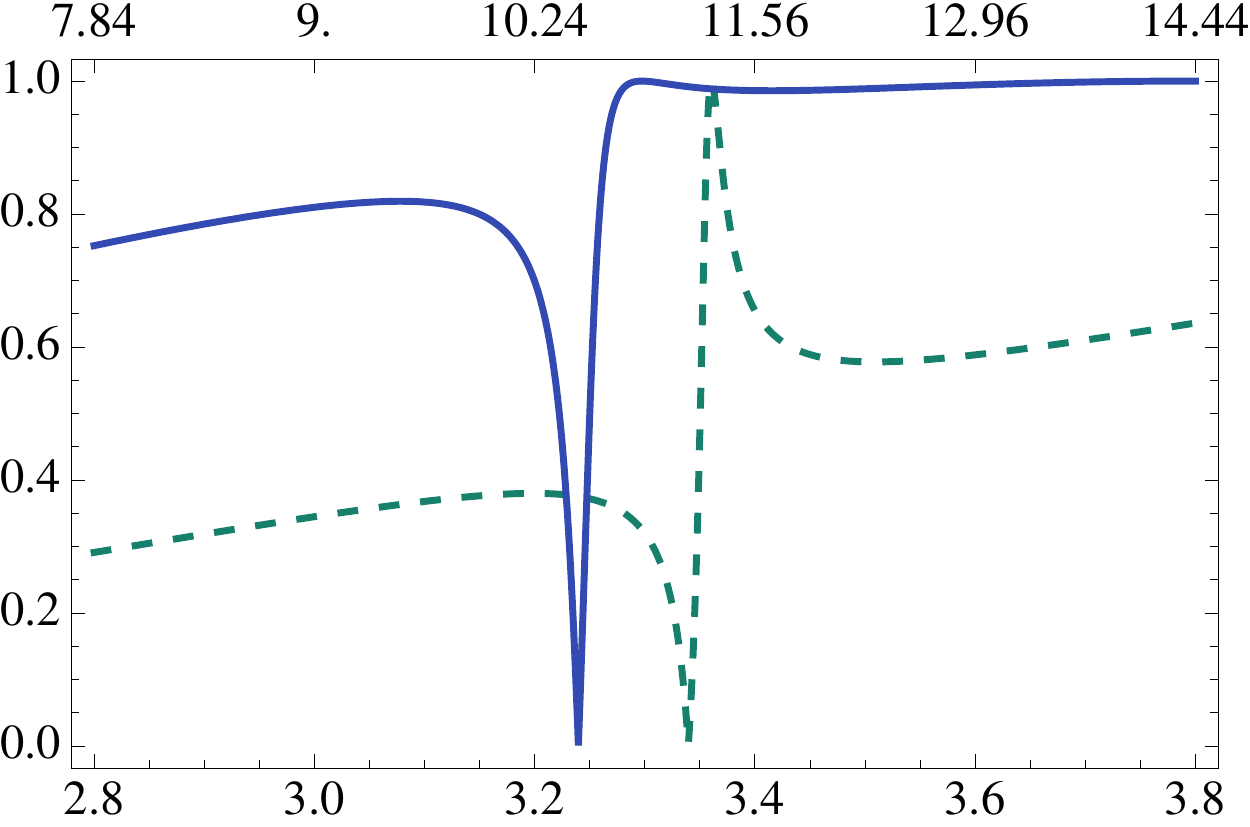}};
\node at (12.00,-1.80) {$\beta_e$};
\node at (12.00,1.90) {$\omega$};
\node at (4.9,2.7) {$|E_0^{(e)}|$};
\node at (9.0,-3.0) {(b)};
\draw [red, thick] (0.08,-2.2) rectangle (1.4,2.1);

\end{tikzpicture}
\caption{
(a) Flexural wave scattering by 
(i) an uncoated inclusion with parameter values: $\rho_i=0.05$, $\rho_e=1.0$, $\nu_i=\nu_e=0.3$, $D_e=1.0$, $D_i=2.5 \times 10^{-4}$, $h=1.0$, $a_i=0.50$ (green/dashed), 
(ii) a coated inclusion with parameter values as in (i) but with material properties of the coating reading:
$\rho_c= 0.005$, $\nu_c=0.3$, $D_c=2.5 \times 10^{-2}$ and $a_c=0.77$ (blue/solid). 
(b) Flexural wave scattering for the parameters of (i) and (ii) in (a), but for a range of $\beta_e$ highlighted by the red box in (a).
We see strong variation in $|E_0^{(e)}|$ for an uncoated inclusion in this range for $\beta_e$, but it is clear that with an appropriate choice of parameter values, we can create a region where $|E_0^{(e)}|$ remains fairly flat.}
\label{E0_coated_inclusion2_Plates}
\end{figure}

\begin{figure}[H]
\begin{center}
\subfigure[]{\label{E0_flat_tot_ex}\includegraphics[width=.33\textwidth]{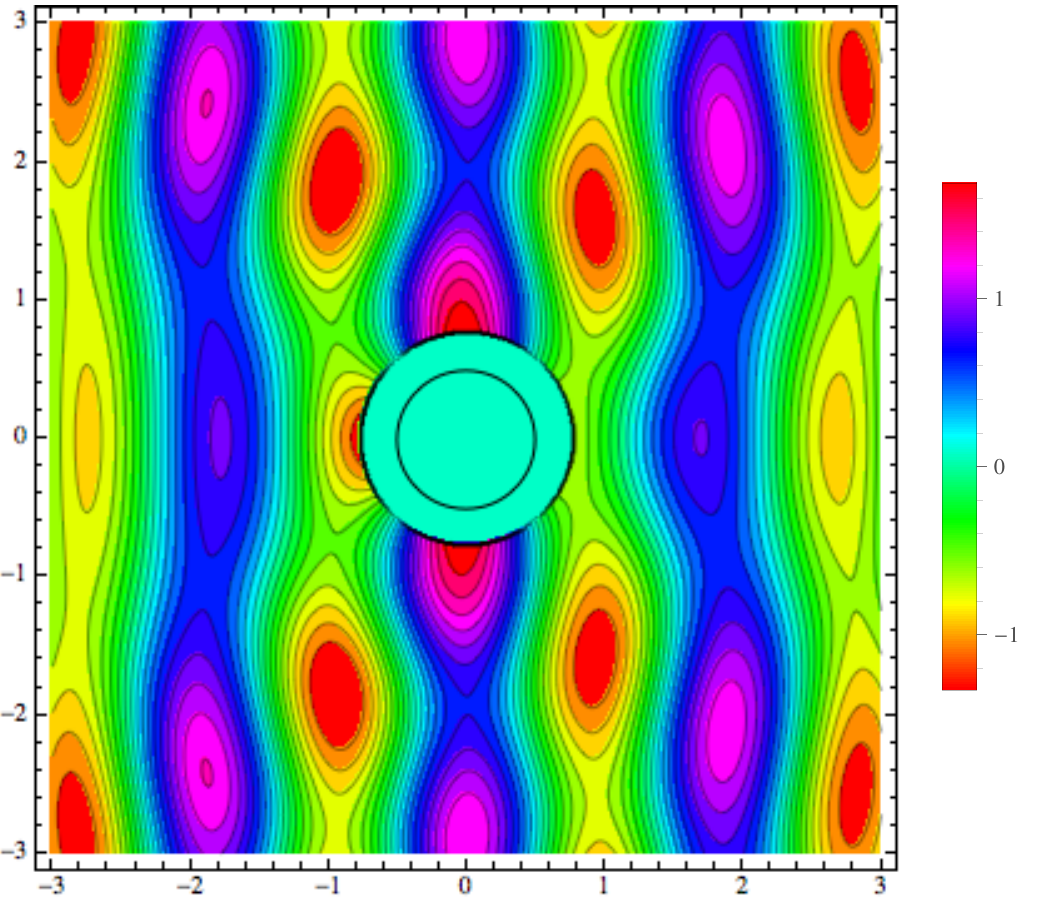}}~~
\subfigure[]{\label{E0_flat_tot_ex}\includegraphics[width=.33\textwidth]{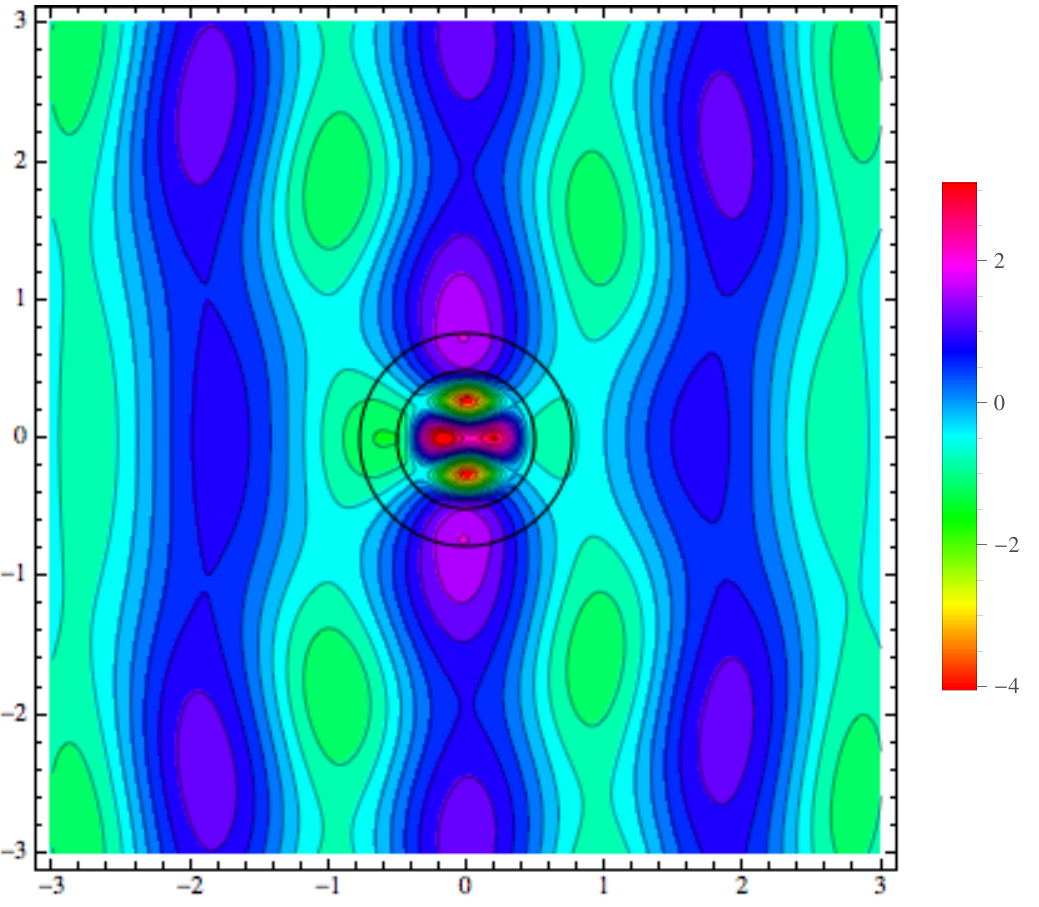}}~~
\subfigure[]{\label{E0_flat_tot_ex_zoom}\includegraphics[width=.33\textwidth]{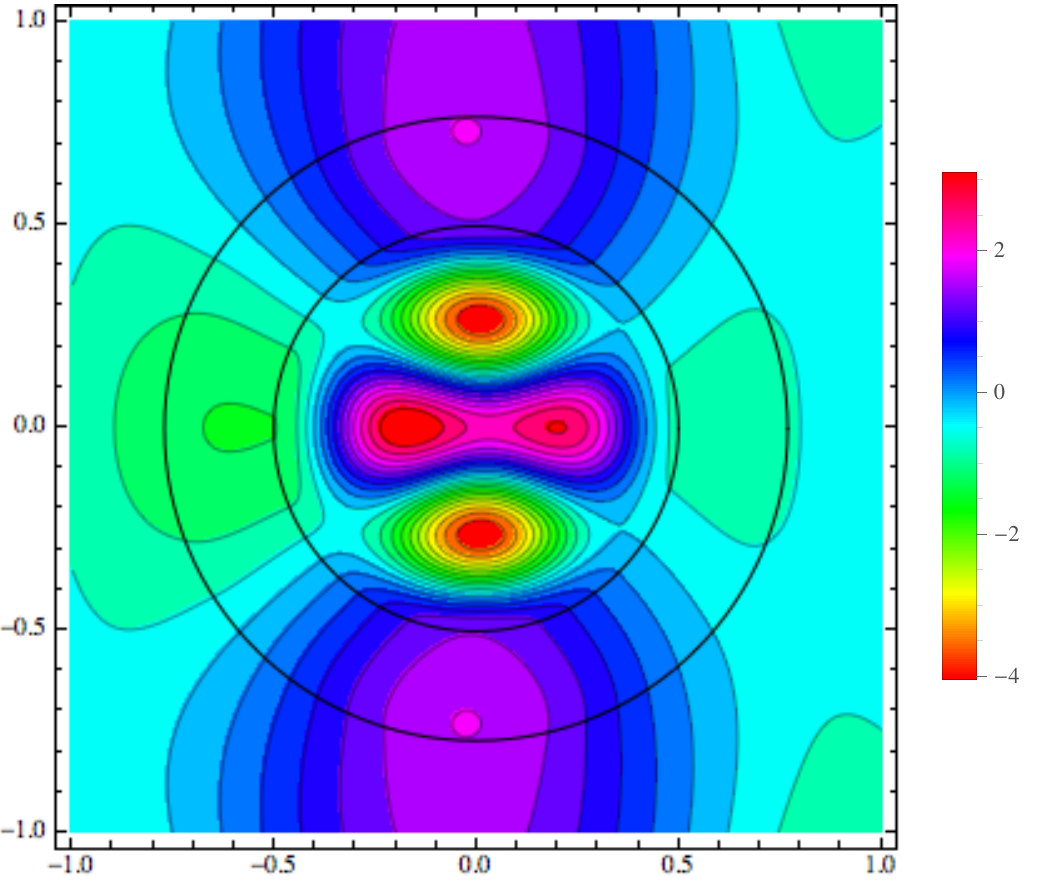}}~~
\caption{Flexural wave scattering: (a) Total displacement field exterior to a coated inclusion with parameter values as in Fig. \ref{E0_coated_inclusion2_Plates} and $\omega= 11.15$. (b) Total displacement field of a coated inclusion with parameter values as in (a) plotted for the whole plate, including the coating and the inclusion. Note that the displacements inside the coating and the inclusion dominate in this case and thus the detail in the exterior is more difficult to observe. (c) A zoomed-in version of the displacement inside the inclusion, the coating and in the close vicinity of the coating. }
\label{E0eflatPlate}
\end{center}
\end{figure}

\begin{figure}[H]
\begin{center}
\begin{tikzpicture}
\node[inner sep=0pt] (figleft) at (0.0,0.0)
{\includegraphics[width=.50\textwidth]{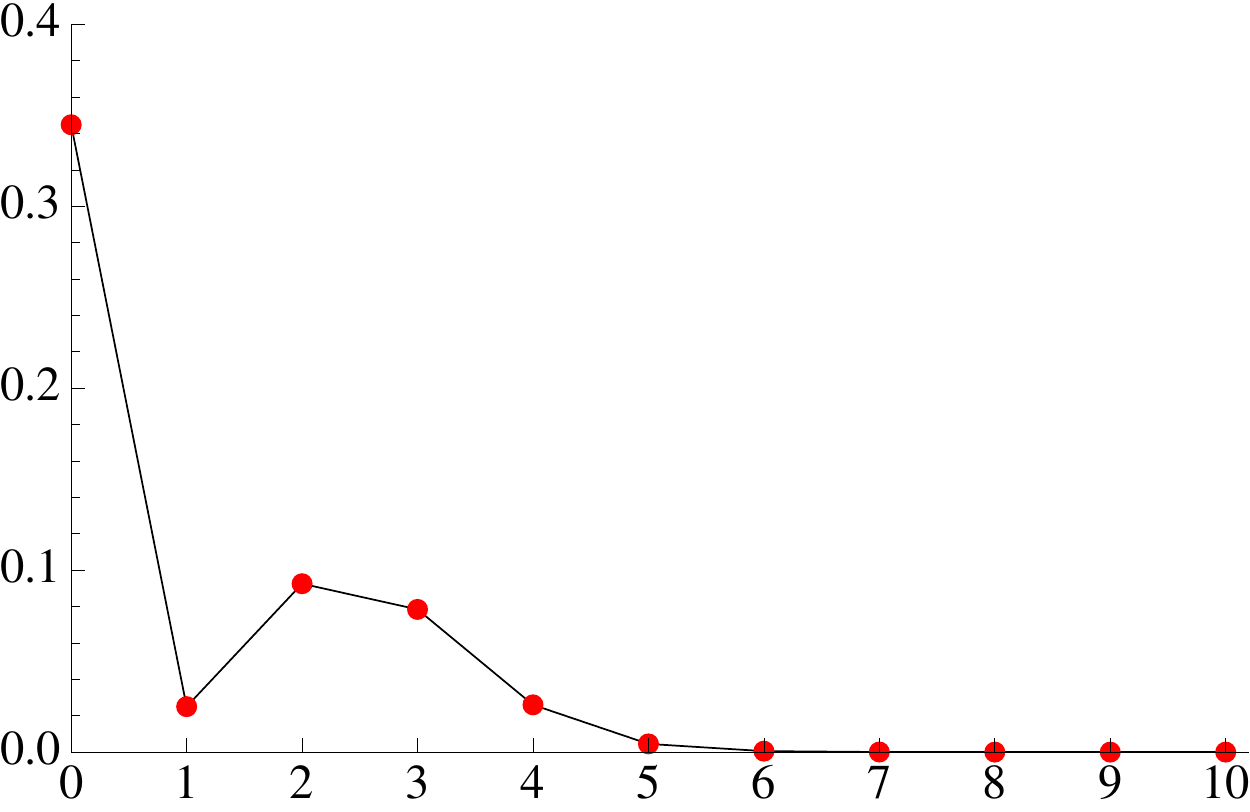}};
\node at (4.4,-2.1) {$n$};
\node at (-3.85,3.0) {$|E_n^{(e)} H_n^{(1)} (\beta_e a_s)|$};
\end{tikzpicture}
\end{center}
\caption{Flexural wave scattering: absolute value of the normalised outgoing wave components $E_n^{(e)} H_n^{(1)} (\beta_e a_s)$ against the multipole order $n$ for $\omega=11.15$. Parameter values as for (blue/solid) curve in Fig. \ref{E0_coated_inclusion2_Plates}: $\rho_i=0.05$,  $\rho_c= 0.005$,   $\rho_e=1.0$,  $\nu_i=\nu_e=\nu_c=0.3$,  $D_i=2.5 \times 10^{-4}$, $D_c=2.5 \times 10^{-2}$, $D_e=1.0$, $h=1.0$,  $a_i=0.5$, $a_c=0.77$, $a_s=1.57$.}
\label{EneVersusnPlate}
\end{figure}

\newpage

\subsection{Active sources: cloaking of flexural waves in resonance regimes}

Here we implement the algorithm of Section \ref{algebsys} for active cloaking in regimes, where the frequency is sufficiently high, so that the monopole coefficient may vary rapidly as a function of frequency.
In this case,  the coating is introduced to provide a smooth behaviour of $E^{(e)}_0$ in the required frequency range. We also comment on the low-frequency regime already discussed above for the case of membrane waves. Explicit analytical representations for amplitudes of active sources are included in 
the Supplementary Material.

\subsubsection{Transition from the low-frequency regime}

As noted in the text above, the scattering produced by the elastic inclusion, in the limit when the wavelength of the incident plane wave tends to infinity compared to the size of the inclusion, is very weak.
In such a regime, the scattering pattern is also dominated by the Helmholtz type waves discussed earlier.

However, as the frequency of the incident wave continues to increase, the role played by the modified Helmholtz waves becomes more important, and we show in this section that a small number of active sources is capable of producing a substantial reduction of the scattered field.

\begin{figure}[H]
\begin{center}
\subfigure[]{\label{foursources_plateA}\includegraphics[width=.33\textwidth]{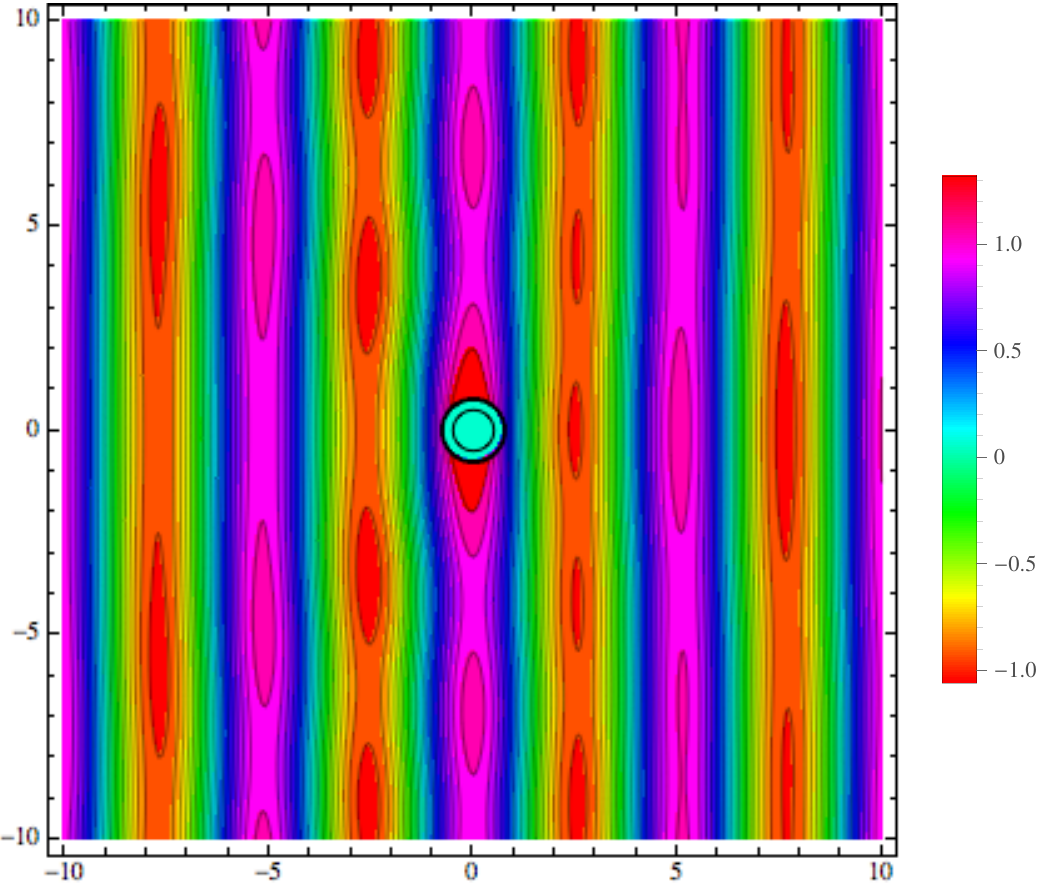}}~~
\subfigure[]{\label{foursources_plateB}\includegraphics[width=.33\textwidth]{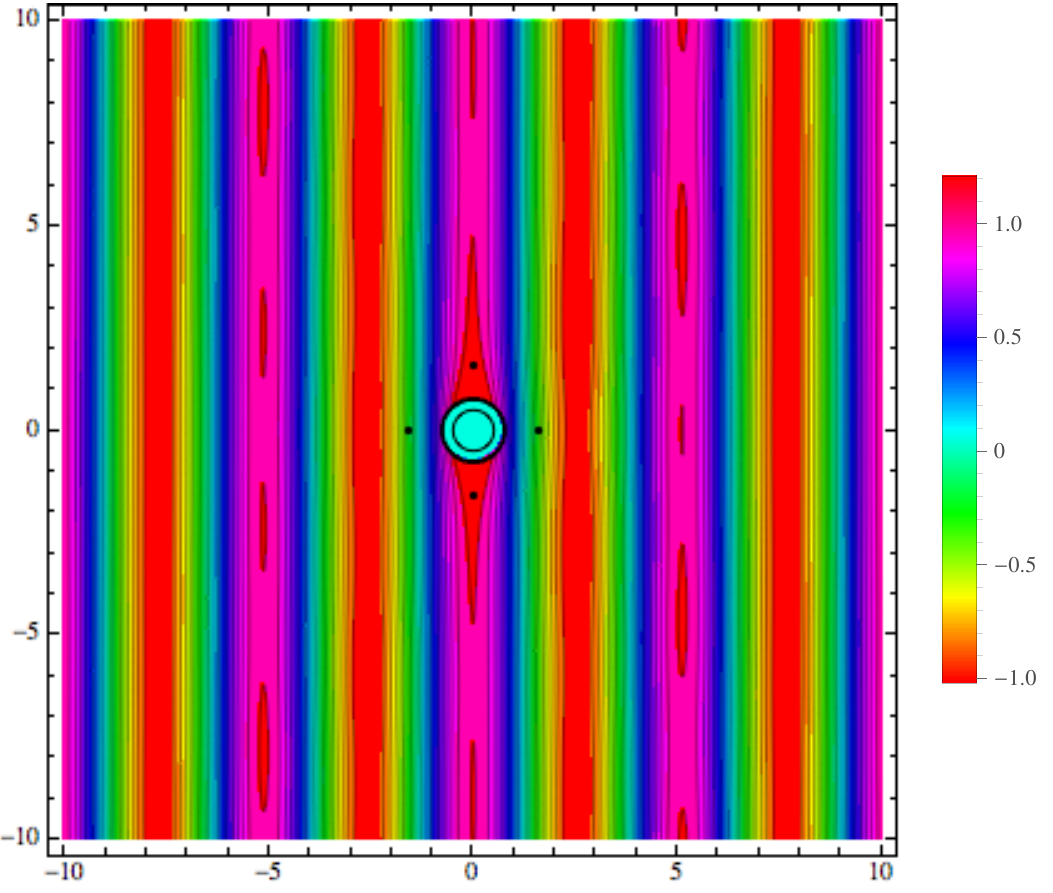}}~~
\caption{ Flexural wave scattering: (a) Total displacement field exterior to a coated inclusion with material properties  as in Fig. \ref{E0_coated_inclusion2_Plates} but at the smaller frequency of $\omega = 1.50$. (b) Total displacement field for the parameter values as in (a) but in the presence of 4 active sources located on the axes (black dots), all 1.57 units away from the origin.}
\label{foursources_plate}
\end{center}
\end{figure}

In Fig. \ref{foursources_plateA}, we present the field plot, where $\omega = 1.50$, and hence $\beta_e = 1.2247$, the wavelength of the incident plane wave is evidently comparable with the diameter of the elastic
inclusion. The geometrical and elastic parameters of the inclusion and the coating are the same as in Fig. \ref{E0eflatPlate}. Figure \ref{foursources_plateA}  shows a non-negligible perturbation of the incident wave by the inclusion. As described in Section \ref{algebsys}, four active sources can be introduced to compensate for the action produced by the coated inclusion on the incident field. The symmetric configuration of sources, shown in Fig. \ref{foursources_plateB}, is used in the simulation, and the amplitudes of the active sources (as outlined in Section \ref{algebsys}) are derived in an explicit closed form (see
Supplementary Material).
The cloaking with four sources, as in Fig. \ref{foursources_plateB}, is quite successful in compensating for the scattering by the coated inclusion for $\omega = 1.50$; beyond this frequency four sources will no longer be adequate.

In the following section, we introduce larger number of sources and demonstrate how the larger-order multipole terms can be suppressed in higher-frequency regimes.

\subsubsection{Higher-frequency regime}

The multipole representation (\ref{total_field}) for the total field is used to produce the numerical data discussed below. The sources are placed on the circle of radius $a_s= 1.57$, outside the coated inclusion, whose  external radius is $a_c= 0.77$.  
The simulations are presented for the cases of $4$, $8$ and $12$ sources, together with  tabular data on the values of high-order multipole coefficients.

\begin{figure}[H]
\begin{center}
\subfigure[]{\label{E0_flat_tot_ex_large_scalePlate}\includegraphics[width=.33\textwidth]{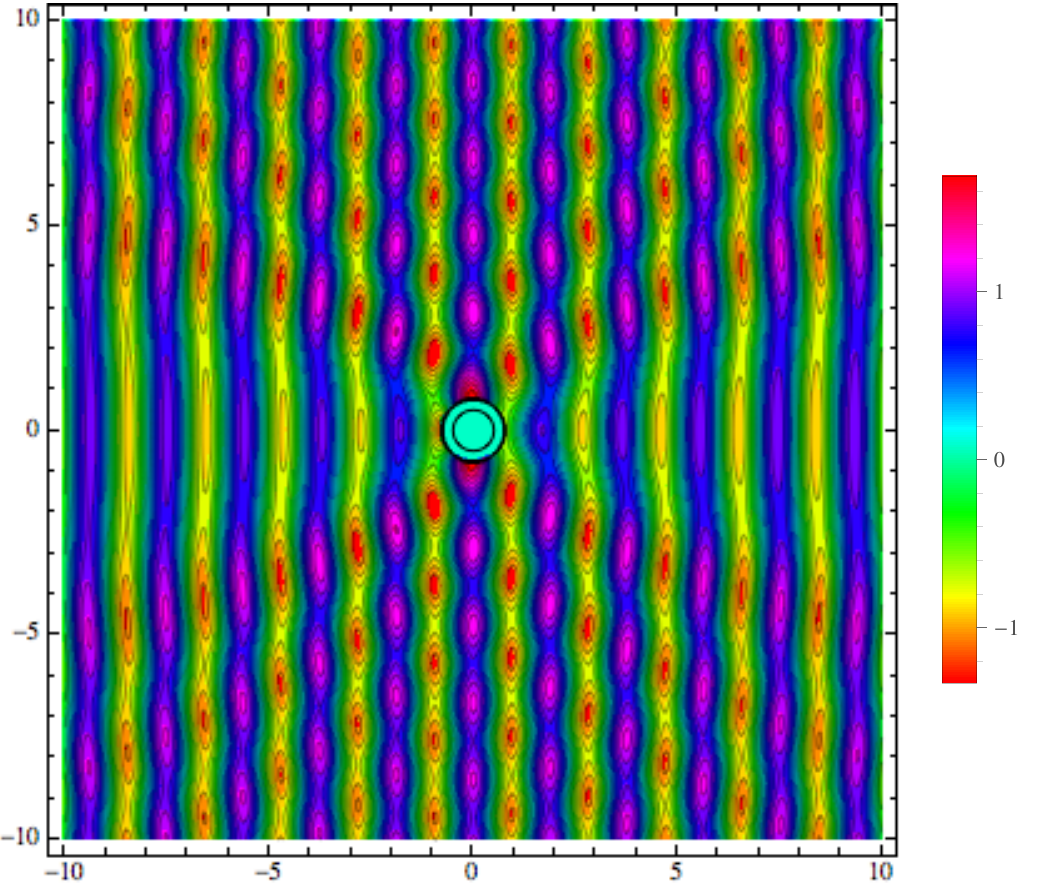}}~~
\subfigure[]{\label{E0_flat_tot_ex_withsources157Plate}\includegraphics[width=.33\textwidth]{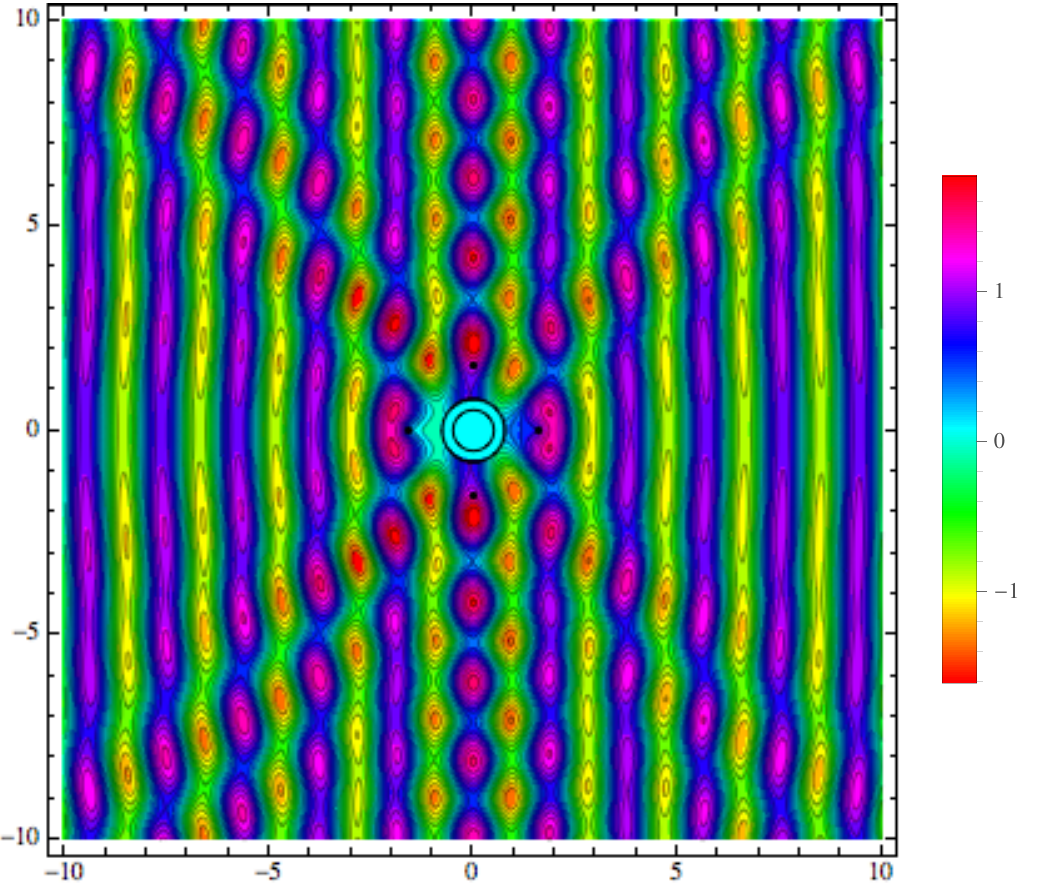}}~~
\subfigure[]{\label{E0_flat_tot_ex_with8sources157Plate}\includegraphics[width=.33\textwidth]{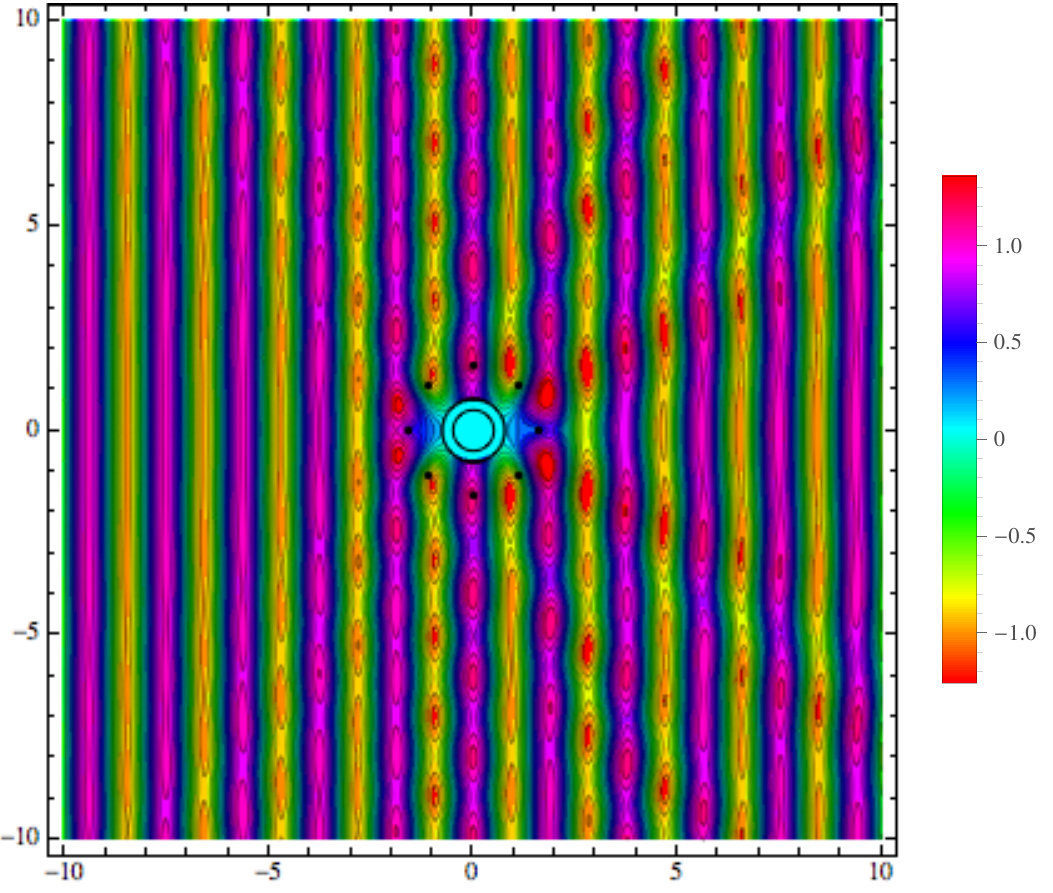}}~~
\caption{Flexural wave scattering: (a) Total displacement field exterior to a coated inclusion with parameter values as in Fig. \ref{E0_coated_inclusion2_Plates}.  (b) Total displacement field in the presence of 4 active sources located on the axes (black dots), 1.57 units away from the origin. (c) Total displacement field in the presence of 8 active sources located  symmetrically 1.57 units away from the origin.}
\label{E0eflatwithsourcesPlate}
\end{center}
\end{figure}

\begin{figure}[H]
\begin{center}
\subfigure[]{\label{Plates_TwelveSources_Exterior}\includegraphics[width=.32\textwidth]{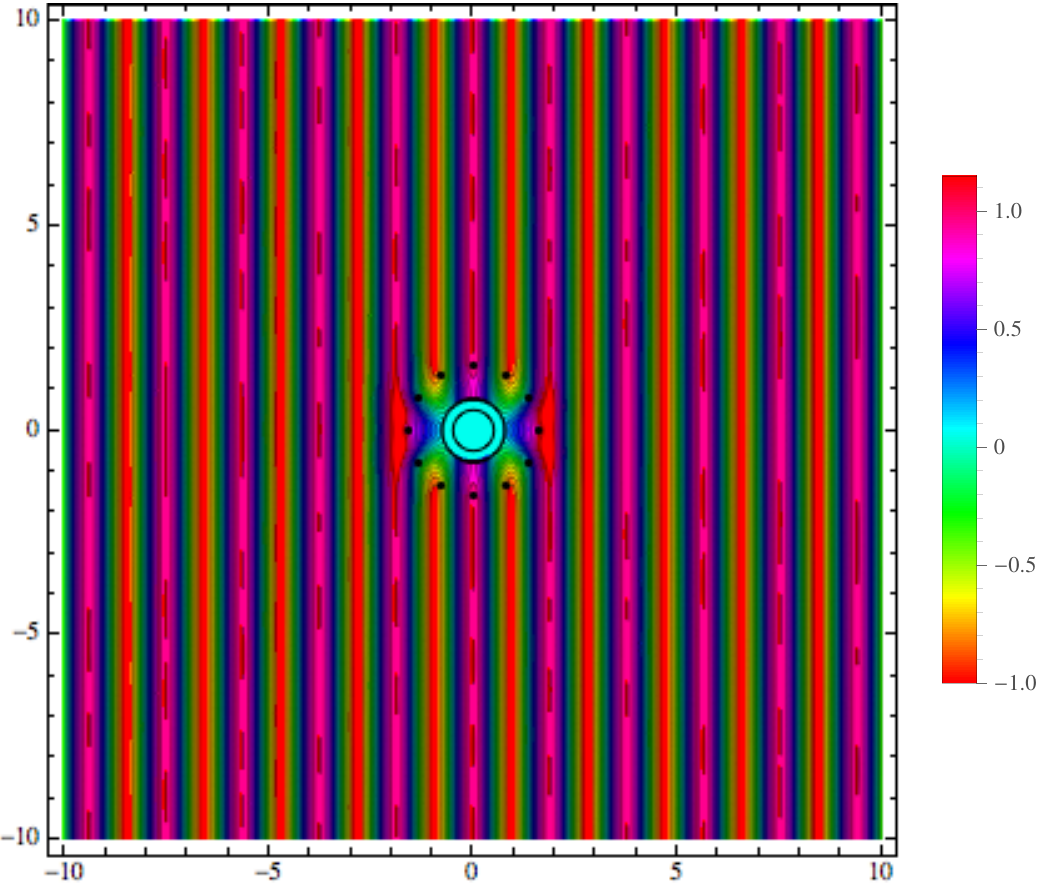}}~~
\subfigure[]{\label{Plates_TwelveSources_TotalField}\includegraphics[width=.32\textwidth]{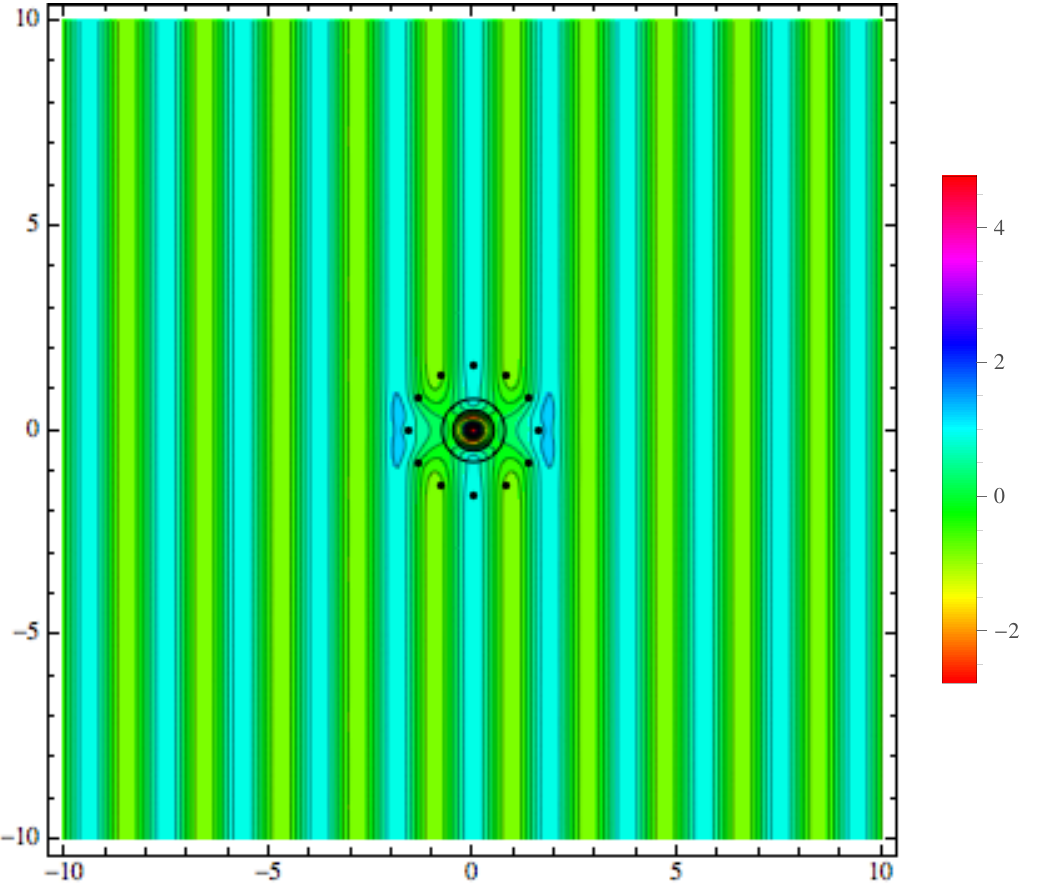}}~~
\subfigure[]{\label{Plates_TwelveSourcesZoom}\includegraphics[width=.32\textwidth]{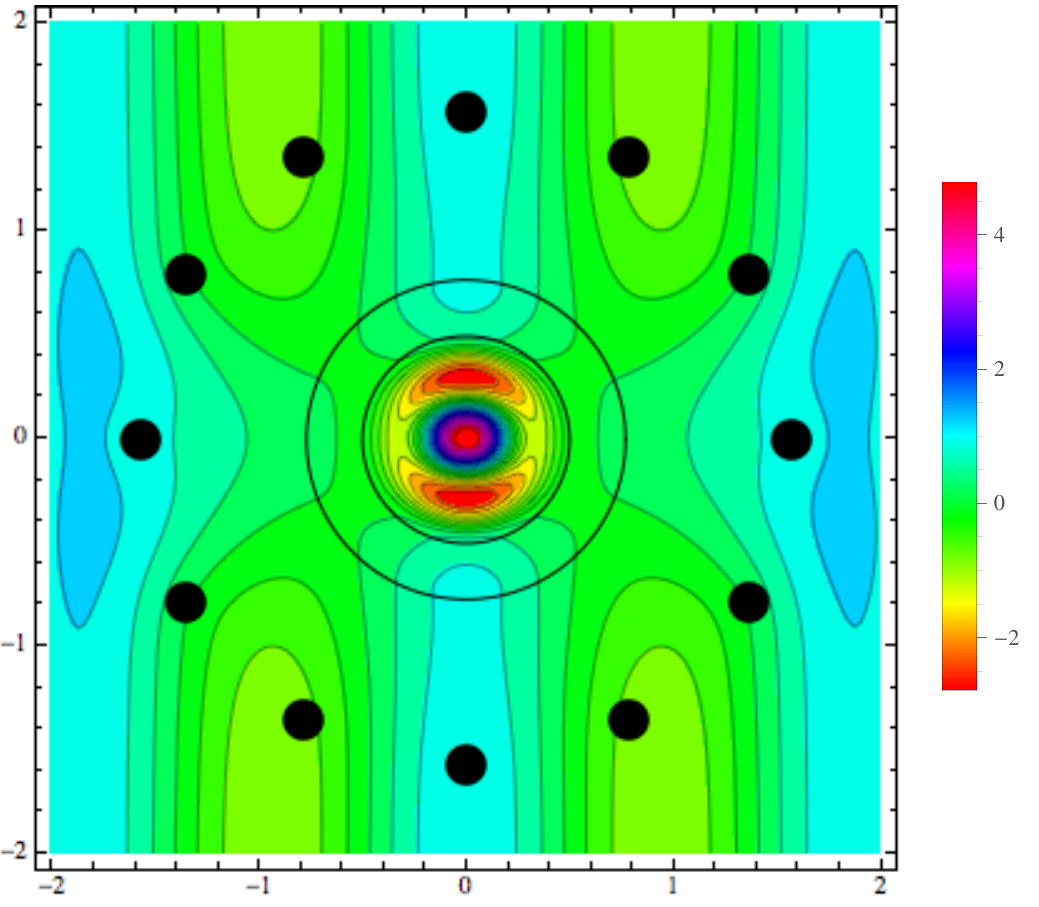}}
\caption{Flexural wave scattering: (a) Total displacement field exterior to a coated inclusion with parameter values as in Fig. \ref{E0_coated_inclusion2_Plates}. (b) Total displacement field in the presence of 12 active sources positioned 1.57 units away from the origin. (c) Same as (b) but over a smaller range of the plate detailing localised fields in the inclusion.}
\label{E0eflatwithsourcesPlateA}
\end{center}
\end{figure}

In Fig. \ref{E0eflatwithsourcesPlate}, we present three cases corresponding to an  incident plane wave scattered by a coated inclusion.  The parameters of the coating are chosen to be the same as
in the simulation depicted in Figure \ref{E0eflatPlate}. The case when active sources are absent (Fig. \ref{E0eflatwithsourcesPlate}a) is compared with the two cases incorporating four and eight active sources, respectively. Although the scattering pattern has been reduced, in this higher-frequency regime it is clear from Fig. \ref{EneVersusnPlate} that around twelve sources are required to make the significant multipole coefficients vanish. The required combination of twelve sources, with the amplitudes computed according to (\ref{fourier_coeff_eqn}) is given in Fig. \ref{E0eflatwithsourcesPlateA}, which shows exemplary quality of cloaking, with the scattered field being suppressed.

It is also noted that Figs. \ref{E0eflatwithsourcesPlateA}a and \ref{E0eflatwithsourcesPlateA}b use a different colour code scheme, so that the diagram (a) plots the field
in the exterior of the ambient matrix only, while the diagrams (b) and (c) show the field in the inclusion, which also indicates the field localisation.

The data in Table \ref{OGW_coeff_plates} for multipole coefficients demonstrates that the core region of multipole orders runs from $n \in [-3,3]$. Our procedure reduces all the amplitudes of targeted orders to zero, while increasing the moduli of adjacent orders. By inspecting the amplitudes of orders $\pm4$ for four sources and $\pm 5$ for eight sources, we can see that in both cases we have an insufficient number of sources for successful cloaking. For twelve sources, both the entries in Table \ref{OGW_coeff_plates} and the displacement field components shown in Fig. \ref{E0eflatwithsourcesPlateA} confirm good quality cloaking.

\begin{table}[H]
\resizebox{6.5in}{!}{
\footnotesize
\begin{tabular}{|c||r|c|c|c|c|} \hline
& $n$ & {\bf no sources} & 4 sources & 8 sources & 12 sources \\ \hline
\multirow{17}{*}{$\tilde{E}_n^{(e)}$}
& $-8$
&
$-3.98\times10^{-12} + 2.00\times10^{-6} \, i$
& 
$1.50\times10^{-7} - 0.075 \, i$
& 
$1.49\times10^{-7} - 0.075 \, i$
& 
$7.66\times10^{-9} - 0.0038 \, i$
\\ \cline{2-6}
& $-7$
&
$-0.000048 - 2.31\times10^{-9} \, i$
& 
$-0.015 - 7.16\times10^{-7}  \, i$
& 
$-0.015 - 7.16\times10^{-7} \, i$
& 
$-0.0022 - 1.05\times10^{-7} \, i$
\\ \cline{2-6}
& $-6$
&
$6.70\times10^{-7} - 0.00082 \, i$
& 
$ 0.00027 - 0.33 \, i$
& 
$ 0.00027- 0.33 \, i$
& 
{\cellcolor{gray!25}$-3.43\times10^{-17} + 5.55\times10^{-17} \,  i$}
\\ \cline{2-6}
& $-5$
&
$0.0092 + 0.000085 \, i$
& 
$ -0.055 - 0.00051  \, i$
& 
$0.24 + 0.0022 \, i$
& 
{\cellcolor{gray!25}$-1.39\times10^{-17} - 8.02\times10^{-17} \,i$}
\\ \cline{2-6}
& $-4$
&
$ -0.0039 + 0.0629 \, i$
&
$ 0.072 - 1.15 \, i$
&
{\cellcolor{gray!25}$-1.96\times10^{-17} + 1.11\times10^{-16} \, i$}
& 
{\cellcolor{gray!25}$ -4.84\times10^{-18} - 1.11\times10^{-16} \, i$}
\\ \cline{2-6}
& $-3$
& 
$-0.20 - 0.043 \, i$
& 
$-0.26 - 0.055 \, i$
&
{\cellcolor{gray!25}$-6.94\times10^{-17} + 1.08\times10^{-16} \, i$}
&
{\cellcolor{gray!25}$ 1.39\times10^{-17} - 1.71\times10^{-18} \, i$}
\\ \cline{2-6}
& $-2$
&
$0.066 - 0.25 \, i$
&  
{\cellcolor{gray!25}$-9.26\times10^{-18} + 8.33\times10^{-17} \, i$}
& 
{\cellcolor{gray!25}$-4.87\times10^{-17} + 1.53\times10^{-16} \, i$}
& 
{\cellcolor{gray!25}$ 4.49\times10^{-18} + 1.25\times10^{-16} \, i $}
\\ \cline{2-6}
& $-1$
& 
$ -0.071 + 0.0051 \, i$
& 
{\cellcolor{gray!25}$ -1.39\times10^{-17} + 4.50\times10^{-18} \, i $}
&
{\cellcolor{gray!25}$-4.16\times10^{-17} + 8.71\times10^{-17} \, i$}
& 
{\cellcolor{gray!25}$-4.16\times10^{-17} +  6.31\times10^{-17} \, i$}
\\ \cline{2-6}
& $0$
& 
$-0.98 - 0.13 i$
& 
{\cellcolor{gray!25}$1.58\times10^{-16} - 5.55\times10^{-17} \, i$}
& 
{\cellcolor{gray!25}$-7.88\times10^{-17}$}
& 
{\cellcolor{gray!25}$-9.27\times10^{-17} - 2.78\times10^{-17} \, i$}
\\ \cline{2-6}
& $1$
& 
$ 0.071 - 0.0051 \, i$
&
{\cellcolor{gray!25}$ 1.39\times10^{-17} - 4.50\times10^{-18} $}
&
{\cellcolor{gray!25}$4.16\times10^{-17} - 8.71\times10^{-17} \, i$}
& 
{\cellcolor{gray!25}$4.16\times10^{-17} -  6.31\times10^{-17} \, i$}
\\ \cline{2-6}
& $2$
& 
$0.066 - 0.25 \, i$
& 
{\cellcolor{gray!25}$-9.26\times10^{-18} + 8.33\times10^{-17}$}
& 
{\cellcolor{gray!25}$-4.87\times10^{-17} + 1.53\times10^{-16} \, i$}
& 
{\cellcolor{gray!25}$ 4.49\times10^{-18} + 1.25\times10^{-16} \, i $}
\\ \cline{2-6}
& $3$
& 
$0.20 + 0.043 \, i$
&
$0.26 + 0.055 \, i$
& 
{\cellcolor{gray!25}$6.94\times10^{-17} - 1.08\times10^{-16} \, i$}
& 
{\cellcolor{gray!25}$ -1.39\times10^{-17} + 1.71\times10^{-18} \, i$}
\\ \cline{2-6}
& $4$
&
$ -0.0039 + 0.0629 \, i$
& 
$ 0.072 - 1.15 \, i$
& 
{\cellcolor{gray!25}$-1.96\times10^{-17} + 1.11\times10^{-16} \, i$}
&
{\cellcolor{gray!25}$ -4.84\times10^{-18} - 1.11\times10^{-16} \, i$}
\\ \cline{2-6}
& $5$
& 
$-0.0092 - 0.000085 \, i$
&
$ 0.055 + 0.00051  \, i$
&
$-0.24 - 0.0022 \, i$
& 
{\cellcolor{gray!25}$1.39\times10^{-17} + 8.02\times10^{-17} \,i$}
\\ \cline{2-6}
& $6$
& 
$6.70\times10^{-7} - 0.00082 \, i$
&
$ 0.00027 - 0.33 \, i$
&
$ 0.00027- 0.33 \, i$
& 
{\cellcolor{gray!25}$-3.43\times10^{-17} + 5.55\times10^{-17} \,  i$}
\\ \cline{2-6}
& $7$
& 
$0.000048 + 2.31\times10^{-9} \, i$
&
$0.015 + 7.16\times10^{-7}  \, i$
&
$0.015 + 7.16\times10^{-7} \, i$
& 
$0.0022 + 1.05\times10^{-7} \, i$
\\ \cline{2-6}
& $8$
& 
$-3.98\times10^{-12} + 2.00\times10^{-6} \, i$
&
$1.50\times10^{-7} - 0.075 \, i$
&
$1.49\times10^{-7} - 0.075 \, i$
& 
$7.66\times10^{-9} - 0.0038 \, i$
\\ \cline{1-6}
\end{tabular}
}
\caption{The coefficients $\tilde{E}_n^{(e)}$ of $H_n^{(1)} (\beta_e r)$ terms for a configuration with zero, four, eight and twelve control sources positioned symmetrically on a circle of radius of 1.57 units from the origin. The frequency of the incident plane wave is $\omega = 11.15$. }
\label{OGW_coeff_plates}
\end{table}

\section{Concluding remarks}

This paper has presented a novel idea of active cloaking  for coated inclusions subjected to incident waves of frequencies ranging from zero up to higher frequency regimes, where the monopole term in the outgoing scattered field can  be a rapidly varying function of frequency. The technique involves choosing the coating to diminish the frequency sensitivity of the scattering problem, and then applying active control sources.

A comparison between the membrane waves and flexural waves in Kirchhoff plates has revealed interesting connections for low frequencies. In particular, we have identified the cases when the flexural wave scattering is driven by the Helmholtz waves. On the other hand, we have also identified important differences where the evanescent waves which are present in the expressions for the solutions to the fourth-order problem contribute significantly.

The utility of the proposed algorithm is high, as the amplitudes of active sources are given in the explicit analytical form, and the accuracy of the cloaking of coated inclusions in resonant regimes 
has been well demonstrated.

\section*{Acknowledgements}

J. O'Neill would like to greatly acknowledge the support from the EPSRC through the grant EP/L50518/1. \"O. Selsil, R.C. McPhedran and N.V. Movchan acknowledge the financial support of the European Commission's Seventh Framework Programme under the contract number PIAPP-GA-284544-PARM-2.  
A.B. Movchan acknowledges the support from EPSRC Programme Grant EP/L024926/1.

\end{document}